\documentclass[pre,twocolumn,amsmath,showpacs,superscriptaddress,floatfix]{revtex4-1}

\usepackage{amsmath}
\usepackage{amsfonts}
\usepackage{amsbsy}
\usepackage{amscd}
\usepackage{amsopn}
\usepackage{amstext}
\usepackage{amsxtra}
\usepackage{epsfig}
\usepackage{multirow}
\usepackage[caption=false]{subfig}

\newcommand{\sref}[1]{\protect\subref{#1}}

\DeclareMathOperator{\vol}{vol}
\DeclareMathOperator{\pr}{pr}

\newcommand{\wbar}{\overline}

\begin{document}

\title{Think Locally, Act Locally: The Detection of Small, Medium-Sized, and \\ Large Communities in Large Networks}

\author{Lucas G. S. Jeub}
\affiliation{Oxford Centre for Industrial and Applied Mathematics, Mathematical Institute, University of Oxford, OX2 6GG, UK}
\author{Prakash Balachandran}
\affiliation{Department of Mathematics \& Statistics, Boston University, Boston, MA 02215, USA}
\author{Mason A. Porter}
\affiliation{Oxford Centre for Industrial and Applied Mathematics, Mathematical Institute, University of Oxford, OX2 6GG, UK}
\affiliation{CABDyN Complexity Centre, University of Oxford, Oxford, OX1 1HP, UK}
\author{Peter J. Mucha}
\affiliation{Department of Applied Physical Sciences, University of North Carolina, Chapel Hill, NC 27599-3216, USA}
\affiliation{Institute for Advanced Materials, Nanoscience \& Technology, University of North Carolina, Chapel Hill, NC 27599-3216, USA}
\author{Michael W. Mahoney}
\affiliation{International Computer Science Institute, Berkeley, CA 94704}
\affiliation{Department of Statistics, University of California at Berkeley, Berkeley, CA 94720}





\begin{abstract}

It is common in the study of networks to investigate intermediate-sized 
(or ``meso-scale'') features to try to gain an understanding of network 
structure and function. 
For example, numerous algorithms have been developed to try to identify 
``communities,'' which are typically construed as sets of nodes with 
denser connections internally than with the remainder of a network.  
In this paper, we adopt a complementary perspective that ``communities'' are 
associated with bottlenecks of locally-biased dynamical processes
that begin at 
seed sets of nodes, and we employ several 
different community-identification procedures (using diffusion-based and geodesic-based dynamics) to investigate community quality as a 
function of community size. 
Using several empirical and synthetic networks, we identify several distinct 
scenarios for ``size-resolved community structure'' that can arise in real 
(and realistic) networks: 
(i) the best small groups of nodes can be {better than} the best large 
groups (for a given formulation of the idea of a good community); 
(ii) the best small groups can have a quality that is {comparable to} the best 
medium-sized and large groups; and
(iii) the best small groups of nodes can be {worse than} the best large 
groups.
As we discuss in detail, which of these three cases holds for a given 
network can make an enormous difference when investigating and making claims 
about network community structure, and it is important to take this into 
account to obtain reliable downstream conclusions. 
Depending on which scenario holds, one may or may not be able to successfully identify 
 ``good'' communities in a given network (and good 
communities might not even exist for a given community quality measure), the 
manner in which different small communities fit together to form meso-scale 
network structures can be very different, and processes such as viral 
propagation and information diffusion can exhibit very different dynamics.
In addition, our results suggest that, for many large realistic 
networks, the output of locally-biased methods that focus on communities 
that are centered around a given seed node might have better conceptual grounding and 
greater practical utility than the output of global 
community-detection methods. 
They also illustrate subtler structural properties that are important to 
consider in the development of better benchmark networks to test methods for 
community detection.
\end{abstract}

\maketitle


\section{Introduction}
\label{sxn:intro}

Many physical, technological, biological, and social systems can be modeled 
as networks, which in their simplest form are represented by graphs.  
A (static and single-layer) graph consists of a set of entities (called 
``vertices'' or ``nodes'') and pairwise interactions (called ``edges'' or 
``links'') between those 
vertices~\cite{Jackson08,EasleyKleinberg10,Newman:2010ur}. 
Graphical representations of data have led to numerous insights in the 
natural, social, and information sciences; and the study of networks has in 
turn borrowed ideas from all of these areas~\cite{Barabasi:2005fy}.

In general, networks can be described using a combination of local, global, 
and ``meso-scale'' perspectives.  
To investigate meso-scale structures---i.e., intermediate-sized structures 
that are responsible for ``coupling'' local properties, such as whether 
triangles close, and global properties such as graph diameter---a fundamental 
primitive in many applications entails partitioning graphs into meaningful 
and/or useful sets of nodes~\cite{Newman:2010ur}.  
The most popular form of such a partitioning procedure, in which one 
attempts to find relatively dense sets of nodes that are relatively sparsely 
connected to other sets, is known as ``community 
detection''~\cite{Porter:2009we,Fortunato:2010iw,Girvan:2002vm}. 
Myriad methods have been developed to algorithmically detect communities~\cite{Porter:2009we,Fortunato:2010iw}; and  
these efforts have led to insights in applications such as committee and 
voting networks in political 
science~\cite{Porter:2005p2163,Mucha:2010p2164,macon2012}, friendship 
networks at universities and other 
schools~\cite{Traud:2012ft,Traud:2011fs,Gonzalez:2007do}, protein-protein 
interaction networks~\cite{Lewis:2010ui}, granular 
materials~\cite{dani-granular2012}, amorphous materials~\cite{zohar2011}, 
brain and behavioral networks in 
neuroscience~\cite{dani2011,wymbs2012,dani2013}, collaboration 
patterns~\cite{evans2011}, human communication 
networks~\cite{Onnela:2007tg,expert2011}, human mobility 
patterns~\cite{Grady:2012iy}, and so on.  

The motivation for the present work is the observation that it can be very 
challenging to find meaningful medium-sized or large communities in large 
networks~\cite{Leskovec:2008vo,Leskovec:2009fy,Leskovec:2010uj}. 
Much of the large body of work on algorithmically identifying communities in 
networks has been applied successfully either to find communities in small 
networks or to find small communities in large 
networks~\cite{Leskovec:2009fy,Porter:2009we,Fortunato:2010iw},
but it has been much less successful at finding meaningful medium-sized and 
large communities in large networks~\footnote{We 
do not wish to tie ourselves too closely to particular numbers, 
because such numbers can vary dramatically in different classes 
of networks.  As a rule of thumb, however, we think of ``small,'' 
``medium-sized,''  and ``large'' communities as those that contain, 
respectively, tens to about a hundred, hundreds to thousands, and 
thousands or more nodes. Similarly, we think of ``small'' networks as 
having tens to hundreds of nodes and ``large'' networks as having 
thousands, tens of thousands, or even more nodes.}. 
There are many reasons that it is difficult to find ``good'' large communities 
in large networks.  
We discuss several such reasons in the following paragraphs.

First, although it is typical to think about communities as sets of nodes with 
``denser'' interactions among its members than between its members and the 
rest of a network, the literature contains neither a consensus definition of 
community nor a consensus on a precise formalization of what constitutes a 
``good'' community~\cite{Porter:2009we,Fortunato:2010iw}. 

Second, the popular formalizations of a ``community'' are computationally 
intractable, and there is little precise understanding or theoretical control 
on how closely popular heuristics to compute communities approximate the exact 
answers in those formulations~\cite{Lancichinetti:2009bb,Good:2010ks}.
Indeed, community structure itself is typically ``defined'' operationally via 
the output of a community-detection algorithm, rather than as the solution 
to a precise optimization problem or via some other mathematically precise 
notion~\cite{Porter:2009we,Fortunato:2010iw}.

Third, many large networks are extremely sparse~\cite{Leskovec:2009fy} and 
thus have complicated structures that pose significant challenges for the 
algorithmic detection of communities via the optimization of objective 
functions~\cite{Good:2010ks}.  
This is especially true when attempting to develop algorithms that scale well 
enough to be usable in practice on large 
networks~\cite{Blondel:2008do,Leskovec:2009fy,Fortunato:2010iw}. 

Fourth, the fact that it is difficult to visualize large networks 
complicates the validation of community-detection methods in such 
networks.  
One possible means of validation is to compare algorithmically-obtained 
communities with known ``ground truth'' communities. 
However, notions of ground truth can be weak in large 
networks~\cite{Leskovec:2009fy,Traud:2011fs,Yang:2012tb}, and one rarely 
possesses even a weak notion of ground truth for most networks.  
Indeed, in many cases, one should not expect a real (or realistic) large 
network to possess a single feature that (to leading order) dominates 
large-scale latent structure in a network. 
Thus, comparing the output of community-detection algorithms to ``ground 
truth'' in practice is most appropriate for obtaining coarse insights into 
how a network might be organized into social or functional groups of 
nodes~\cite{Traud:2012ft}.  
Alternatively, different notions and/or formalizations of ``community'' 
concepts might be appropriate in different 
contexts~\cite{Porter:2009we,Fortunato:2010iw,Lambiotte:2008p2891,Schaub2012,Schaub2012b}, 
so it is desirable to formulate flexible methods that can incorporate 
different perspectives.  

Fifth, community-detection algorithms often have subtle and counterintuitive 
properties as a function of sizes of their inputs and/or outputs.
For example, the community-size ``resolution limit'' of the popular 
modularity objective function is a fundamental consequence of the additive 
form of that objective function, but it only became obvious to people after 
it was explicitly pointed out~\cite{Fortunato:2007js}. 

Motivated by these observations, we consider the question of community 
quality as a function of the size (i.e., number of nodes) of a purported 
community.
That is, we are concerned with questions such as the following. 
(1) What is the relationship between communities of different sizes in a given 
network?  
In particular, for a given network and a given community-quality objective, 
are larger communities ``better'' or ``worse'' than smaller communities?
(2) What is an appropriate way to think about medium-sized and large 
communities in large networks?  
In particular, how do smaller communities ``fit together'' into medium-sized 
and larger communities? 
(3) More generally, what effect do the answers to these questions have on 
downstream tasks that are of primary concern when modeling data using 
networks? 
For example, what effect do they have on processes such as viral propagation 
or the diffusion of information on networks?

By considering a suite of networks and using several related 
notions of community quality, we identify several scenarios that can arise 
in realistic networks.
\begin{enumerate}
\item
\textbf{Small communities are better than large communities.}
In this first scenario, for which there is an upward-sloping \emph{network 
community profile} (NCP; see the discussion below), a network has small 
groups of nodes that correspond more closely than any large groups to 
intuitive ideas of what constitutes a good community.
\item
\textbf{Small and large communities are similarly good or bad.}
In this second scenario, for which an NCP is roughly flat, the most 
community-like small groups of nodes in a network have similar community 
quality scores to the most community-like large groups.
\item
\textbf{Large communities are better than small communities.}
In this third scenario, for which an NCP is downward-sloping, a network has 
large groups of nodes that are more community-like (i.e., ``better'' in some 
sense) than any small groups.
\end{enumerate}
Although the third scenario is the one that has an intuitive isoperimetric 
interpretation and thus corresponds most closely with peoples' intuition 
when they develop and validate community-detection algorithms, one of our 
main conclusions is that most large realistic networks correspond to the 
first or second scenarios.  
This is consistent with recent results on network community structure using 
related approaches~\cite{Leskovec:2008vo,Leskovec:2009fy,Leskovec:2010uj}
as well as somewhat different approaches~\cite{jure2013,Yang:2012tb}, and 
it also helps illustrate the importance of considering community structures 
with groups that have large overlaps.  
For more on this, see our discussions below.

One of the main tools that we use to justify the above observations and to 
interpret the implications of 
community structure in a network is a \emph{network community profile 
(NCP)}, which was originally introduced in Ref.~\cite{Leskovec:2009fy}.
Given a community ``quality'' score---i.e., a formalization of the idea of a
``good'' community---an NCP plots the score of the best community of a given 
size as a function of community size. 
The authors of Ref.~\cite{Leskovec:2008vo,Leskovec:2009fy} considered the 
community quality notion of conductance and employed various algorithms to 
approximate it.  In subsequent work~\cite{Leskovec:2010uj}, many other notions of community 
quality have also been used to compute NCPs.

In the present paper, we compute NCPs using three different 
procedures to identify communities.
\begin{enumerate}
\item
\textbf{Diffusion-based dynamics.}
First, we consider a diffusion-based dynamics 
(called the \textsc{AclCut} method; see the discussion below)
from the original NCP analysis~\cite{Leskovec:2009fy}
that has an interpretation that good communities correspond to bottlenecks 
in the associated dynamics. 
\item
\textbf{Spectral-based optimization.}
Second, we consider
a spectral-based optimization rule 
(called the \textsc{MovCut} method; see below)
that is a locally-biased analog of 
the usual global spectral graph partitioning 
problem~\cite{Mahoney:2012wl}.
\item
\textbf{Geodesic-based dynamics.}
Finally, we consider
a geodesic-based spreading process 
(called the \textsc{EgoNet} method; see the discussion below)
that has an interpretation that 
nodes in a good community are connected by short paths that emanate from 
a seed node~\cite{Bagrow:2005et}. 
\end{enumerate}
We describe these three procedures in more detail in 
Appendix\nobreakspace \ref {sxn:measures}. 
For now, we note that the first and the third procedures have a 
natural interpretation as defining communities operationally as the output 
of an underlying dynamics, and the first and second procedures allow us to 
compare this operational approach with an optimization-based approach.

Viewed from this perspective, the computation of network community structure 
depends fundamentally on three things: actual network structure, the 
dynamics or application of interest, and the initial conditions or network 
region of interest. 
Although there are differences between the aforementioned three 
community-identification methods, these methods all take the perspective 
that a network's community structure depends not only on the connectivity of 
its nodes but also on (1) the region of a large network in which one is 
interested and (2) the application of interest. 
The perspective in point (1) contrasts with the prevalent view of community 
structure as arising simply from network 
structure~\cite{Porter:2009we,Fortunato:2010iw}, but it is consistent with 
the notion of dynamical systems depending fundamentally on their initial 
conditions, and it is crucial in many applications (e.g., both 
social~\cite{christakis2007,fowlerreview} and biological 
contagions~\cite{ccc,kuperman2010,gleeson2013PRX}).  

For example, Facebook's Data Team and its collaborators have demonstrated 
that one can view Facebook as a collection of egocentric networks that have 
been patched together into a network whose global structure is very
sparse~\cite{facebook-egonet,carter-egonet}. 
The above three community-identification methods have the virtue of 
combining the prevalent structural perspective with the idea that one is 
often interested in structure that is located ``near'' (in terms of both network 
topology and edge weights) an exogenously-specified ``seed set'' of 
nodes~\cite{HM13_TR}.  
The perspective in point (2) underscores the fact that one should not expect 
answers to be ``universal.'' 
The differences between the aforementioned three methods lie in the specific 
dynamical processes that underlie them.  
We also note that, although we focus on the measure of community quality 
known as ``conductance'' (which is intimately related to the problem of 
characterizing the mixing rates of random walks~\cite{Jerrum:1988iv}), one can view other 
quality functions (e.g., based on non-conservative 
dynamics~\cite{Lerman:2012cq,ghosh2012,Ghosh:2014wj} or geodesic-based dynamics~\cite{Bagrow:2005et}) as solving other problems, and they thus can reveal different aspects of 
community structure in networks.

The global NCPs that we compute from the three community-identification 
procedures are rather similar in some respects, suggesting that the 
characteristic features of NCPs are actual features of networks and not just 
artifacts of a particular way of sampling local communities. 
However, we observe significant differences in their local behaviors because 
they are based on different dynamical processes. 
In concert with other recent work 
(e.g.,~\cite{localreview,oslom,Schaub2012b}), our results with these three 
procedures suggest that ``local'' methods that focus on finding communities 
centered around an exogenously-specified seed node (or seed set of nodes) 
might have better theoretical grounding and more practical utility than 
other methods for community detection.

Our ``local'' (and ``size-resolved'') perspective on community structure 
also yields several other interesting insights.  
By design, it allows us to discern how community structure depends both on 
the seed node and on the size scales and time scales of a dynamical process 
running on a network.  
Similar perspectives were discussed in recent work on detecting communities 
in networks using Markov 
processes~\cite{Rosvall:2008fi,Lambiotte:2008p2891,Mucha:2010p2164,Lambiotte:2011gk,Schaub2012,Schaub2012b,Ziv:2005hg}, 
and our approach is in the spirit of research on dynamical systems more 
generally, as bottlenecks to diffusion and other dynamics depend 
fundamentally on initial conditions. Local information algorithms are also an important approach for many other optimization problems and for practical purposes such as friend recommendation systems in online social networks \cite{chayes2012}. Moreover, taking a local perspective on community structure is also consistent with the sociological idea of egocentric networks (and 
with real-world experience of individuals, such as users of 
Facebook~\cite{facebook-egonet,carter-egonet}, who experience their personal neighborhood 
of a social network). The local community experienced by a given node should 
be similarly locally-biased, and we demonstrate this feature quantitatively 
for several real networks.  
Using our perspective, we also demonstrate subtle yet fundamental 
differences between different networks: some networks have high-quality 
communities of different sizes (especially small ones), whereas others do 
not possess communities of any size that give bottlenecks to diffusion-based 
dynamics.  
This is consistent with, and helps explain, prior direct observations of 
networks in which algorithmically computed communities seemed to have 
little or no effect on several dynamical processes \cite{Melnik:2011fn}.  
More generally and importantly, whether small or large communities are 
``better'' with respect to some measure of community quality has significant
consequences not only for algorithms that attempt to identify communities 
but also for the dynamics of processes such as viral propagation and 
information diffusion.

The rest of this paper is organized as follows. 
Because our approach to examining network communities is uncommon in the physics literature, we start in Section\nobreakspace \ref {sxn:prelim} with an informal 
description of our approach.  
We then introduce NCPs in Section\nobreakspace \ref {sxn:ncp}.  
In Section\nobreakspace \ref {sxn:main-empirical}, we present our main empirical results 
on community quality as a function of size, and we provide a detailed 
comparison of our three community-identification procedures when applied to 
real networks.
This illustrates the three distinct scenarios of community quality versus 
community size that we described above. 
In Section\nobreakspace \ref {sxn:benchmarks}, we illustrate the behavior of these methods on the well-known LFR benchmark 
networks that are commonly used to evaluate the performance of 
community-detection techniques.  We find that their NCPs have a characteristic shape for a wide range of parameter values and are unable to reproduce the different scenarios that one observes for real networks. 
We then conclude in Section\nobreakspace \ref {sxn:conc} with a discussion of our results. 
In Appendix\nobreakspace \ref {sxn:expanders}, we provide a brief discussion of expander 
graphs (a.k.a. ``expanders''). 
In Appendix\nobreakspace \ref {sxn:measures}, we describe the three specific procedures 
that we use to identify communities in detail.
Appendices\nobreakspace \ref {sxn:NCP-MOV} and\nobreakspace  \ref {sxn:NCP-EGO} contain empirical results for the two methods that we mentioned but did not discuss in detail in Section\nobreakspace \ref {sxn:main-empirical}.


\section{Background and Preliminaries} \label{sxn:prelim}

In this section, we describe some background and 
preliminaries that provide the framework that we use to interpret our results on size-resolved community structure 
in Sections\nobreakspace \ref {sxn:main-empirical} and\nobreakspace  \ref {sxn:benchmarks}.
 We start in Section\nobreakspace \ref {sxn:prelim-notation} by defining the notation that we use throughout this paper, and we continue in  Section\nobreakspace \ref {sxn:prelim-looklike} with a brief discussion 
of possible ways that a network might ``look like'' if one is interested in 
its meso-scale or large-scale structure. To convey the basic idea of our approach, much of 
our discussion in this section is informal. In later sections, we will make these ideas more precise.


\subsection{Definitions and Notation}
\label{sxn:prelim-notation}

We represent each of the networks that we study as an undirected graph.  We consider both weighted and unweighted graphs.  

Let $G=(V,E,w)$ be a connected and undirected graph with node set $V$, edge set $E$, and set $w$ of weights on the edges.  Let $n=|V|$ denote the number of nodes, and let $m=|E|$ denote the number of edges. The edge $\{i,j\}$ has weight $w_{ij}$.
Let $A = A_G \in \mathbb{R}^{n \times n}$ denote the
(weighted) adjacency matrix of $G$.  Its components are $A_G(i,j)=w_{ij}$ if $\{i,j\}\in E$ and $A_G(i,j)=0$ otherwise. The matrix $D = D_G \in \mathbb{R}^{n \times n}$ denotes the diagonal degree matrix of $G$.  Its components are $D_G(i,i)=d_i = \sum_{\{i,j\} \in E} w_{ij}$, where $d_i$ is called the ``strength'' or ``weighted degree'' of node $i$.  The combinatorial Laplacian of $G$ is  $L_G = D_G-A_G$, and the normalized Laplacian of $G$ is $\mathcal{L}_G = D_G^{-1/2}L_GD_G^{-1/2}$. 

A path $P$ in $G$ is a sequence of edges 
$P=\left\{\{i_k,j_k\}\right\}_{k=1}^s$, such that $j_k=i_{k+1}$ for 
$k=1,\ldots,s-1$. The length of path $P$ is 
$|P|=\sum_{\{i,j\}\in P} l_{ij}$, where $l_{ij}$ is the length of the edge that connects nodes $i$ and $j$. For an unweighted network, $l_{ij}=1$ for all edges. For weighted networks, $w_{ij}$ is a measure of closeness of the tie between nodes $i$ and $j$, a common choice for $l$ is $l_{ij}=\frac{1}{w_{ij}}$. Let $\mathcal P_{ij} $ be the set of all paths between $i$ and $j$. The geodesic distance $\Delta_{ij}=\min_{P\in \mathcal P_{ij}} |P|$ between nodes $i$ and $j$ is the length of a shortest path between $i$ and $j$. The $k$-neighborhood $N_k(i)=\{j\in V \colon \Delta_{ij}\leq k\}$ of $i$ is the set of all nodes that are at most a distance $k$ away from $i$, and the $k$-neighborhood of a set of nodes $S$ is $N_k(S)=\bigcup_{i\in S} N_k(i)$.

\subsection{What Can Networks ``Look Like''?}
\label{sxn:prelim-looklike}

Before examining real networks, we start with the following question: What are possible ways 
that a network can ``look like,'' very roughly if one ``squints'' at it?
This question is admittedly vague, but the answer to it governs how 
small-scale network structure ``interacts'' with large-scale network 
structure, and it informs researchers' intuitions and the design decisions 
that they make when analyzing networks (and when developing methods to analyze networks).
As an example of this idea, it should be intuitively clear that if one 
``squints'' at the nearest-neighbor $\mathbb{Z}^{2}$ network (i.e., the uniform lattice of pairs of integers on the 
Euclidean plane), then they ``look like'' the Euclidean plane $\mathbb{R}^{2}$.
Distances are approximately preserved, and up to boundary conditions
and discretization effects, dynamical processes on one approximate the 
analogous dynamic processes on the other.
In the fields of geometric group theory and coarse geometry, this intuitive connection 
between $\mathbb{Z}^{2}$ and $\mathbb{R}^{2}$ has been made precise using 
the notions of coarse embeddings and quasi-isometries~\cite{Bridson:1999tu}.

\begin{figure}[tb]
\subfloat[Low-dimensional structure]{
\label{fig:stylized-hotdog}
\includegraphics[width=0.22\linewidth]{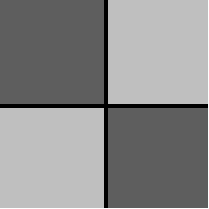}
\includegraphics[width=0.22\linewidth]{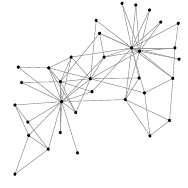}
}
\hfill
\subfloat[Core-periphery structure]{
\label{fig:stylized-coreper}
\includegraphics[width=.22\linewidth]{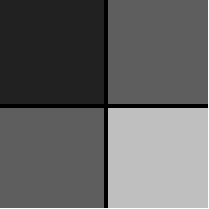}
\includegraphics[width=.22\linewidth]{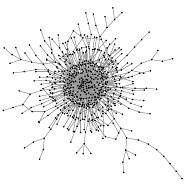}
}\\

\subfloat[Expander or complete graph]{
\label{fig:stylized-expander}
\includegraphics[width=.22\linewidth]{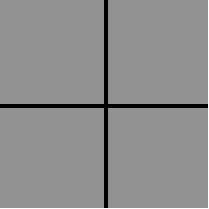}
\includegraphics[width=.22\linewidth]{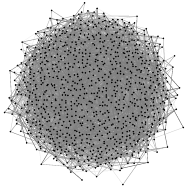}
}
\hfill
\subfloat[Bipartite structure]{
\label{fig:stylized-bipartite}
\includegraphics[width=.22\linewidth]{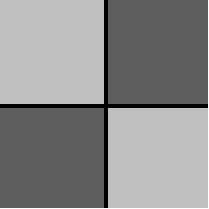}
\includegraphics[width=.22\linewidth]{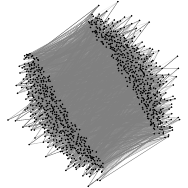}
}
\caption{Idealized block models of network adjacency matrices; darker blocks 
correspond to denser connections among its component nodes.
Figure\nobreakspace \ref {fig:stylized-hotdog} illustrates a low-dimensional ``hot dog'' or
``pancake'' structure; 
Fig.\nobreakspace \ref {fig:stylized-coreper} illustrates a ``core-periphery'' structure;
Fig.\nobreakspace \ref {fig:stylized-expander} illustrates an unstructured expander or 
complete graph; and 
Fig.\nobreakspace \ref {fig:stylized-bipartite} illustrates a bipartite graph. 
Our example networks are the Zachary Karate Club~\cite{Zachary:1977tj} in Fig.\nobreakspace \ref {fig:stylized-hotdog} and a realization of a 
random-graph block model in Figs.\nobreakspace  \ref {fig:stylized-coreper}--\ref {fig:stylized-bipartite}. For Fig.\nobreakspace \ref {fig:stylized-coreper} we only show the largest connected component (LCC), whereas the networks in Figs.\nobreakspace \ref {fig:stylized-expander} and\nobreakspace  \ref {fig:stylized-bipartite} are connected.
The parameters for the block models are as follows: 
(b) $\alpha_{11}=0.3$, $\alpha_{22}=0.001$, $\alpha_{12}=0.005$, $n_1=50$ nodes, and $n_2=950$ nodes (the LCC has 615 nodes);
(c) $\alpha_{11}=\alpha_{22}=\alpha_{12}=0.01$, and $n_1+n_2=1000$ nodes;
(d) $\alpha_{11}=\alpha_{22}=0$, $\alpha_{12}=0.02$, and $n_1=n_2=500$ nodes.
}
\label{fig:stylized}
\end{figure}

Establishing quasi-isometric relationships on networks that are 
expander graphs (a.k.a. ``expanders''; see Appendix\nobreakspace \ref {sxn:expanders}) is technically brittle~\cite{Chen:2012wg}.  Thus, for the present
informal discussion, we rely on a simper notion.
Suppose that we are interested in the ``best fit'' of the adjacency matrix $A$ 
to a $2 \times 2$ block matrix:
\begin{align*}
A = \left( \begin{array}{cc}
            A_{11} & A_{12}^{T} \\
            A_{12} & A_{22} \end{array} 
    \right) \,,
\end{align*}
where $A_{ij} = \alpha_{ij} \vec{1}\vec{1}^T$, where the ``1-vector'' $\vec{1}$ is a column vector of the appropriate dimension that contains a $1$ in every entry and $\alpha_{ij}\in \mathbb{R}^{+}$.  Thus,  each block in $A$ has uniform values for all its elements, and larger values of $\alpha_{ij}$ correspond to stronger interactions between nodes.  The structure of $A$ is then determined based on the relative sizes of $\alpha_{11}$, $\alpha_{12}$, and $\alpha_{22}$.  The various relative sizes of these three scalars have a strong bearing on the structure of the network associated with $A$.
We illustrate several examples in Fig.\nobreakspace \ref {fig:stylized}. For the block models that we use for three of its panels, one block has $n_1$ nodes and the second block has $n_2$ nodes, and a node in block $i$ is connected to a node in block $j$ with probability $\alpha_{ij}$~\cite{block-model}.
\begin{itemize}
\item
\textbf{Low-dimensional structure.}
In Fig.\nobreakspace \ref {fig:stylized-hotdog}, we illustrate the case in which 
$\alpha_{11} \approx \alpha_{22} \gg \alpha_{12}$.
In this case, each half of the network interacts with itself 
more densely than it interacts with the other half of the network.
This ``hot dog'' or ``pancake'' structure corresponds to the situation in which 
there are two (or any number, in the case of networks more generally) dense communities of nodes that are reasonably well-balanced in the sense that each community has roughly the same number of nodes. In this case, the network embeds relatively well in a one-dimensional, 
two-dimensional, or other low-dimensional space. Spectral clustering or other
clustering methods often find meaningful communities in such networks, and one can often readily construct meaningful and interpretable visualizations of network structure.
\item
\textbf{Core-periphery structure.}
In Fig.\nobreakspace \ref {fig:stylized-coreper}, we illustrate the case in which
$\alpha_{11} \gg \alpha_{12} \gg \alpha_{22}$. 
This is an example of a network with a density-based ``core-periphery'' 
structure~\cite{cp-review,Borgatti:2000cq,Rombach:2012uy,Leskovec:2008vo,Leskovec:2009fy}. 
In these cases, there is a core set of nodes that are relatively 
well-connected amongst themselves as well as to a peripheral set of nodes 
that interact very little amongst themselves.
\item
\textbf{Expander or complete graph.}
In Fig.\nobreakspace \ref {fig:stylized-expander}, we illustrate the case in which
$\alpha_{11} \approx \alpha_{12} \approx \alpha_{22}$.
This corresponds to a network with little or no discernible structure.
For example, if $\alpha_{11} = \alpha_{12} = \alpha_{22} = 1$, then the 
graph is a clique (i.e., the complete graph).
Alternatively, if the graph is a constant-degree expander, then
$\alpha_{11} \approx \alpha_{12} \approx \alpha_{22} \ll 1$.
As discussed in Appendix\nobreakspace \ref {sxn:expanders}, constant-degree expanders 
are the metric spaces that embed least well in low-dimensional Euclidean 
spaces. In terms of the idealized block model in Fig.\nobreakspace \ref {fig:stylized},
 they ``look like'' complete graphs, and partitioning them would not yield network structure that one should expect to construe as meaningful.  Informally, they are largely unstructured 
when viewed at large size scales. 
\item
\textbf{Bipartite structure.}
In Fig.\nobreakspace \ref {fig:stylized-bipartite}, we illustrate the case in which 
$\alpha_{12} \gg \alpha_{11} \approx \alpha_{22}$.
This corresponds to a bipartite or nearly-bipartite graph.  Such networks arise, e.g., 
when there are two different types of nodes, such that one type of node connects only to (or predominantly to) nodes of the other type~\cite{Newman:2006hj}.
\end{itemize}

Most methods for algorithmic detection of communities have been developed and 
validated using the intuition that networks have some sort of 
low-dimensional structure \cite{Porter:2009we,Leskovec:2009fy,jure2013}.  
As an example, consider the infamous Zachary Karate Club 
network~\cite{Zachary:1977tj}, which we show in Fig.\nobreakspace \ref {fig:stylized-hotdog}. 
This well-known benchmark graph, which seems to be an almost obligatory 
example to discuss in papers that discuss community 
structure \cite{puck-karate,tumble}, clearly ``looks like'' it has a nice low-dimensional structure.
For example, there is a clearly identifiable left half and right half, and 
two-dimensional visualizations of the network (such as that 
in Fig.\nobreakspace \ref {fig:stylized-hotdog}) highlight that bipartition.
Indeed, the Zachary Karate Club network possesses well-balanced and (quoting Herbert 
Simon \cite{Simon:1962vg}) ``nearly decomposable'' communities; and the 
nodes in each community are more densely connected to nodes in the same 
community than they are to nodes in the other community.  
Relatedly, reordering the nodes of the Zachary Karate Club appropriately yields an 
adjacency-matrix representation with an almost block-diagonal structure with two blocks 
(as typified by the cartoon in Fig.\nobreakspace \ref {fig:stylized-hotdog}); and any 
reasonable community-detection algorithm should be able to find (exactly 
or approximately) the two communities.

Another well-known network that (slightly less obviously) ``looks like'' it has a 
low-dimensional structure is a so-called caveman network, which we 
illustrate later (in Fig.\nobreakspace \ref {fig:ncp-cavemangraph}).  
Arguably, a caveman network has many more communities than the Zachary Karate Club, so details such as whether an algorithm ``should'' split it into two or a somewhat larger 
number of reasonably well-balanced communities might be different than in 
the Zachary Karate Club network. However, a caveman network also has a natural well-balanced partition that 
respects intuitive community structure. Reasonable two-dimensional 
visualizations of this network (such as the one that we present in Fig.\nobreakspace \ref {fig:ncp-cavemangraph}) 
shed light on that structure; and any reasonable community-detection 
algorithm can be adjusted to find (exactly or approximately) the expected 
communities. In this paper, we will demonstrate that most realistic networks do \emph{not} ``look like'' these small examples. Instead, realistic networks are often poorly-approximated by low-dimensional structures (e.g., with a small number of relatively well-balanced communities, 
each of which is more densely connected internally than it is with the rest of the 
network). 
Realistic networks often include substructures that more closely resemble 
core-periphery graphs or expander graphs (see Fig.\nobreakspace \ref {fig:stylized-coreper} 
and Fig.\nobreakspace \ref {fig:stylized-expander}); and networks that partition into nice 
nearly-decomposable communities tend to be the exception rather than 
typical~\cite{Leskovec:2008vo,Leskovec:2009fy,jure2013}.


\section{Network Community Profiles (NCPs) and Their Interpretation}
\label{sxn:ncp}

Recall from Section\nobreakspace \ref {sxn:intro} that an NCP measures the quality of the best possible community of a given 
size as a function of the size of the purported community \cite{Leskovec:2008vo,Leskovec:2009fy,Leskovec:2010uj}. In this section, we provide a brief description of NCPs and how we will use it.


\subsection{The Basic NCP: Measuring Size-Resolved Community Quality}

We start with the definition of conductance and the original 
conductance-based definition of an NCP from Ref.~\cite{Leskovec:2009fy}, 
and we then discuss our extensions of such ideas.
For more details on conductance and NCPs, see 
Refs.~\cite{Chung:1997tk,Chung:2006vs,Leskovec:2009fy,Mahoney:2012wl}.
If $G=(V,E,w)$ is a graph with weighted adjacency matrix $A$, then the ``volume'' 
between two sets $S_1$ and $S_2$ of nodes (i.e., $S_i \subset V$) 
equals the total weight of edges with one end in $S_1$ and one end in $S_2$.
That is,
\begin{equation}
\label{eqn:volume2}
	\vol(S_1,S_2)=\sum_{i \in S_1} \sum_{ j\in S_2} A_{ij}\,.
\end{equation}
In this case, the ``volume'' of a set $S \subset V$ of nodes is
\begin{equation}
\label{eqn:volume}
	\vol(S)=\vol(S,V)=\sum_{i \in S} \sum_{j \in V} A_{ij}\,.
\end{equation} 
In other words, the set volume equals the total weight of edges that are attached to nodes in the set. 
The volume $\vol(S, \wbar S)$ between a set $S$ and its complement $\wbar S$ has a natural interpretation as the ``surface area'' of the ``boundary'' between $S$ and $\wbar S$.  In this study, a set $S$ is a hypothesized community. Informally, the conductance of a set $S$ of nodes is the ``surface area'' of that hypothesized community divided by ``volume'' (i.e., size) of that community. From this perspective, studying community structure amounts to an exploration of the isoperimetric structure of $G$. 

Somewhat more formally, the \emph{conductance of a set of nodes} $S \subset V$ is 
\begin{equation}
\label{eqn:conductance-set}
	\phi(S)=\frac{\vol(S,\wbar S)}{\min(\vol(S),\vol(\wbar S))}\,.
\end{equation}
Thus, smaller values of conductance correspond to better communities.
The \emph{conductance of a graph} $G$ is the minimum conductance of any 
subset of nodes:
\begin{equation}
\label{eqn:conductance-graph}
	\phi(G)=\min_{S\subset V} \phi(S)\,.
\end{equation}
Computing the conductance $\phi(G)$ of an arbitrary graph is an intractable 
problem (in the sense that the associated decision problem is 
NP-hard~\cite{ARV_JACM09}), but this quantity can be approximated by the second smallest
eigenvalue $\lambda_2$ of the normalized Laplacian
\cite{Chung:1997tk,Chung:2006vs}.

If the ``surface area to volume'' (i.e., isoperimetric) interpretation captures the 
notion of a good community as a set of nodes that is connected more densely internally than with the remainder of a network, then 
computing the solution to Eq.\nobreakspace \textup {(\ref {eqn:conductance-graph})} 
leads to the ``best'' (in this sense) community of any size in the network.

Instead of defining a community quality score in terms of the best community 
of any size, it is useful to define a community quality score in terms of 
the best community of a given size $k$ as a function of the size $k$. 
To do this, Ref.~\cite{Leskovec:2009fy} introduced the idea of a 
\emph{network community profile (NCP)} as the lower envelope of the 
conductance values of communities of a given size:
\begin{equation}\label{ncp}
	\phi_k(G) = \min_{S\subset V, |S|=k} \phi(S)\,.
\end{equation}
An NCP plots a community quality score (which, as in Ref.~\cite{Leskovec:2009fy}, we
take to be the set conductance of communities) of the best possible community of 
size $k$ as a function of $k$.
Clearly, it is also intractable to compute the quantity $\phi_k(G)$ in 
Eq.\nobreakspace \textup {(\ref {ncp})} exactly. 
Previous work has used spectral-based and flow-based approximation algorithms
to approximate it~\cite{Leskovec:2008vo,Leskovec:2009fy,Leskovec:2010uj}.

\begin{figure}[tb]
\subfloat[Three possible NCPs]{
\label{fig:ncp-possiblencps}
\includegraphics[width=.47\linewidth]{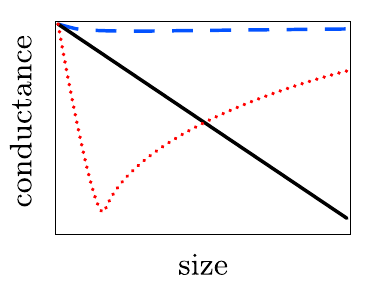}
}
\hfill
\subfloat[Realistic NCP from~\cite{Leskovec:2009fy}]{
\label{fig:ncp-bigncp}
\includegraphics[width=.47\linewidth]{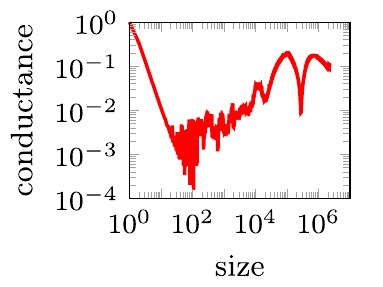}
}
\hfill
\subfloat[A caveman network]{
\label{fig:ncp-cavemangraph}
\includegraphics[width=.47\linewidth]{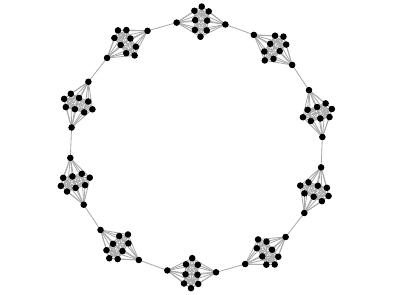}
}
\hfill
\subfloat[NCP of caveman network]{
\label{fig:ncp-cavemanncp}
\includegraphics[width=.47\linewidth]{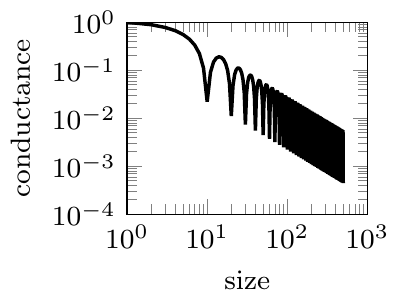}
}
\caption{Illustration of network community profiles (NCPs).
\sref{fig:ncp-possiblencps}
Stylized versions of possible shapes for an NCP: downward-sloping (black, solid), 
upward-sloping (red, dotted), and flat (blue, dashed).
\sref{fig:ncp-bigncp}
NCP of a \textsc{LiveJournal} network that illustrates the characteristic 
upward-sloping NCP that is typical for many large empirical social and information
networks~\cite{Leskovec:2009fy}. 
\sref{fig:ncp-cavemangraph}
A toy ``caveman network'' with 10 cliques of 10 nodes each, where one edge from each clique has been rewired to create a ring~\cite{Watts:1999vk}. 
\sref{fig:ncp-cavemanncp}
NCP for a similar caveman network with 100 cliques of 10 nodes each (the NCP for the network in panel \sref{fig:ncp-cavemangraph} is identical for communities with fewer than 50 nodes), illustrating the characteristic downward-sloping NCP that is typical of networks that are embedded in a low-dimensional space.
\label{fig:ncp}}
\end{figure}

To gain insight into how to understand an NCP and what it reveals
about network structure, consider Fig.\nobreakspace \ref {fig:ncp}.
In Fig.\nobreakspace \ref {fig:ncp-possiblencps}, we illustrate three possible ways that an 
NCP can behave.  
In each case, we are using conductance as a measure of community quality.
\begin{itemize}
\item
\textbf{Upward-sloping NCP.} 
In this case, small communities are ``better'' than large communities.
\item
\textbf{Flat NCP.} 
In this case, community quality is independent of size. 
(As illustrated in this figure, the quality tends to be comparably poor for 
all sizes.)
\item
\textbf{Downward-sloping NCP.} 
In this case, large communities are ``better'' than small communities.
\end{itemize}
For ease of visualization and computational considerations, we only show 
NCPs for communities up to half of the size of a network. 
An NCP for very large communities that we do not show in figures as a result 
of this choice roughly mirrors that for small communities, as the complement 
of a good small community is a good large community because of the inherent 
symmetry in conductance (see Eq.\nobreakspace \textup {(\ref {eqn:conductance-set})}).

In Fig.\nobreakspace \ref {fig:ncp-bigncp}, we show an NCP of a LiveJournal network from 
Ref.~\cite{Leskovec:2009fy}.  
It demonstrates an empirical fact about a wide range of large social and 
information networks: there exist good small conductance-based communities, 
but there do not exist any good large conductance-based communities in many 
such networks.  
See Refs.~\cite{Chung:1997tk,Chung:2006vs,Leskovec:2008vo,Leskovec:2009fy,Leskovec:2010uj,Mahoney:2012wl}) 
for more empirical evidence that large social and information networks tend 
not to have large communities with low conductances.
On the contrary, Fig.\nobreakspace \ref {fig:ncp-cavemangraph} illustrates a small toy 
network---a so-called ``caveman network''---formed from several small 
cliques connected by rewiring one edge from each clique to create a 
ring~\cite{Watts:1999vk}.
As illustrated by its downward-sloping NCP in Fig.\nobreakspace \ref {fig:ncp-cavemanncp}, 
this network possesses good conductance-based communities, and large 
communities are better than small ones.  
One obtains a similar downward-sloping NCP for the Zachary Karate Club 
network~\cite{Zachary:1977tj} as well as for many other networks for which 
there exist meaningful visualizations~\cite{Leskovec:2009fy}. 
The wide use of networks that have interpretable visualizations (such as 
the Zachary Karate Club and planted partition models~\cite{condon2001} with 
balanced communities) to help develop and evaluate methods for community 
detection and other procedures can lead to a strong selection bias when 
evaluating the quality of those methods.

\begin{figure}[tb]
\subfloat[Low-dimensional structure \label{sfig:NCP_karate}]{\includegraphics[width=0.47\linewidth]{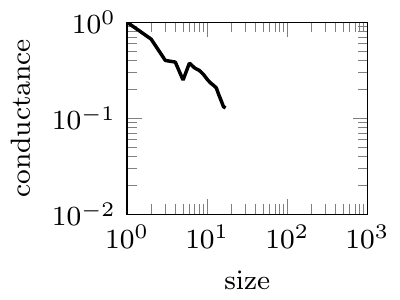}}\hfill
\subfloat[Core-periphery structure\label{sfig:NCP_core_periphery}]{\includegraphics[width=0.47\linewidth]{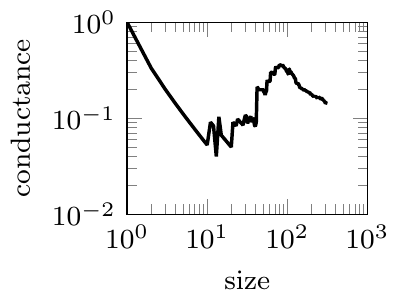}}\hfill
\subfloat[Expander or complete graph\label{sfig:NCP_expander}]{\includegraphics[width=0.47\linewidth]{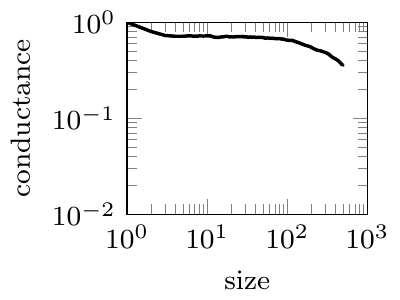}}\hfill
\subfloat[Bipartite structure\label{sfig:NCP_bipartite}]{\includegraphics[width=0.47\linewidth]{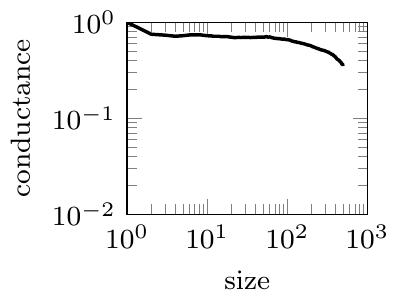}}\hfill
\caption{Network community profiles (NCPs) of the idealized example networks from Fig.\nobreakspace \ref {fig:stylized}. 
\sref{sfig:NCP_karate} NCP for the Zachary Karate Club network.
\sref{sfig:NCP_core_periphery} NCP for an example network generated from a block model with core-periphery structure.
\sref{sfig:NCP_expander} NCP for an Erd\H{o}s-R\'enyi graph.
\sref{sfig:NCP_bipartite} NCP for an example network generated from a bipartite block model.
\label{fig:stylized_NCP} }
\end{figure}

We now consider the relationship between the phenomena illustrated in 
Fig.\nobreakspace \ref {fig:ncp} and the idealized block models of Fig.\nobreakspace \ref {fig:stylized}. As a concrete example, Fig.\nobreakspace \ref {fig:stylized_NCP} shows the NCPs for the example networks in the right panels of Fig.\nobreakspace \ref {fig:ncp}.

First, note that the best partitions consist roughly of well-balanced 
communities in the low-dimensional case of Figs.\nobreakspace \ref {fig:stylized-hotdog} and\nobreakspace  \ref {sfig:NCP_karate}, and 
the ``lowest'' point on an NCP tends to be for large community sizes. 
Thus, an NCP tends to be downward-sloping. 

Networks with pronounced core-periphery structure---i.e., networks that ``look like'' the example network in Fig.\nobreakspace \ref {fig:stylized-coreper}---tend to have many good small communities but no equally good or better large 
communities. This situation arises in many large, extremely sparse networks~
\cite{Leskovec:2008vo,Leskovec:2009fy,Leskovec:2010uj}. The good small communities in such networks are sets of connected nodes in the extremely sparse periphery, and they do not combine to form good, large communities, as they are only connected via a set of core nodes with denser connections than the periphery. Thus, an NCP of a network with core-periphery structure tends to be upward-sloping, as illustrated in Figs.\nobreakspace \ref {fig:stylized-coreper} and\nobreakspace  \ref {sfig:NCP_core_periphery}.  However, this observation does not apply to all networks with well-defined density-based core-periphery structure. If the periphery is sufficiently well-connected (though still much sparser than the core), then one no longer observes good, small communities. Such networks act like expanders from the perspective of the behavior of random walkers, so they have a flat NCP. One can generate examples of such networks by modifying the parameters of the block-model that we used to generate the example network in Fig.\nobreakspace \ref {fig:stylized-coreper}~\cite{block-model}.

For a complete graph or a degree-homogeneous expander 
(see Figs.\nobreakspace \ref {fig:stylized-expander} and\nobreakspace  \ref {sfig:NCP_expander}), all communities tend to have poor 
quality, so an NCP is roughly flat.  (See Appendix\nobreakspace \ref {sxn:expanders} for a discussion of expander graphs.) 

Finally, bipartite structure itself does not have any characteristic influence on an NCP. Instead, an NCP of a bipartite network reveals other structure present in a network. For the example network in Fig.\nobreakspace \ref {fig:stylized-bipartite}, the two types of nodes are connected uniformly at random, so its NCP (Fig.\nobreakspace \ref {sfig:NCP_bipartite}) has the characteristic flat shape of an expander.


\subsection{Robustness and Information Content of NCPs}

It is important to discuss the robustness properties of NCPs.  
These are not obvious a priori, as the NCP is an extremal diagnostic.  
Importantly, though, the qualitative property of being downward-sloping, 
upward-sloping, or roughly flat is very robust to the removal of nodes and 
edges, variations in data generation and preprocessing decisions, and 
similar sources of 
perturbation~\cite{Leskovec:2008vo,Leskovec:2009fy,Leskovec:2010uj}.  
For example, upward-sloping NCPs typically have many small communities of 
good quality, so losing some communities via noise or some other 
perturbations has little effect on a realistic NCP.  
Naturally, whether a particular set of nodes achieves a local minimum is not 
robust to such modifications.  
In addition, one can easily construct pathological networks whose NCPs are 
not robust.

It is also important to consider the robustness of a network NCPs with 
respect to the use of conductance versus other measures of community 
quality. 
(Recall that many other measures have been proposed to capture the criteria 
that a good community should be densely-connected internally but sparsely 
connected to the rest of a network~\cite{Porter:2009we,Leskovec:2009fy}.)  
Indeed, it has been shown that measures that capture both criteria of 
community quality (internal density and external sparsity) behave in a 
roughly similar manner to conductance-based NCPs, whereas measures that 
capture only one of the two criteria exhibit qualitatively different 
behavior, typically for rather trivial reasons~\cite{Leskovec:2010uj}.

Although the basic NCP that we have been discussing yields numerous insights about 
both small-scale and large-scale network structure, it also has important 
limitations. For example, an NCP gives no information on the number or density of 
communities with different community quality scores.  
(This contributes to the robustness properties of NCP with respect to 
perturbations of a network.) Accordingly, the communities that are revealed by 
an NCP need not be representative of the majority of communities
in a network. However, the extremal features that are revealed by an NCP have important
system-level implications for the behavior of 
dynamical processes on a network: they are responsible for the most severe
bottlenecks for associated dynamical processes on networks~\cite{Mihail:1989vv}. 

Another property that is not revealed by an NCP is the internal structure of 
communities. 
Recall from Eq.\nobreakspace \textup {(\ref {eqn:conductance-set})} that the conductance of a community 
measures how well (relative to its size) that it is separated from the 
remainder of a network, but it does not consider the internal structure of a 
community (except for size and edge density). 
In an extreme case, a community with good conductance might even consist of 
several disjoint pieces. 
Recent work has addressed how spectral-based approximations to optimizing 
conductance also approximately optimize measures of internal 
connectivity~\cite{ZLM13}.

We augment the information from basic NCPs with some additional computations. 
To obtain an indication of a community's internal structure, we compute the 
internal conductance of the communities that form an NCP. 
The \emph{internal conductance} $\phi_\text{in}(C)$ of a community $C$ is 
\begin{equation}
 	\phi_\text{in}(C)= \phi(G|_C)  \,,
\label{eqn:internal_conductance}
\end{equation}
where $G|_C$ is the subgraph of $G$ induced by nodes in the community $C$.  
The internal conductance is equal to the conductance of the best partition 
into two communities of the network $G|_C$ viewed as a graph in isolation. 
Because a good community should be well-separated from the remainder of a 
network and also relatively well-connected internally, we expect good 
communities to have low conductance but high internal conductance. 
We thus compute the \emph{conductance ratio} 
\begin{equation}
	\Phi(C)=\frac{\phi(C)}{\phi_\text{in}(C)}
\end{equation} 
to quantify this intuition.  
A good community should have a small conductance ratio, and thus we also plot 
so-called \emph{conductance ratio profiles (CRPs)}~\cite{Leskovec:2009fy} to 
illustrate how conductance ratio depends on community size in networks.


\subsection{Our Application and Extension of NCPs}

In this paper, we examine the small-scale, medium-scale, and large-scale 
community structure using conductance-based NCPs and CRPs. 
We employ three different methods, which we introduce in detail in Appendix\nobreakspace \ref {sxn:measures}, for 
sampling an NCP: one based on local diffusion dynamics (the \textsc{AclCut} method), one based on a local 
spectral optimization (the \textsc{MovCut} method), and one based on geodesic distance from a seed node (the \textsc{EgoNet} method). 
In each case, we find communities of different sizes, and we then plot the 
conductance of the best community for each size as a function of size. 

An NCP provides a signature of community structure in a network, and we can thereby compare community structure across different networks.  This helps one to discern which properties are attributable predominantly to network structure and which are attributable predominantly to choice of algorithms for community detection.  Our approach of comparing community structures in networks using NCPs and CRPs is very general: one can of course follow a similar procedure with other community-quality diagnostics on the vertical axis, other procedures for community generation, and so on.




\section{Empirical Results on Real Networks}
\label{sxn:main-empirical}

In this section, we present the results of our empirical evaluation of the 
small-scale, medium-scale, and large-scale community structure in our 
example networks.

\subsection{Example Network Data Sets}
\label{sxn:prelim-networkdata}

We will examine six empirical networks in depth. They fall into three classes: coauthorship networks, Facebook 
networks, and voting similarity networks. 
For each class, we consider two networks of two different sizes. 
\begin{itemize}
\item
\textbf{Collaboration graphs.}
The two (unweighted) coauthorship networks were constructed from papers submitted to the 
{\tt arXiv} preprint server in the areas of general relativity and quantum 
cosmology (\textsc{CA-GrQc}) and Astrophysics (\textsc{CA-AstroPh}). 
In each case, two authors are connected by an edge if they coauthored at 
least one paper, so a paper with $k$ authors appears as a $k$-clique
(i.e., a complete $k$-node subgraph) in the network. 
These network data are available as part of the Stanford Network Analysis 
Package (SNAP), and they were examined previously in 
Refs.~\cite{Leskovec:2008vo,Leskovec:2009fy,Leskovec:2010uj}.
\item
\textbf{Facebook graphs.}
The two (unweighted) Facebook networks are anonymized data sets that consist of a snapshot of 
``friendship'' ties on one particular day in September 2005 for two 
United States (U.S.) universities: Harvard (\textsc{FB-Harvard1}) and 
Johns Hopkins (\textsc{FB-Johns55}).  
They form a subset of the \textsc{Facebook100} data set from 
Refs.~\cite{Traud:2012ft,Traud:2011fs}.
In addition to the friendship ties, note that we possess node labels for 
gender and class year as well as numerical identifiers for student or some other (e.g., faculty) status, major, and high school.
\item
\textbf{Congressional voting graphs.}
The two (weighted) Congressional voting networks represent similarities in voting 
patterns among members of the U.S. House of Representatives 
(\textsc{US-House}) and U.S. Senate (\textsc{US-Senate}).
Our construction follows prior 
work~\cite{Mucha:2010p2164,Mucha:2010th}. 
In particular, we represent these two data sets as  ``multilayer'' temporal
networks~\cite{Mucha:2010p2164,mikko-review}.
Each layer corresponds to a single two-year Congress, and edge weights 
within a layer represent the voting similarity between two legislators 
during the corresponding Congress.  
In layer $s$, this yields adjacency elements of
$A_{ij}^{(s)}=\frac{1}{b_{ij}(s)} \sum_{k} \gamma_{ijk}$,
where $\gamma_{ijk}=1$ if both legislators voted the same way on the 
$k^{th}$ bill, $\gamma_{ijk}=0$ if they voted in different ways on that bill, $b_{ij}(s)$ is the number of 
bills on which both legislators voted during that Congress, and the sum is 
over bills. A tie between the same legislator in consecutive Congresses is represented 
by an interlayer edge with weight $\omega$~\cite{Mucha:2010p2164}.
(We use $\omega = 1$; the effect of changing $\omega$ has been investigated 
previously~\cite{Mucha:2010th,Bassett:2013fh}.)
We represent each multilayer voting network using a single ``supra-adjacency matrix'' (see Refs.~~\cite{mikko-review,DeDomenico:2013tt,DeDomenico:2013td,Gomez:2013tk}) in which the 
different Congresses correspond to diagonal blocks and interlayer edges 
correspond to off-block-diagonal terms in the matrix. Note that throughout this paper we treat the Congressional voting graphs at the level of this supra-adjacency matrix, without any additional labeling or distinguished treatment of inter- and intra-layer edges (cf.~\cite{Mucha:2010p2164}).
\end{itemize}

We chose these three sets of networks because (as we will see in later
sections) they have \emph{very} different properties with respect to their 
large-scale versus small-scale community structures.
We thus emphasize that, with respect to the topic of this paper, 
these six networks are representative of several broad classes of 
previously-studied networks:
\textsc{CA-GrQc} and \textsc{CA-AstroPh} are representative of the SNAP 
networks that were examined previously in 
Refs.~\cite{Leskovec:2008vo,Leskovec:2009fy,Leskovec:2010uj}; 
both \textsc{FB-Harvard1} and \textsc{FB-Johns55} (aside from a few very small communities in \textsc{FB-Harvard1}) are representative of the \textsc{Facebook100} networks that were examined 
previously in Refs.~\cite{Traud:2012ft,Traud:2011fs}; and \textsc{US-House} 
and \textsc{US-Senate} give examples of networks (that are larger than 
the Zachary Karate Club and caveman networks) on which conventional notions of and algorithms for community detection have been validated successfully~\cite{Mucha:2010p2164,Mucha:2010th}. 

In Table\nobreakspace \ref {tab:data_summary}, we provide summary statistics for each of 
the six networks. We give the numbers of nodes and edges in the largest 
connected component, the mean degree/strength ($\langle k_i\rangle$), 
the second-smallest eigenvalue ($\lambda_2$) of the normalized Laplacian matrix, 
and mean clustering coefficient ($\langle C_i\rangle$). We use the 
local clustering coefficient 
$ C_i = \frac{1}{k_i(k_i-1)} \sum_{j,k} \left( \hat w_{ij} \hat w_{ik} \hat w_{jk} \right)^\frac{1}{3} $,
where $\hat{w}_{ij} = \frac{w_{ij}}{\max_{ij} w_{ij}}$, which reduces to the 
usual expression for local clustering coefficients in unweighted networks~\cite{Onnela:2005de,Saramaki:2007de,Watts:1998vc}. The high values for mean clustering coefficient in both the U.S. Congress and coauthorship networks are unsurprising, given how those networks have been 
constructed. However, the latter is noteworthy, as the coauthorship networks are much sparser than the Facebook networks.

\begin{table*}[tb]
\begin{ruledtabular}
\begin{tabular}{l|r|r|r|r|r|c|l}
&\multicolumn{1}{c|}{Nodes} & \multicolumn{1}{c|}{Edges} &\multicolumn{1}{c|}{$\langle k\rangle$}&\multicolumn{1}{c|}{$ \lambda_2$}& \multicolumn{1}{c|}{$\langle$C$\rangle$} & Refs. & Description \\
\hline
\hline
\textsc{CA-GrQc} &4\,158&13\,422& 6.5 &0.0019&0.56 & 
\cite{Leskovec:2008vo,Leskovec:2009fy,Leskovec:2010uj} &
Coauthorship network: {\tt arXiv} general relativity \\
\hline
\textsc{CA-AstroPh} &17\,903 & 196\,972 & 22.0& 0.0063&0.63 & 
\cite{Leskovec:2008vo,Leskovec:2009fy,Leskovec:2010uj} &
Coauthorship network: {\tt arXiv} astrophysics \\
\hline
\hline
\textsc{FB-Johns55} &5\,157 &186\,572 &72.4& 0.1258&0.27 & 
\cite{Traud:2012ft,Traud:2011fs} &
Johns Hopkins Facebook network \\
\hline
\textsc{FB-Harvard1} & 15\,086 &824\,595&109.3&0.0094&0.21 & 
\cite{Traud:2012ft,Traud:2011fs} &
Harvard Facebook network \\
\hline
\hline
\textsc{\textsc{US-Senate}} &8\,974 &422\,335 &60.3&0.0013&0.50 & 
\cite{Mucha:2010p2164,Mucha:2010th,Bassett:2013fh} &
Network of voting patterns in U.S. Senate \\
\hline
\textsc{\textsc{US-House}} &36\,646 & 6\,930\,858 &240.5&0.0002&0.58 & 
\cite{Mucha:2010p2164,Mucha:2010th,Bassett:2013fh} &
Network of voting patterns in U.S. House \\
\end{tabular}
\end{ruledtabular}
\caption{Six medium-sized networks. 
For each network, we show the number of nodes and edges in the largest connected component 
(LCC), the mean degree/strength ($\langle k_i\rangle$), the second-smallest
eigenvalue ($\lambda_2$) of the normalized Laplacian matrix, the mean 
clustering coefficient ($\langle C_i\rangle$), prior references that used 
these networks, and a brief description.
}
\label{tab:data_summary}
\end{table*}

\begin{figure}[tb]
\includegraphics[width=\linewidth]{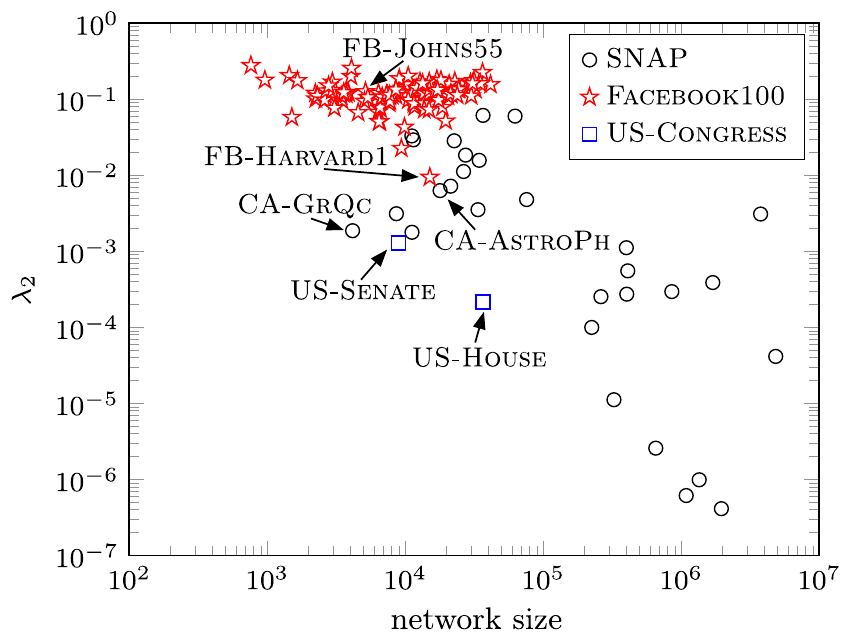}
\caption{Scatter plot of the second smallest eigenvalue ($\lambda_2$) of the 
normalized Laplacian versus size of the network for: 
the networks from the SNAP data ~\cite{snap} that were studied 
in~\cite{Leskovec:2008vo,Leskovec:2009fy}; 
all 100 networks in the \textsc{Facebook100} data 
set~\cite{Traud:2011fs,Traud:2012ft}; 
and the two \textsc{US-Congress} temporal 
networks~\cite{Mucha:2010p2164,Mucha:2010th,Bassett:2013fh}.
}
\label{fig:lambda_2}
\end{figure}

Recall that the second smallest eigenvalue $\lambda_2$ of the normalized 
Laplacian provides a qualitative notion of connectivity that can be used to 
bound the mixing time of diffusion-based dynamics on 
networks~\cite{Jerrum:1988iv} (where larger values of $\lambda_2$ imply that 
there are fewer bottlenecks to mixing) and that can also be used to 
partition a graph into 
communities~\cite{Newman:2010ur,Newman:2006hj,Chung:2006vs} (where smaller 
values of $\lambda_2$ correspond to better communities).
We show the values of $\lambda_2$ for our six networks in 
Table\nobreakspace \ref {tab:data_summary}. 
For comparison, we show in Fig.\nobreakspace \ref {fig:lambda_2} a scatter plot of $\lambda_2$ 
versus the size of the network (i.e., the number of nodes in the network) for 
these six networks; for the remaining networks from the Stanford Network 
Analysis Project (SNAP)~\cite{snap} (black circles) that were also studied 
in~\cite{Leskovec:2008vo,Leskovec:2009fy}; and for the remaining 98 networks 
from the \textsc{Facebook100} data set (red stars) studied 
in~\cite{Traud:2011fs,Traud:2012ft}.

The first point to note about Fig.\nobreakspace \ref {fig:lambda_2} is that $\lambda_2$ for 
nearly all of the \textsc{Facebook100} graphs is much larger than those for 
the two collaboration graphs and the two voting graphs.
Figure\nobreakspace \ref {fig:lambda_2} and previous empirical results (from 
Refs.~\cite{Leskovec:2008vo,Leskovec:2009fy}) clearly demonstrate that 
the $\lambda_2$ values for the two collaboration graphs are representative 
of (and, in many cases, higher than) those of the other SNAP graphs studied 
empirically in Refs.~\cite{Leskovec:2008vo,Leskovec:2009fy}.
That is, nearly all of the networks have $\lambda_2$ values that are much 
smaller than those in the \textsc{Facebook100} graphs.
This implies, in particular, that those graphs contain more substantial 
bottlenecks to mixing.
(Note, though, that the value of $\lambda_2$ says nothing about the size or 
cardinality of the set of nodes that achieves the minimum.)
In order to understand these differences, we study two networks from the 
\textsc{Facebook100} data set in detail: one (\textsc{FB-Johns55}) with a 
typical value of $\lambda_2$ and another (\textsc{FB-Harvard1}) that is an 
``outlier,'' in that it has the lowest value of $\lambda_2$ in the entire 
\textsc{Facebook100} data set.  
(The \text{FB-Caltech36} network is the smallest network in the 
\textsc{Facebook100} data set---it has 762 nodes in its largest connected 
component (LCC)---and it has the largest value of $\lambda_2$.)

The second point to note about Fig.\nobreakspace \ref {fig:lambda_2} and 
Table\nobreakspace \ref {tab:data_summary} is that they suggest that \textsc{FB-Johns55} (and possibly also \textsc{FB-Harvard1}) are better connected than the other four 
networks, and that the connectivity properties of the two collaboration 
graphs and the two voting graphs (and perhaps also \textsc{FB-Harvard1})
might be very similar.  As we will see below, however, the situation is considerably more subtle.

\begin{figure}[tb]
\subfloat[\textsc{CA-GrQc}]{
\label{fig:spyplots-gr}
\includegraphics[width=0.29\linewidth]{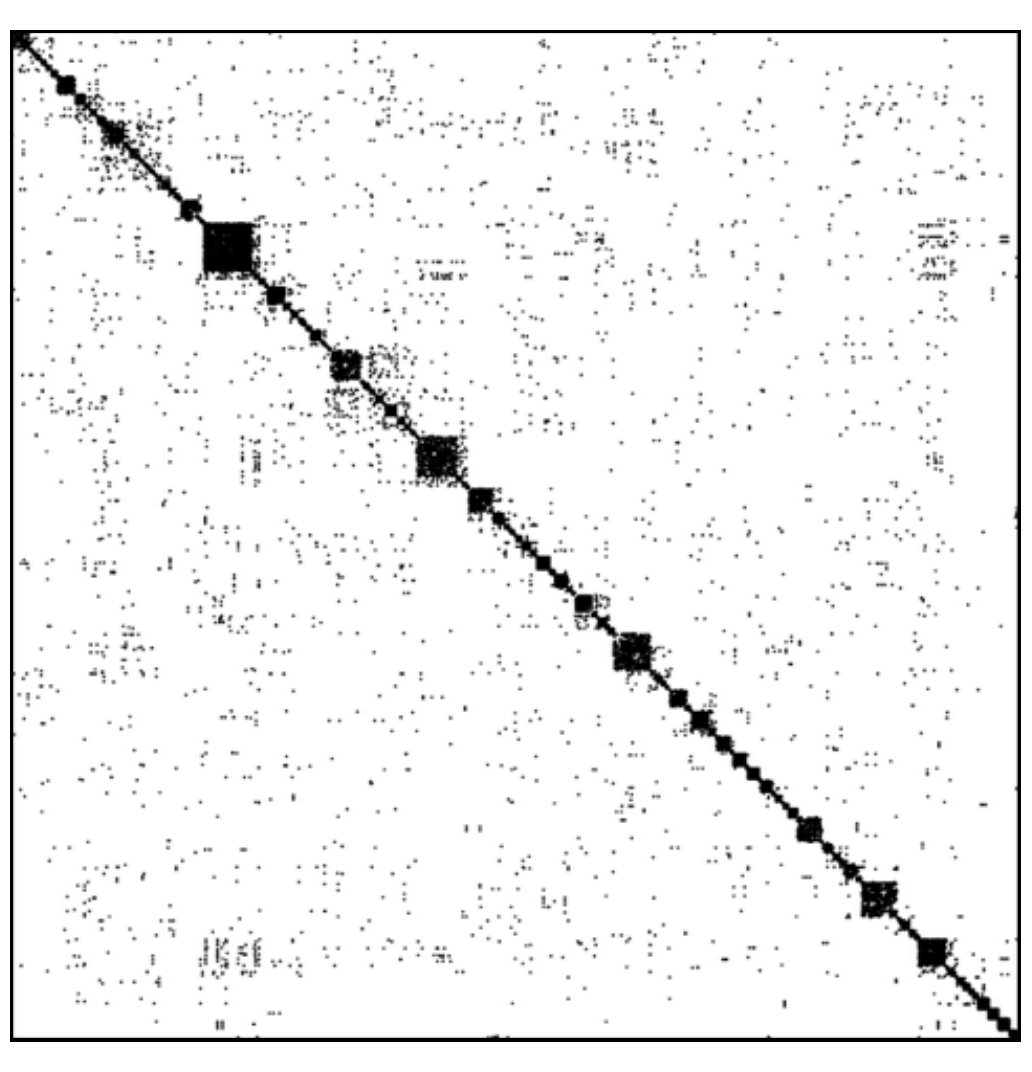}
} 
\hfill
\subfloat[\textsc{FB-Johns55}]{
\label{fig:spyplots-johns}
\includegraphics[width=0.29\linewidth]{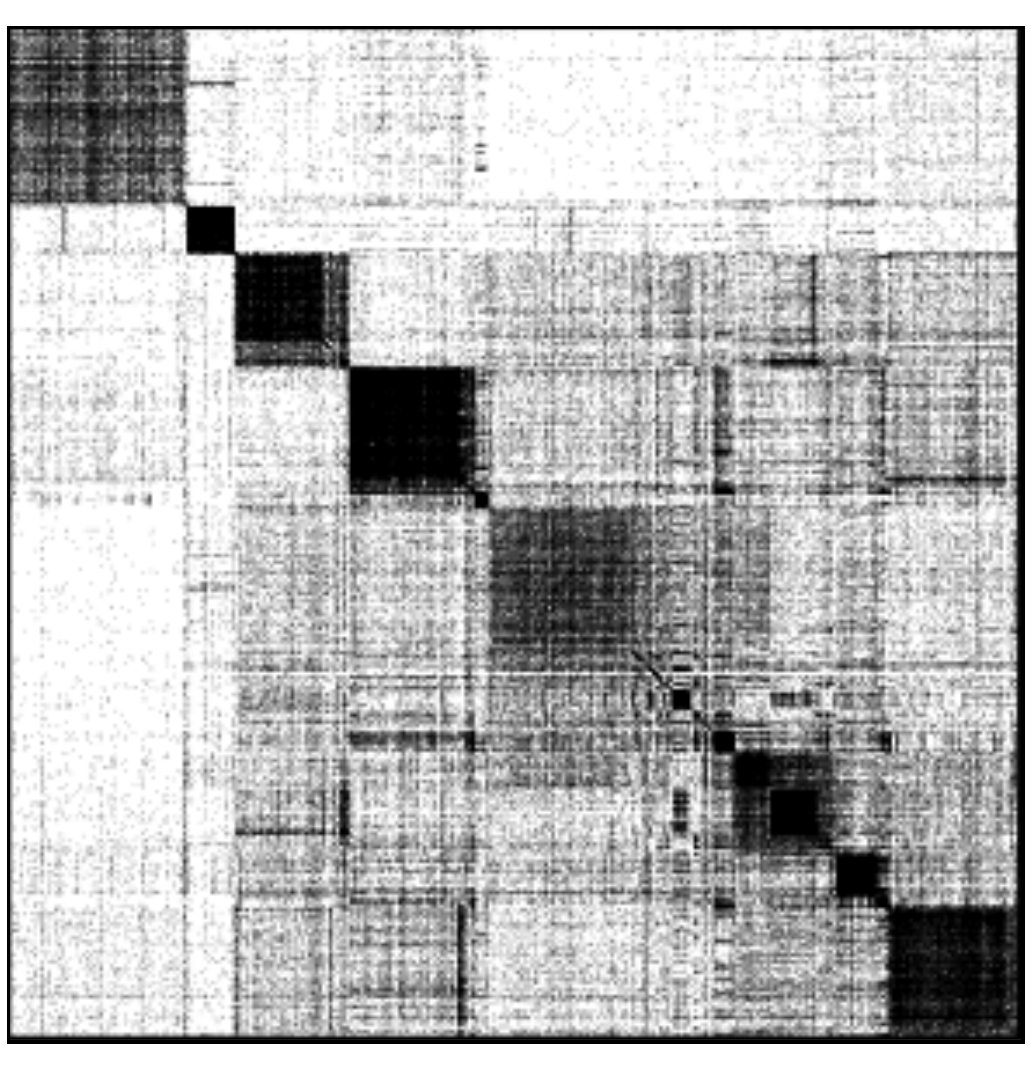}
}
\hfill
\subfloat[\textsc{US-Senate}]{
\label{fig:spyplots-senate}
\includegraphics[width=0.29\linewidth]{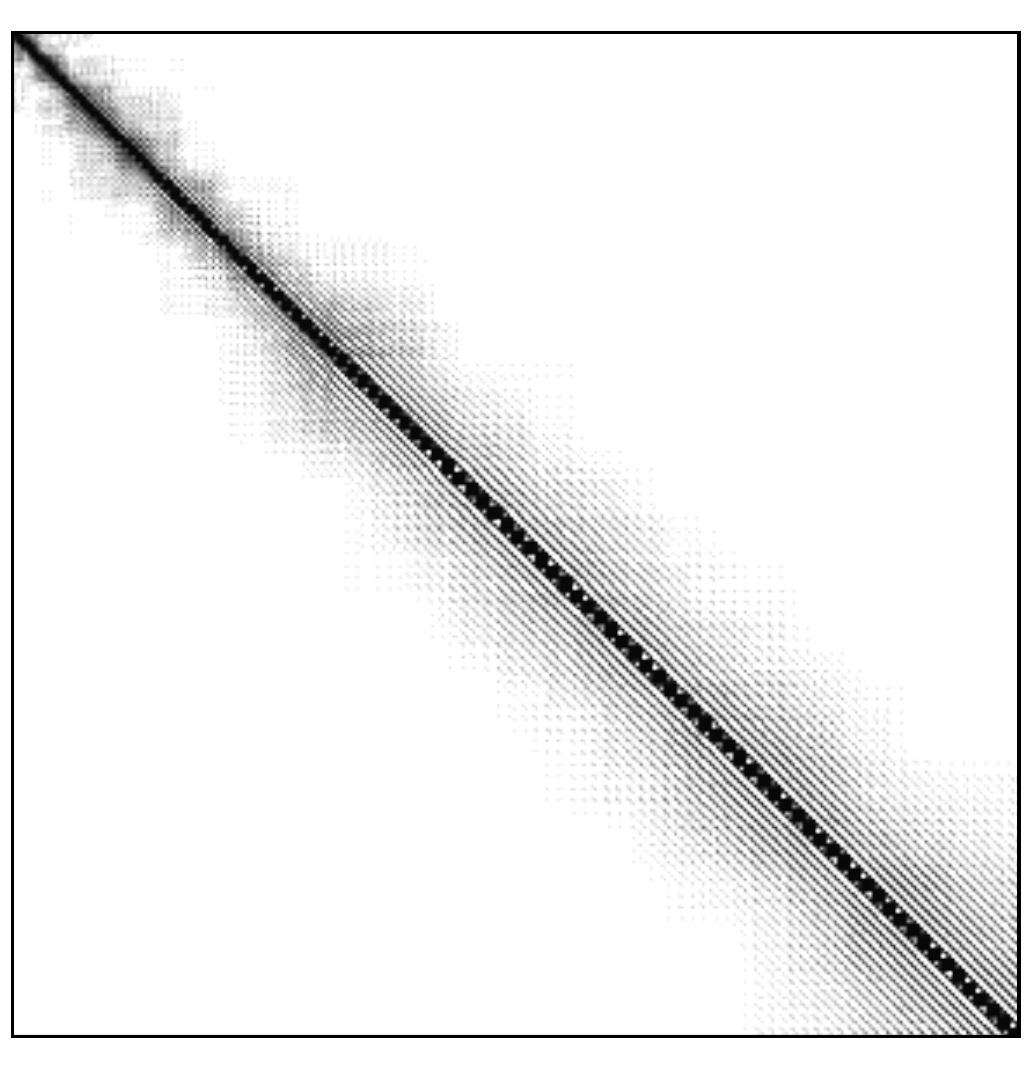}
}\\
\subfloat[\textsc{CA-AstroPh}]{
\label{fig:spyplots-astro}
\includegraphics[width=0.29\linewidth]{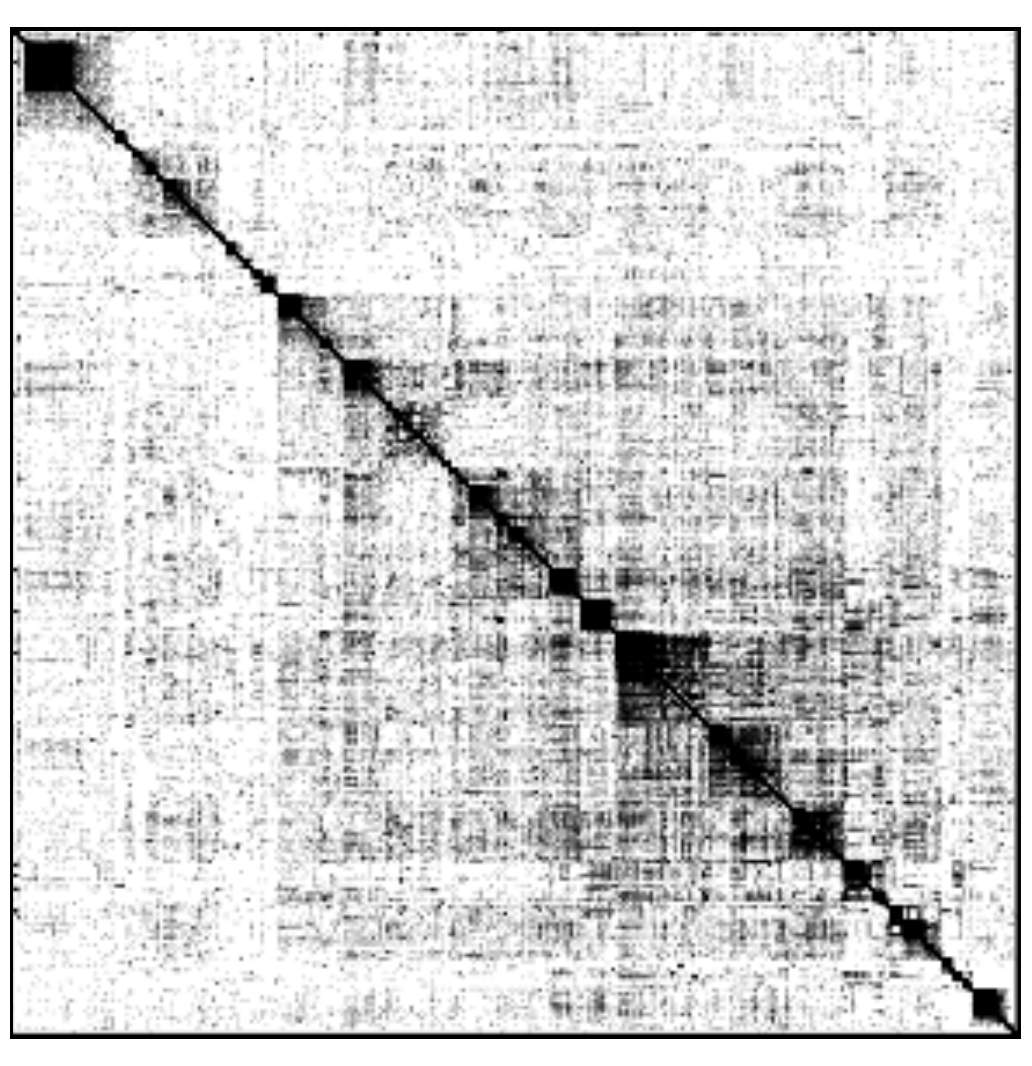}
}
\hfill
\subfloat[\textsc{FB-Harvard1}]{
\label{fig:spyplots-harvard}
\includegraphics[width=0.29\linewidth]{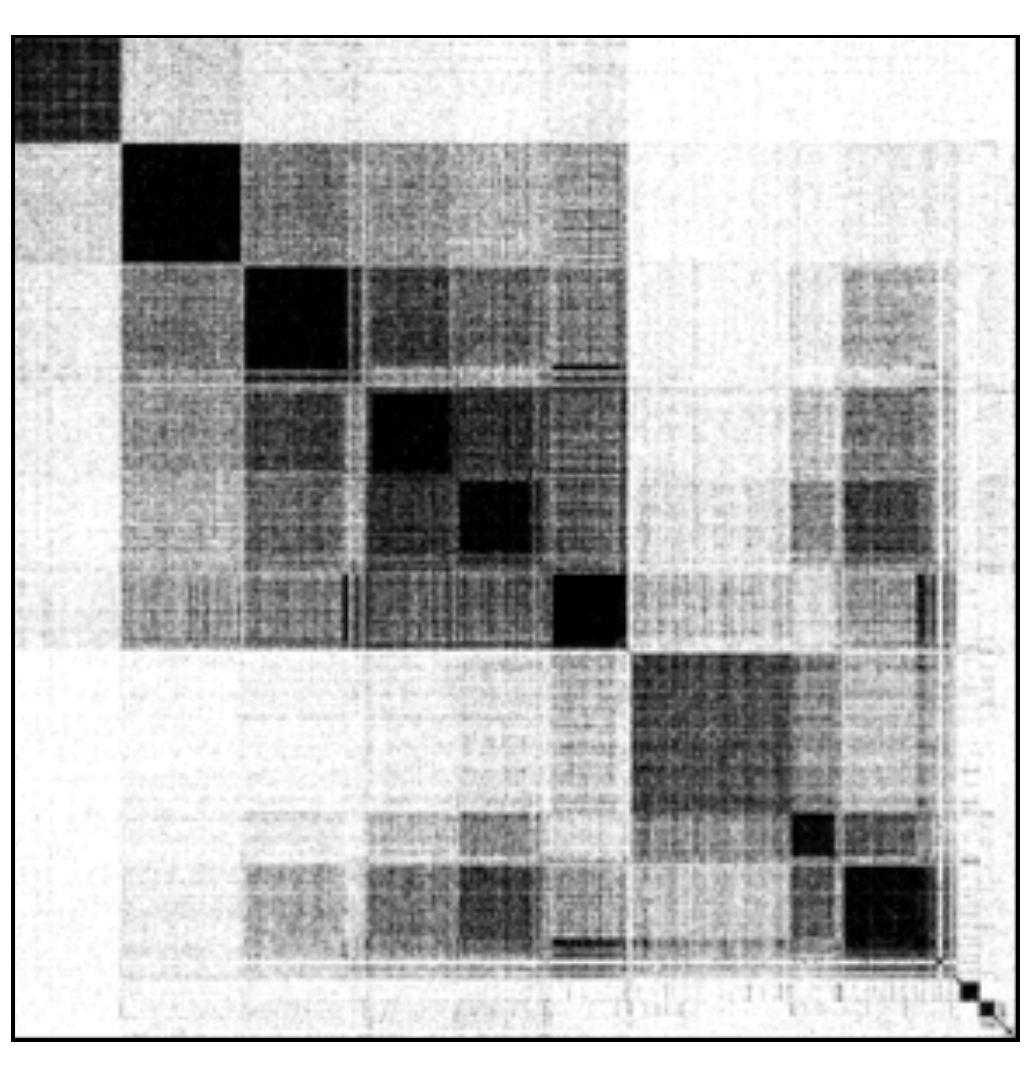}
} 
\hfill
\subfloat[\textsc{US-House}]{
\label{fig:spyplots-house}
\includegraphics[width=0.29\linewidth]{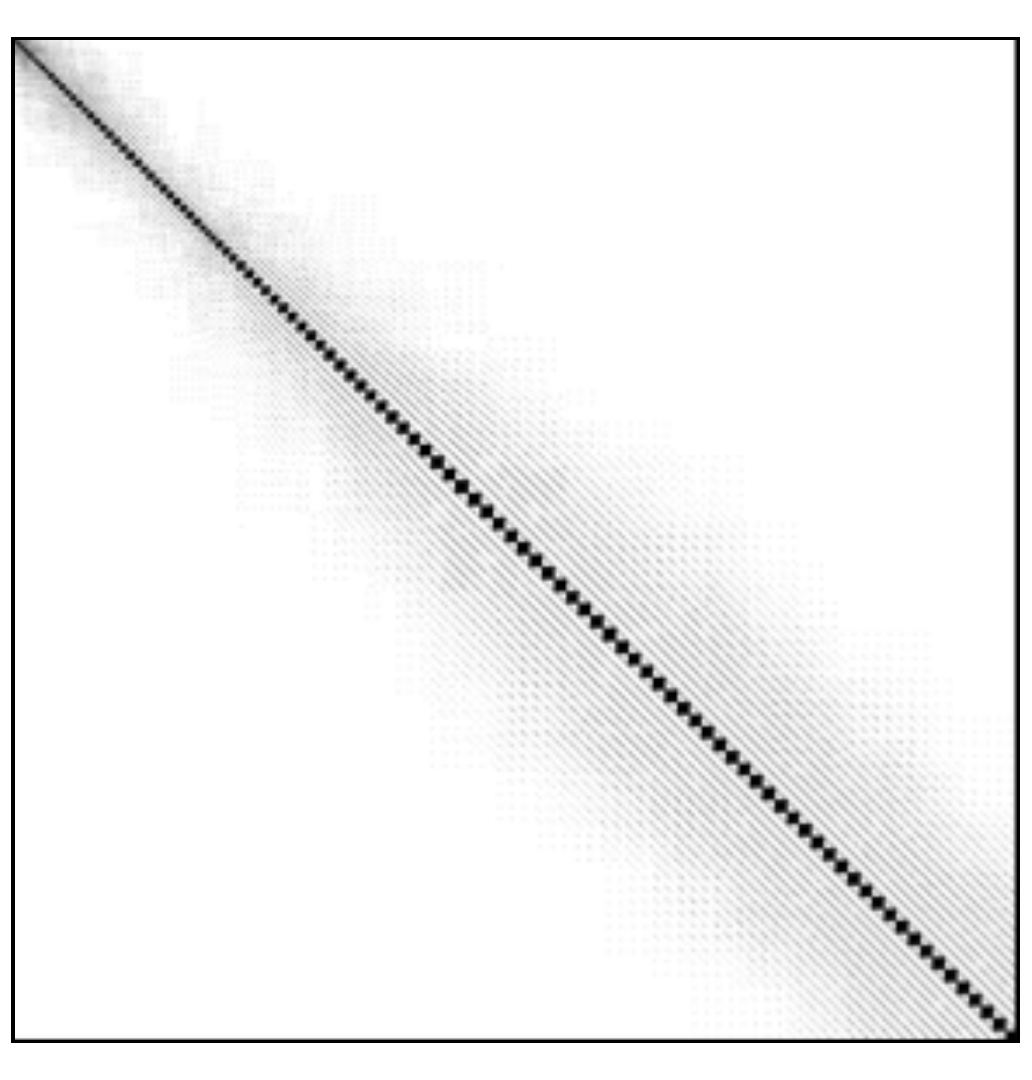}
}
\caption{Sparsity-pattern (Spy) plots for the largest connected component of 
each of our six example networks. The coauthorship networks (\textsc{CA-GrQc} and 
\textsc{CA-AstroPh}) and Facebook networks (\textsc{FB-Johns55} and 
\textsc{FB-Harvard1}) are arranged by communities that we obtained using 
an implementation \cite{AgeneralizedLouvai:2012uh} of a Louvain-like heuristic for modularity optimization~\cite{Blondel:2008do}. For \textsc{US-Congress}, we preserve the temporal order of the nodes starting with the first Congress in the top left and ending with the 110th Congress in the bottom right. 
}
\label{fig:spyplots}
\end{figure}

In Fig.\nobreakspace \ref {fig:spyplots}, we visualize the adjacency matrices of each of these 
networks using a sparsity-pattern (Spy) plot. 
We draw the nonzero entries of the adjacency matrix as black dots. 
The grayscale visualization in Fig.\nobreakspace \ref {fig:spyplots} is a result of coarsening 
the dpi-resolution and illustrates the density of connections in an area of 
the adjacency matrix. This yields a visualization comparable to the idealized block models in Fig.\nobreakspace \ref {fig:stylized}.
The node order in a Spy plot is arbitrary and, by permuting the nodes, can sometimes yield visualizations that are suggestive of structural features in a network. 
For the coauthorship and Facebook networks, we use results from a single run 
of an implementation~\cite{AgeneralizedLouvai:2012uh} of a Louvain-like 
heuristic~\cite{Blondel:2008do} for modularity optimization to partition 
these networks into communities. 
We then sorted nodes by community assignment: we chose the order of 
communities manually to suggest potential large-scale structures. 
For the voting similarity networks, time provides a natural order for the 
nodes.  
We started with nodes from the 1st Congress and ended with the nodes from 
the 110th Congress. 
The small blocks on the diagonal are the individual Congresses, which are 
almost fully connected internally, and the off-diagonals result from the 
interlayer coupling between the same individuals from different Congresses.

While certainly not definitive, Fig.\nobreakspace \ref {fig:spyplots} suggests several 
hypotheses about the relationship between small-scale structure and the 
large-scale structure---and, in particular, between small communities and 
large communities---in these six networks.
First, from Figs.\nobreakspace \ref {fig:spyplots-senate} and\nobreakspace  \ref {fig:spyplots-house}, it appears that 
the large-scale structure in \textsc{US-Senate} and \textsc{US-House} 
corresponds to that of a ``banded'' 
matrix~\footnote{Banded matrices arise, for example, in the study of finite-difference equations.  In banded matrices, the non-zero entries are confined to diagonal bands near the main diagonal.  Using our informal terminology from Section\nobreakspace \ref {sxn:prelim-looklike}, this is an extreme example of a``hot dog''.}. 
This banded structure is a result of the interlayer edges in these networks.
Second, from Figs.\nobreakspace \ref {fig:spyplots-gr} and\nobreakspace  \ref {fig:spyplots-astro}, it appears that 
\textsc{CA-GrQc} and \textsc{CA-AstroPh} both have many small-scale 
communities.  
It appears that they have a large-scale structure that is roughly banded;
but there also appear to be many ``long-range'' off-diagonal interactions 
between distant nodes in the depicted ordering.
Third, from Figs.\nobreakspace \ref {fig:spyplots-johns} and\nobreakspace  \ref {fig:spyplots-harvard}, we observe that 
both \textsc{FB-Johns55} and \textsc{FB-Harvard1} appear to have roughly 
$10$ communities that are both relatively large and relatively good.  

From these visuals, it appears that nearly all of these communities have 
dense internal connections and sparse connections to other communities. 
Given the usual notion that communities are sets of nodes with denser
connections among its constituent nodes than with the rest of the network, 
the visualizations in  Fig.\nobreakspace \ref {fig:spyplots} appear to suggest that there 
might be interesting large-scale structure that might be exploitable in 
\textsc{FB-Johns55} and \textsc{FB-Harvard1} but not in the other networks; 
and, in particular, that \textsc{FB-Johns55} and \textsc{FB-Harvard1} seem to 
be examples of the case $\alpha_{11} \approx \alpha_{22} \gg \alpha_{12}$ 
illustrated in Fig.\nobreakspace \ref {fig:stylized-hotdog}.

The focus of the present investigation is to test the extent to which the 
above hypotheses about the relationship between small-scale structure and 
large-scale structure in these six networks is correct.  
As we have discussed, intuition like what we have illustrated in 
Fig.\nobreakspace \ref {fig:spyplots} is common in the development and validation of methods 
for community detection, so it is useful to delve into great depth on a set 
of networks to explore the connections between small-scale and large-scale 
connections in networks.
As we will see in the next several sections, the situation is considerably 
more subtle than these figures (and commonly-employed intuition) might suggest.
For example, with the exception of the small communities in 
\textsc{CA-GrQc}/\textsc{CA-AstroPh} and the large-scale structure (i.e., the 
one-dimensional temporal ordering) in \textsc{US-Senate} and \textsc{US-House}, 
these intuitive hypotheses about the relationship between the local structure 
and the global structure in these networks are not unambiguously supported by 
other evidence.
Similarly, many communities that appear to be ``good'' based on the usual 
intuition and visualizations like that in Fig.\nobreakspace \ref {fig:spyplots} often are 
judged to be largely artifactual from the perspective of quantitative 
measures of community quality.


\begin{figure*}[tb]
\subfloat[\label{NCP_ACL_small}NCP]{
\includegraphics[width=.47\linewidth]{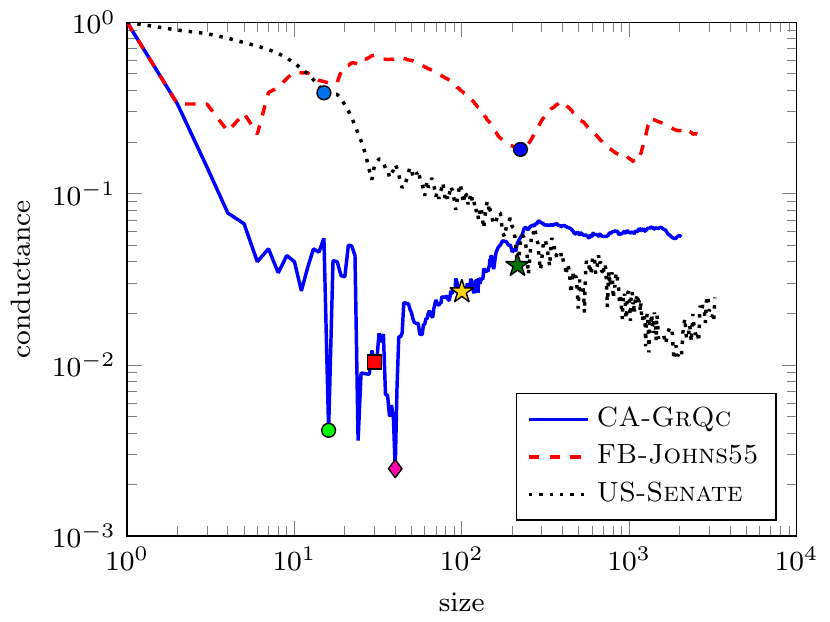}
}
\hfill
\subfloat[\label{condratio_ACL_small}CRP]{
\includegraphics[width=.47\linewidth]{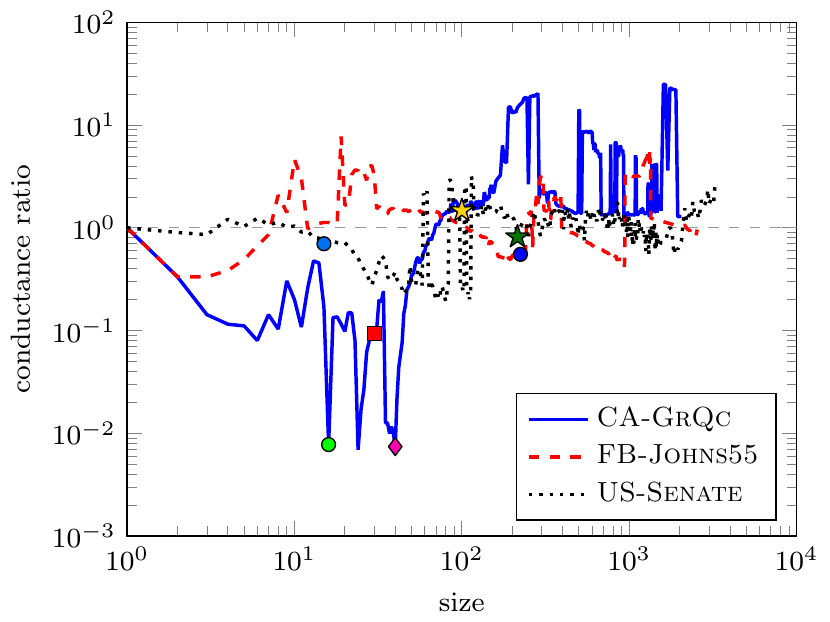}
}\\
\subfloat[\label{spring_ACL_GR}\textsc{CA-GrQc}]{
\includegraphics[height=5cm]{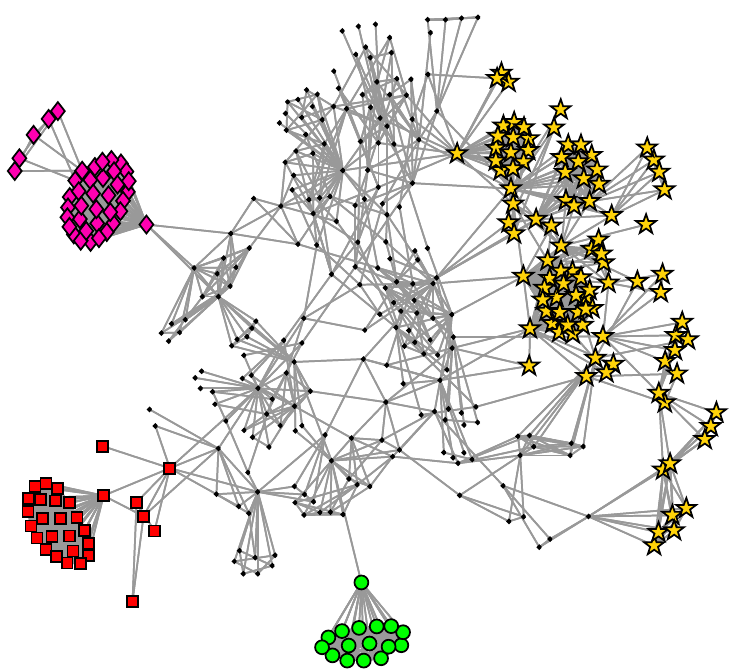}
}\hfill
\subfloat[\label{spring_ACL_Johns55}\textsc{FB-Johns55}]{
\includegraphics[height=5cm]{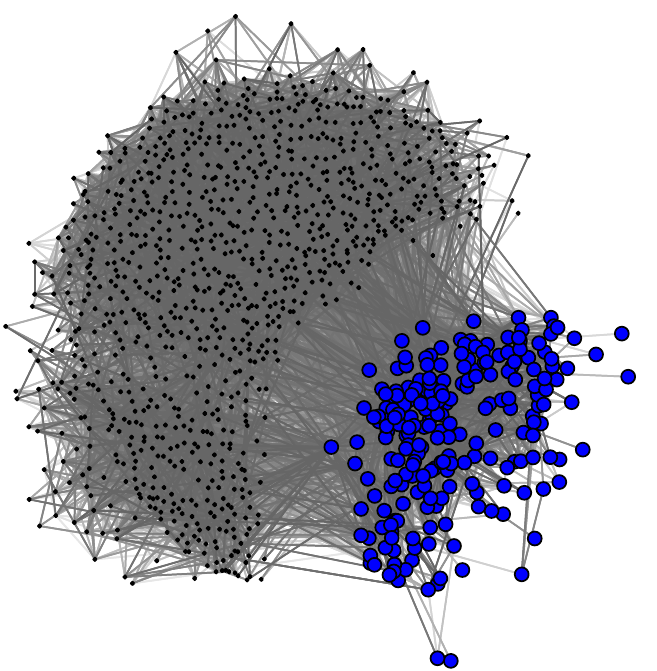}
}\hfill
\subfloat[\label{spring_ACL_Senate}\textsc{US-Senate}]{
\includegraphics[height=5cm]{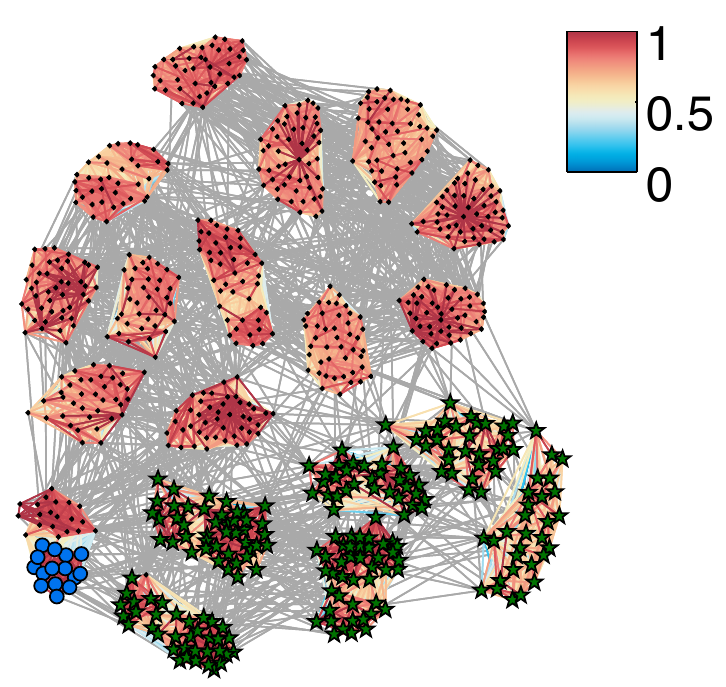}
}
\caption{NCP plots [in panel \sref{NCP_ACL_small}] and conductance ratio 
profile (CRP) plots [in panel \sref{condratio_ACL_small}] for 
\textsc{CA-GrQc}, \textsc{FB-Johns55}, and \textsc{US-Senate} (i.e., the 
smaller network in each of the three pairs of networks from 
Table\nobreakspace \ref {tab:data_summary}) generated using the \textsc{AclCut} method.
In panels \sref{spring_ACL_GR}--\sref{spring_ACL_Senate}, 
we show modified Kamada-Kawai~\cite{Kamada:1989dj} spring-embedding 
visualizations that emphasize community structure~\cite{viscoms} of 
corresponding (color-coded) communities and their neighborhoods 
(2-neighborhood for \textsc{CA-GrQc}, a 1-neighborhood for 
\textsc{FB-Johns55}, and all Senates that have at least one Senator in common with the communities for \textsc{US-Senate}). 
We find good small communities but no good large communities in 
\textsc{CA-GrQc}; some weak large-scale structure in \textsc{FB-Johns55} 
that does \emph{not} create substantial bottlenecks to the random-walk 
dynamics; and signatures of low-dimensional structure (i.e., good large communities but no good small communities) for \textsc{US-Senate}, which results from the 
multilayer structure that encapsulates the network's temporal properties.
}
\label{ACL_small}
\end{figure*}

\begin{figure*}[tb]
\subfloat[\label{NCP_ACL_large}NCP]{
\includegraphics[width=.47\linewidth]{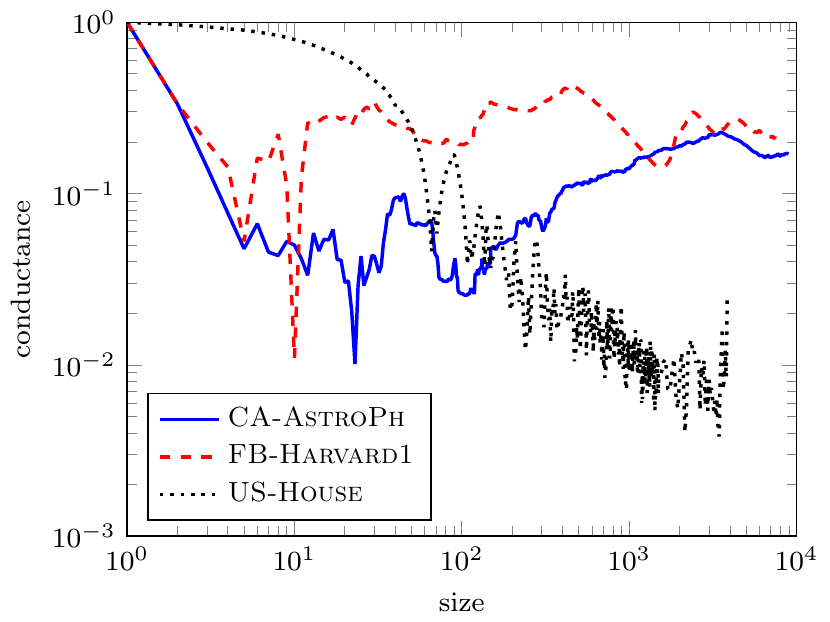}
}
\hfill
\subfloat[\label{condratio_ACL_large}CRP]{
\includegraphics[width=.47\linewidth]{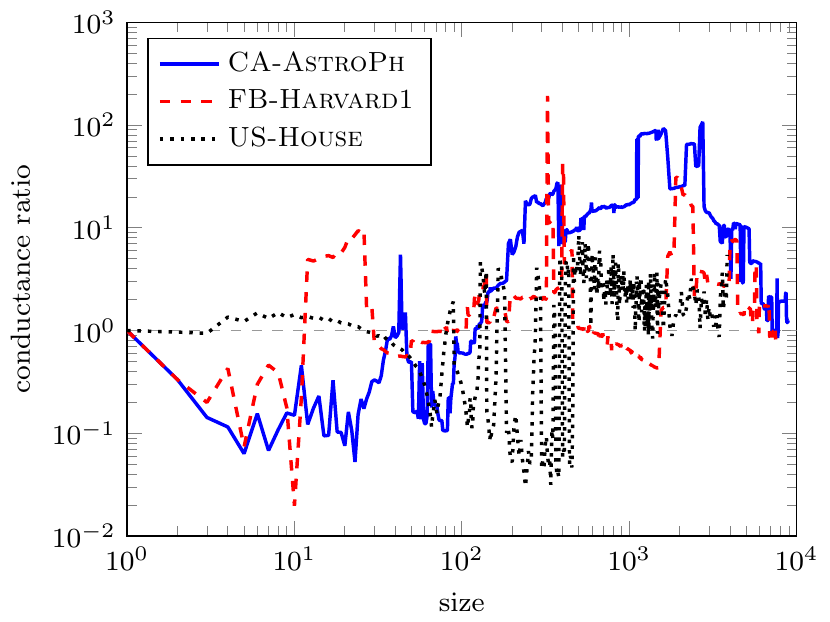}
}
\caption{
NCP plots [in panel \sref{NCP_ACL_large}] and CRP 
plots [in panel \sref{condratio_ACL_large}] for
\textsc{CA-AstroPh}, \textsc{FB-Harvard1}, and \textsc{US-House}
(i.e., the larger network in each of the three pairs of networks 
from Table\nobreakspace \ref {tab:data_summary}) generated using the \textsc{AclCut} method.
}
\label{ACL_large}
\end{figure*}

\subsection{Network Community Profiles}

We start by presenting our main results from using the \textsc{AclCut} 
method (see Figs.\nobreakspace \ref {ACL_small} and\nobreakspace  \ref {ACL_large}). One obtains similar insights about 
global structure using the \textsc{MovCut} (see Appendix\nobreakspace \ref {sxn:NCP-MOV}) and \textsc{EgoNet} (see Appendix\nobreakspace \ref {sxn:NCP-EGO}), although they can exhibit rather different local behavior.

In Fig.\nobreakspace \ref {ACL_small}, we show the NCPs and CRPs for the smaller network from each of the three pairs of networks from Table\nobreakspace \ref {tab:data_summary}. In Fig.\nobreakspace \ref {ACL_large}, we show the results for the corresponding larger networks. Note the logarithmic scale for both the vertical and horizontal axes in these figures as well as in subsequent NCP and CRP plots. 
Observe from Figs.\nobreakspace \ref {NCP_ACL_small} and\nobreakspace  \ref {NCP_ACL_large} that the NCPs for networks of the same type are qualitatively similar, whereas NCPs for networks of different types have qualitatively distinct shapes.
\begin{itemize}
\item
For the co-authorship networks \textbf{\textsc{CA-GrQc}} and \textbf{\textsc{CA-AstroPh}}, the NCPs have a mostly upward-sloping shape, except for the region with fewer than 100 nodes. We conclude that \textsc{CA-GrQc} and \textsc{CA-AstroPh} have good small (e.g., consisting of tens of nodes) communities, but they do not have good large (e.g., 
consisting of hundreds or thousands of nodes) communities.
These results are consistent with the NCPs of LiveJournal from
Fig.\nobreakspace \ref {fig:ncp-bigncp} and with the results of~\cite{Leskovec:2008vo,Leskovec:2009fy,Leskovec:2010uj}. 
Additionally, the high values for the CRPs for the co-authorship networks (see Figs.\nobreakspace \ref {condratio_ACL_small} and\nobreakspace  \ref {condratio_ACL_large}) for communities with hundreds or thousands of nodes reveals that these large communities are loosely connected collections of good, small communities. This feature is also visible in Fig.\nobreakspace \ref {spring_ACL_GR}, which shows selected communities and their neighborhoods for the \textsc{CA-GrQc} network. 
\item
For the Facebook networks \textbf{\textsc{FB-Johns55}} and \textbf{\textsc{FB-Harvard1}}, all of the communities at every size (except for two small ``communities'' with 5 and 10 nodes in \textsc{FB-Harvard1}~\footnote{The two small  ``communities''  in \textsc{FB-Harvard1} are responsible for the low value of $\lambda_2$ for this network compared with all of the other Facebook networks (recall Fig.\nobreakspace \ref {fig:lambda_2}). Removing these two communities from the network increases the value of 
$\lambda_2$ by an order of magnitude (from $0.0094$ to $0.075$), yielding a 
value that is fairly typical for networks in the \textsc{Facebook100} data set. 
In other words, aside from the 15 nodes from these two ``communities,'' 
\textsc{FB-Harvard1} looks like \textsc{FB-Johns55} and the rest of the 
\textsc{Facebook100} networks. This is one of few examples where a feature of an NCP is \emph{not} robust.})
have very large conductances (greater than $10^{-1}$).  This indicates that the communities in this network all have very poor community quality, in sharp contrast (though for different reasons) with both the co-authorship and voting networks. The essentially flat shape for the NCPs of the Facebook networks illustrate that these networks have strong expander-like properties (see Appendix\nobreakspace \ref {sxn:expanders}) and relatedly that there are no substantial bottlenecks to the rapid mixing of random walks on these networks. Both Facebook networks have noticeable dips in their NCPs at larger community sizes (about 220 and 1100 nodes for \textsc{FB-Johns55}, and about 1500 nodes for \textsc{FB-Harvard1}), and the sets of nodes associated with each of these dips correlate strongly with self-reported demographic information~\footnote{For \textsc{FB-Johns55}, the community of size about 220 corresponds closely to a set of students with the same major, and the community associated with the dip near 1100 corresponds to first-year students. Similarly, the community associated with the dip near 1500 for \textsc{FB-Harvard1} also corresponds to first-year students.
Consistent with the results of Ref.~\cite{Traud:2012ft}, we find similar poorly-connected and moderately-large communities that correlate reasonably well with first-year students in most of the other networks in the \textsc{Facebook100} data set.  (As usual, the most notable exception among the \textsc{Facebook100} networks 
is \textsc{Caltech36}, which is known to be influenced much more by dormitory 
(``House'') residence than by class year \cite{Traud:2012ft}.)
  }.
\item
For the voting networks \textbf{\textsc{US-Senate}} and \textsc{US-House}, the NCP has a predominantly downward-sloping 
shape. 
This is characteristic of ``low-dimensional'' networks, in the sense that we 
described informally in Section\nobreakspace \ref {sxn:prelim-looklike}. 
Informally, the reason for the downward-sloping shape is that 
\textsc{US-Senate} and \textsc{US-House} consist of a low-dimensional structure that is evolving 
along a one-dimensional scaffolding (i.e., time), upon which the detailed 
structure of individual Congresses (i.e., a good partition that is nearly along party lines) is 
superimposed.  
(One can examine such structures by using smaller values of the interlayer 
coupling parameter; see Ref.~\cite{Mucha:2010th}.) 
This is consistent with previous results \cite{CM11_TR}.
\end{itemize}

These results, which illustrate that community quality changes very differently with 
size in each of the three pairs of networks, also indicate that these three types of networks
have very different properties with respect to large-scale versus 
small-scale community structure. Moreover, the qualitative similarity in behavior between the two networks in each pair suggests that the coarse behavior of an NCP (downward-sloping, upward-sloping, or flat) is indicative of large classes of networks and not an artifact of our particular choice of example networks. One obtains similar insights about 
global structure using the \textsc{MovCut} (see Appendix\nobreakspace \ref {sxn:NCP-MOV}) and \textsc{EgoNet} (see Appendix\nobreakspace \ref {sxn:NCP-EGO}) methods, although they can exhibit rather different local behavior. We investigate these differences in local behavior in Section\nobreakspace \ref {sxn:local-comp}.


\subsection{Comparison of Results from {\sc AclCut}, {\sc MovCut}, and {\sc EgoNet}}
\label{sxn:local-comp}

The NCPs generated using either \textsc{AclCut} or \textsc{MovCut} (see Appendix\nobreakspace \ref {sxn:NCP-MOV}), and 
to a slightly lesser extent using \textsc{EgoNet} (see Appendix\nobreakspace \ref {sxn:NCP-EGO}), have similar global 
features---i.e., they exhibit the same general trends and have dips at small
size scales that correspond to nearly identical communities---indicating that 
we obtain a broadly similar picture of the large-scale community structure by 
using any of the methods. 
However, the detailed local behavior of the three methods can differ 
considerably. 
Such behavior depends sensitively on the choice of seed node, the choice of 
the parameters in the different methods, and the specific details of each 
method. 
In this section, we discuss the similarities and differences in the results 
from these methods.  
In this section, we only do calculations for the smaller networks from each 
of the three network pairs in Table\nobreakspace \ref {tab:data_summary} (but we have observed
similar results on the larger networks).

To compare different methods, we note that any meaningful difference between 
them should manifest itself as a difference in the rank order of nodes, as 
this determines the assignment of nodes to local communities. 
We quantify rank differences by computing the Spearman rank correlation~\cite{Spearman:1904ua} 
between the (exact for \textsc{MovCut} and approximate for \textsc{AclCut}) PPR and EgoRank ranking vectors. 
To make results from \textsc{AclCut} and \textsc{MovCut} comparable, we exploit the relation between $\gamma$ and $\alpha$ (see Appendix\nobreakspace \ref {sxn:measures}) to parametrize the \textsc{MovCut} method in terms of $\alpha$. We also restrict all comparisons to the support of the corresponding approximate PPR vector that we obtained using the \textsc{AclCut} method. This induces an indirect dependency of the results from \textsc{MovCut} and \textsc{EgoNet} on $\alpha$ and $\epsilon$ (in addition to the direct dependency of \textsc{MovCut} on $\alpha$).


\begin{table*}[tb]
\begin{ruledtabular}
\begin{tabular}{ccr|rrr|rrr|rrr|rrr|rrr}
\multicolumn{3}{c}{}&\multicolumn{15}{c}{\Large$\alpha$}\\
\multicolumn{3}{c}{}&\multicolumn{3}{c|}{0.6}&\multicolumn{3}{c|}{0.7}&\multicolumn{3}{c|}{0.8}&\multicolumn{3}{c|}{0.9}&\multicolumn{3}{c}{0.99}\\
\multicolumn{3}{c|}{}&A-M&A-E&M-E&A-M&A-E&M-E&A-M&A-E&M-E&A-M&A-E&M-E&A-M&A-E&M-E\\
\cline{2-18}
\multirow{12}{1em}[-1em]{\Large$\epsilon$}
&\multirow{3}{*}{$10^{-3}$}
&max&\phantom{$-$}1.00&\phantom{$-$}0.99 &\phantom{$-$}0.99 &\phantom{$-$}1.00&\phantom{$-$}0.98 &\phantom{$-$}0.98 &\phantom{$-$}1.00&\phantom{$-$}0.97 &\phantom{$-$}0.97 &\phantom{$-$}1.00 &\phantom{$-$}0.95 &\phantom{$-$}0.95 &\phantom{$-$}0.98&\phantom{$-$}0.89 &\phantom{$-$}0.86 \\
&&mean&0.98&0.78 &0.77 &0.98&0.76 &0.73 &0.98&0.72 &0.68 &0.97&0.68 &0.62 &0.91&0.60 &0.48 \\
&&min&0.92&0.26 &0.23 &0.91&0.18 &0.14 &0.94&0.21 &0.15 &0.85&$-$0.01 &$-$0.05 &0.74&0.25 &0.06 \\
\cline{2-18}
&\multirow{3}{*}{$10^{-4}$}
&max&1.00&0.97 &0.97 &1.00&0.97 &0.97 &1.00&0.94 &0.94 &0.99&0.89 &0.85 &$\mathbf{0.92}$&0.69 &0.53 \\
&&mean&0.99&0.74 &0.72 &0.98&0.73 &0.68 &0.97&0.70 &0.64 &0.95&0.63 &0.54 &0.89&0.51 &0.36 \\
&&min&0.96&0.16 & 0.10&0.91&$-$0.05 &$-$0.19 &0.84&0.35 &$-$0.02 &0.88&0.43 &0.25 &0.85&0.32 &0.18 \\
\cline{2-18}
&\multirow{3}{*}{$10^{-5}$}
&max&1.00&0.96 &0.95 &0.97&0.92 &0.87 &0.94&0.81 &0.67 &0.89&0.73 &0.55 &0.93&0.75 &0.60 \\
&&mean&0.91&0.74 &0.58 &0.89&0.69 &0.51 &0.85&0.65 &0.42 &0.78&0.62 &0.33 &$\mathbf{0.84}$&0.63 &0.36 \\
&&min&0.24&0.21 &$-$0.20 &0.42&0.30 &$-$0.10 &0.42&0.39 &$-$0.13 &0.25&0.43 &$-$0.10 &$\mathbf{0.49}$&0.44 &0.05 \\
\cline{2-18}
&\multirow{3}{*}{$10^{-6}$}
&max&$\mathbf{0.84}$&0.85 &0.68 &$\mathbf{0.79}$&0.81 &0.49 &$\mathbf{0.70}$&0.79 &0.39 &$\mathbf{0.80}$&0.81 &0.47 &0.93&0.75 &0.60 \\
&&mean&$\mathbf{0.62}$&0.72 &0.25 &$\mathbf{0.57}$&0.69 &0.17 &$\mathbf{0.51}$&0.69 &0.12 &$\mathbf{0.51}$&0.72 &0.13 &$0.85$&0.63 &0.37 \\
&&min&$\mathbf{0.01}$&0.57 &$-$0.36 &$\mathbf{0.07}$&0.52 &$-$0.30 &$\mathbf{-0.06}$&0.50 &$-$0.27 &$\mathbf{-0.13}$&0.57 &$-$0.32 &$0.52$&0.46 &0.08 \\
\end{tabular}
\end{ruledtabular}
\caption{Pairwise comparison of the three methods using the Spearman rank correlation between the rank vectors from the \textsc{AclCut} (A), \textsc{MovCut} (M), and \textsc{EgoNet} (E) methods for \textsc{CA-GrQc}. We use a uniform random sample of 50 nodes for each of several values for the teleportation parameter $\alpha$ and truncation parameter $\epsilon$. We take the maximum, mean, and minimum over the seed nodes. Bold values highlight the largest deviations between \textsc{AclCut} and \textsc{MovCut} methods for a given value of~$\alpha$.
\label{spearman_comp_GR}}
\end{table*}

\begin{table*}[tb]
\begin{ruledtabular}
\begin{tabular}{ccr|rrr|rrr|rrr|rrr|rrr}
\multicolumn{3}{c}{}&\multicolumn{15}{c}{\Large$\alpha$}\\
\multicolumn{3}{c}{}&\multicolumn{3}{c|}{0.6}&\multicolumn{3}{c|}{0.7}&\multicolumn{3}{c|}{0.8}&\multicolumn{3}{c|}{0.9}&\multicolumn{3}{c}{0.99}\\
\multicolumn{3}{c|}{}&A-M&A-E&M-E&A-M&A-E&M-E&A-M&A-E&M-E&A-M&A-E&M-E&A-M&A-E&M-E\\
\cline{2-18}
\multirow{12}{1em}[-1em]{\Large$\epsilon$}
&\multirow{3}{*}{$10^{-3}$}
&max&\phantom{$-$}1.00&\phantom{$-$}1.00 &\phantom{$-$}1.00 &\phantom{$-$}1.00&\phantom{$-$}1.00 &\phantom{$-$}1.00 &\phantom{$-$}1.00&\phantom{$-$}1.00 &\phantom{$-$}1.00 &\phantom{$-$}1.00&\phantom{$-$}1.00 &\phantom{$-$}1.00 &\phantom{$-$}1.00&\phantom{$-$}1.00 &\phantom{$-$}1.00  \\
&&mean&1.00&0.85 &0.84 &1.00&0.82 &0.81 &0.99&0.79 &0.79 &0.98&0.78 &0.77 &0.98&0.78 &0.75 \\
&&min&0.95& 0.61&0.56 &0.94&0.54 &0.47 &0.80&0.55 &0.53 &0.80&0.55 &0.49 &0.77&0.40 &0.39 \\
\cline{2-18}
&\multirow{3}{*}{$10^{-4}$}
&max&1.00&0.89 &0.89 &1.00&0.93 &0.93 &1.00&0.91 &0.91 &1.00&0.89 &0.89 &1.00&0.88 &0.88 \\
&&mean&0.98&0.48 &0.47 &0.97&0.45 &0.44 &0.96&0.43 &0.41 &0.95&0.41 &0.37 &0.94&0.38 &0.33 \\
&&min&0.81&0.06 &0.05 &0.78&0.05 &0.04 &0.74&0.07 &0.06 &0.69&0.07 &0.07 &0.69&0.07 &0.00 \\
\cline{2-18}
&\multirow{3}{*}{$10^{-5}$}
&max&1.00&0.85 &0.85 &1.00&0.86 &0.85 &1.00&0.84 &0.83 &0.99&0.84 &0.82 &1.00&0.81 &0.78 \\
&&mean&0.98& 0.58&0.54 &0.97&0.59 &0.54 &0.97&0.63 &0.57 &0.96&0.61 &0.54 &0.95&0.53 &0.46 \\
&&min&0.90& 0.24&0.18 &0.91&0.26 &0.17 &0.91&0.23 &0.14 &0.89&0.22 &0.07 &0.88&0.18 &0.04 \\
\cline{2-18}
&\multirow{3}{*}{$10^{-6}$}
&max&$\mathbf{0.99}$&0.75 &0.46 &$\mathbf{0.97}$&0.69 &0.46 &$\mathbf{0.91}$&0.71 &0.39 &$\mathbf{0.90}$&0.73 &0.47 &$\mathbf{0.97}$&0.69 &0.57 \\
&&mean&$\mathbf{0.79}$&0.41 &0.20 &$\mathbf{0.75}$&0.45 &0.13 &$\mathbf{0.72}$&0.56 &0.20 &$\mathbf{0.77}$&0.62 &0.32 &$\mathbf{0.88}$&0.59 &0.41 \\
&&min&$\mathbf{0.57}$&0.23 &$-$0.07 &$\mathbf{0.49}$&0.24 &$-$0.06 &$\mathbf{0.43}$& 0.32 &$-$0.02 &$\mathbf{0.49}$ &0.32 &0.01 &$\mathbf{0.60}$&0.26 &0.07 \\
\end{tabular}
\end{ruledtabular}
\caption{Pairwise comparison of the three methods using the Spearman rank correlation between the ranking vectors from the \textsc{AclCut} (A), \textsc{MovCut} (M), and  \textsc{EgoNet} (E) methods for \textsc{FB-Johns55}. We use a uniform random sample of 50 nodes for each of several values for the teleportation parameter $\alpha$ and truncation parameter $\epsilon$. We take the maximum, mean, and minimum over the seed nodes. Bold values highlight the largest deviations between \textsc{AclCut} and \textsc{MovCut} methods for a given value of~$\alpha$.
\label{spearman_comp_Johns55}}
\end{table*}

\begin{table*}[tb]
\begin{ruledtabular}
\begin{tabular}{ccr|rrr|rrr|rrr|rrr|rrr}
\multicolumn{3}{c}{}&\multicolumn{15}{c}{\Large$\alpha$}\\
\multicolumn{3}{c}{}&\multicolumn{3}{c|}{0.6}&\multicolumn{3}{c|}{0.7}&\multicolumn{3}{c|}{0.8}&\multicolumn{3}{c|}{0.9}&\multicolumn{3}{c}{0.99}\\
\multicolumn{3}{c|}{}&A-M&A-E&M-E&A-M&A-E&M-E&A-M&A-E&M-E&A-M&A-E&M-E&A-M&A-E&M-E\\
\cline{2-18}
\multirow{12}{1em}[-1em]{\Large$\epsilon$}
&\multirow{3}{*}{$10^{-3}$}
&max&\phantom{$-$}1.00&\phantom{$-$}1.00 &\phantom{$-$}1.00 &\phantom{$-$}1.00&\phantom{$-$}1.00 &\phantom{$-$}1.00 &\phantom{$-$}1.00&\phantom{$-$}1.00 &\phantom{$-$}1.00 &\phantom{$-$}1.00&\phantom{$-$}1.00 &\phantom{$-$}1.00 &\phantom{$-$}1.00&\phantom{$-$}1.00 &\phantom{$-$}1.00  \\
&&mean&$\mathbf{0.65}$&0.66 &0.40 &0.68&0.71 &0.30 &0.74&0.77 &0.46 &$\mathbf{0.74}$&0.85 &0.57 &$\mathbf{0.80}$&0.80 &0.56 \\
&&min&0.02&0.50 &$-0.05$ &0.38&0.32 &$-0.24$ &0.46&0.53 &$-0.07$ &0.50&0.68&$-0.02$ &$\mathbf{0.51}$&0.62 &$-$0.00 \\
\cline{2-18}
&\multirow{3}{*}{$10^{-4}$}
&max&1.00 &0.93 &0.87 &1.00&0.92 &0.77 &1.00&0.93 &0.82 &1.00&0.89 &0.75 &$\mathbf{0.93}$&0.81 &0.61 \\
&&mean&$0.75$&0.61 &0.36 &$\mathbf{0.54}$&0.59 &0.17 &$\mathbf{0.67}$&0.62 &0.34 &$0.86$&0.50& 0.34&$0.92$&0.36 &0.23 \\
&&min&$\mathbf{-0.03}$&$-0.16$ &$-0.51$ &$\mathbf{-0.32}$&$-$0.11 &$-$0.63 &$\mathbf{-0.44}$&0.15 &$-$0.36 &$\mathbf{0.50}$&$-$0.08 &$-$0.26 &$0.77$&$-$0.20 &$-$0.38 \\
\cline{2-18}
&\multirow{3}{*}{$10^{-5}$}
&max&1.00&0.91 &0.90 &1.00&0.89 &0.87 &1.00&0.85 &0.83 &1.00&0.80 &0.78 &1.00&0.85 &0.84 \\
&&mean&0.99&0.58&0.54 &0.99&0.48 &0.46 &0.99&0.42 &0.40 &0.96&0.44 &0.40 &0.95&0.43&0.39 \\
&&min&0.86&0.03 &0.01 &0.88&$-$0.02 &$-$0.02 &0.88&$-$0.09 &$-$0.13 &0.80&$-$0.17 &$-$0.20 &0.80&0.07 &0.01 \\
\cline{2-18}
&\multirow{3}{*}{$10^{-6}$}
&max&1.00&0.84 &0.84 &1.00&0.84 &0.84 &1.00&0.84 &0.85 &1.00&0.87 &0.86 &1.00&0.96 &0.96 \\
&&mean&0.94&0.54 &0.48 &0.98&0.55 &0.50 &0.99&0.62 &0.60 &1.00&0.69 &0.68 &0.99&0.91 &0.89 \\
&&min&0.85&0.30 &0.22 &0.89&0.08 &0.19 &0.98&0.25 &0.21 &0.98&0.30 &0.29 &0.96&0.83& 0.73 \\
\end{tabular}
\end{ruledtabular}
\caption{Pairwise comparison of the three methods using the Spearman rank correlation between the ranking vectors from the \textsc{AclCut} (A), \textsc{MovCut} (M), and  \textsc{EgoNet} (E) methods for \textsc{US-Senate}. We use a uniform random sample of 50 nodes for each of several values for the teleportation parameter $\alpha$ and truncation parameter $\epsilon$. We take the maximum, mean, and minimum over the seed nodes. Bold values highlight the largest deviations between \textsc{AclCut} and \textsc{MovCut} methods for a given value of~$\alpha$.
\label{spearman_comp_Senate}}
\end{table*}


In Tables\nobreakspace  \ref {spearman_comp_GR}--\ref {spearman_comp_Senate}, we show the results of our calculations of Spearman rank correlations. For each of the three networks, we select 50 seed nodes by sampling uniformly without replacement.
We then compute PPR vectors for these seed nodes using the \textsc{AclCut} 
and \textsc{MovCut} method for different values of the truncation parameter 
$\epsilon$ and teleportation parameter $\alpha$, and we also compute the 
EgoRank vector for each of the seed nodes. 
Recall that \emph{smaller} values of $\alpha$ correspond to \emph{more local} 
versions of the procedures, but that \emph{larger} values of $\epsilon$ 
correspond to \emph{more local} versions of the procedures.

The \textsc{AclCut} and \textsc{MovCut} methods give very similar results 
for most of the 50 seed nodes in our sample, although (as discussed below) 
some seed nodes do yield noticeable differences. 
The two methods give the most similar results for \textsc{FB-Johns55} 
(mean: 0.92, minimum: 0.43), whereas we find larger deviations in both 
\textsc{CA-GrQc} (mean: 0.85, minimum: $-0.13$) and \textsc{US-Senate} 
(mean: 0.86, minimum: $-0.44$). 
Note that we calculated the mean, maximum, and minimum over all sampled 
seed nodes and parameter values.

Interestingly, the larger deviations between the two methods for 
\textsc{CA-GrQc} and \textsc{US-Senate} occur at different values of the 
truncation parameter $\epsilon$. 
For \textsc{CA-GrQc} (and, to a lesser extent, for \textsc{FB-Johns55}), we 
obtain the largest deviations for smaller values (e.g., $\epsilon=10^{-6}$). 
For \textsc{US-Senate}, however, we obtain the largest deviations for 
$\epsilon=10^{-4}$. 
See the bold values in Tables\nobreakspace  \ref {spearman_comp_GR}--\ref {spearman_comp_Senate}. 
This is consistent with the very different isoperimetric properties of 
these three networks, as revealed by their NCPs, as well as with 
well-known connections between conductance and random walks.

There are two potential causes for the differences between the \textsc{AclCut} and \textsc{MovCut} method. 
First, there is a truncation effect, governed by the parameter $\epsilon$, in approximating the PPR vector using the \textsc{AclCut} method. 
As $\epsilon$ becomes smaller, the approximation in \textsc{AclCut} becomes more accurate and this effect diminishes. 
Second, the two methods differ in the precise way that they use a seed vector to represent a seed node. 
Recall that the \textsc{AclCut} method uses an indicator vector $\vec{s}$ to represent a seed node $i$; thus, we use $\vec{s}_i=1$ whenever $i$ is a seed node, and we set all other entries in that vector to $0$.  
In contrast, the \textsc{MovCut} method projects the indicator vector onto the orthogonal complement of the strength vector to ensure that $\vec{s}^T D \vec{1} =0$ (see Appendix\nobreakspace \ref {sxn:measures}). 
This effect decreases as~$\alpha \rightarrow 1$.  

\begin{figure*}[tb!]
\includegraphics[width=\linewidth]{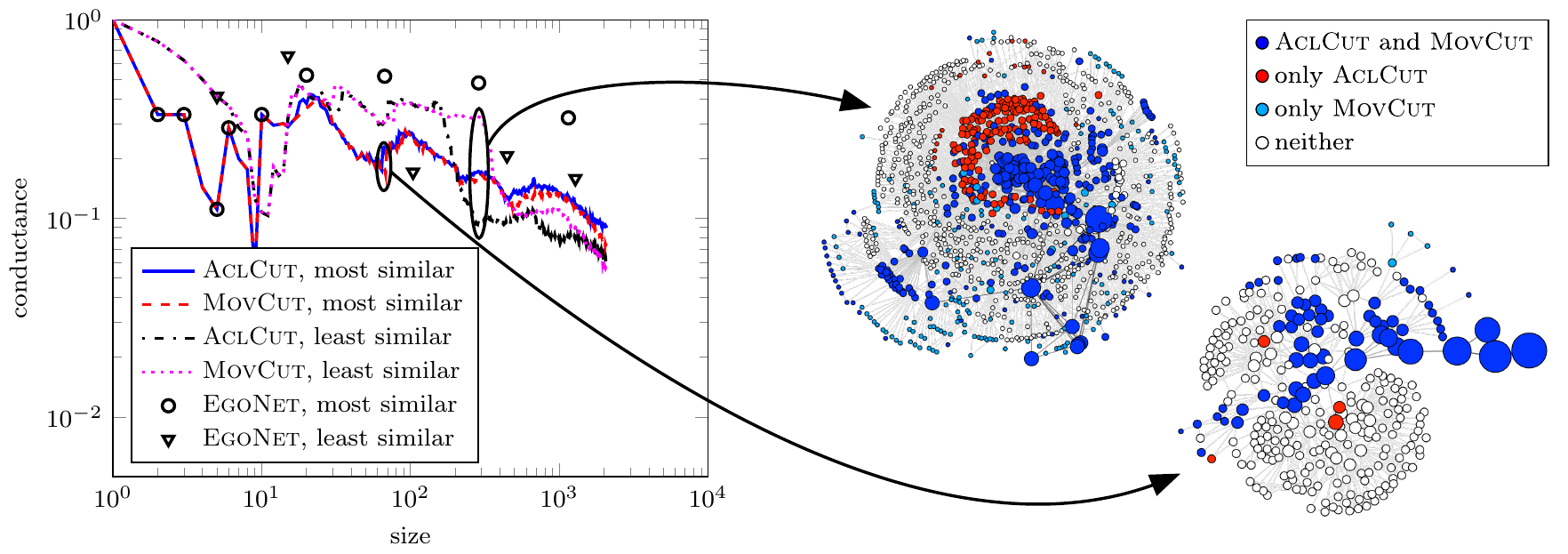}
\subfloat[Local NCP\label{fig:local_ncp_GR}]{\hspace{0.5\linewidth}}
\subfloat[Spring-embedding visualizations\label{fig:spring_local_GR}]{\hspace{0.5\linewidth}}
\caption{\textsc{CA-GrQc}: 
(a) Local NCPs for the seed nodes (out of the 50 nodes that we sampled) with the highest and lowest mean Spearman correlation over the sampled parameter 
values.  These NCPs highlight the difference in behavior for the two methods for 
large communities.
(b) Kamada-Kawai-like spring-embedding visualization \cite{viscoms} of (bottom right) the 9-neighborhood of the seed node with the smallest difference between the two methods and (top left) the 6-neighborhood of the seed node with the largest difference.
In these two visualizations, the node size decreases as a function of geodesic distance from the seed node. 
We color the nodes according to whether they belong to the local 
community that we obtained using the \textsc{AclCut} method, the one we obtained using the \textsc{MovCut} method, or both methods. 
\label{fig:local_GR} 
}
\end{figure*}

\begin{figure*}[tb!]
\includegraphics[width=\linewidth]{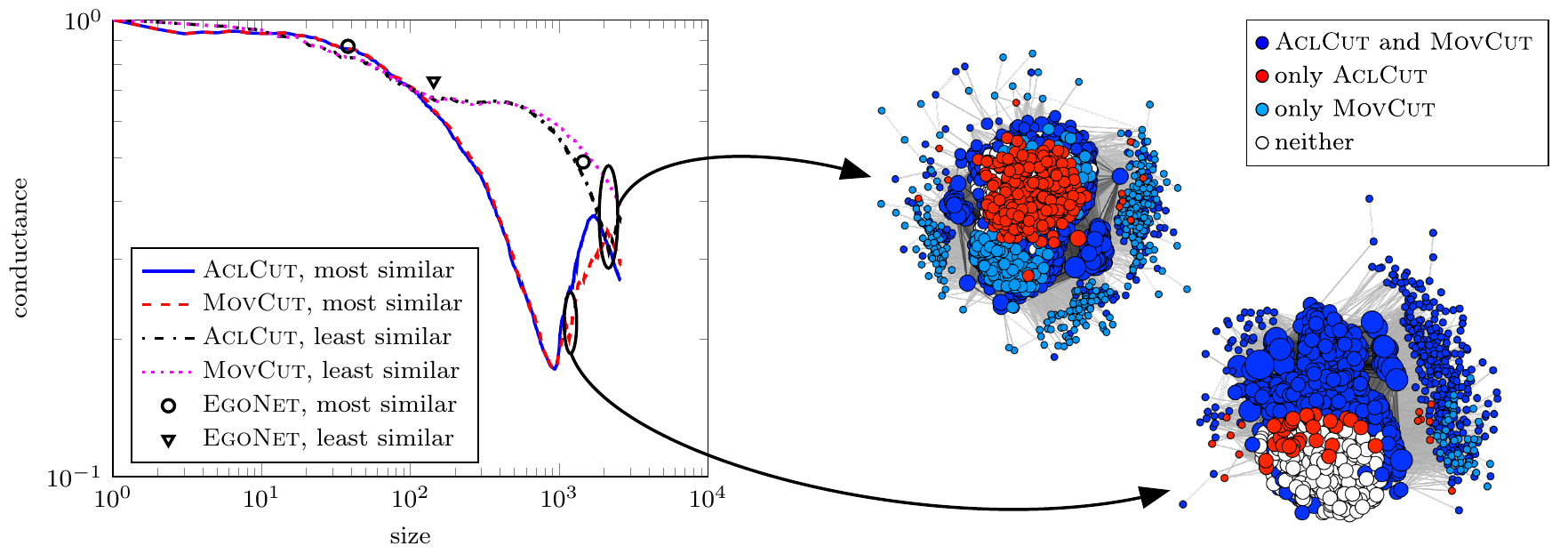}
\subfloat[Local NCP\label{fig:local_ncp_johns}]{\hspace{0.5\linewidth}}
\subfloat[Spring-embedding visualizations\label{fig:spring_local_johns}]{\hspace{0.5\linewidth}}
\caption{\textsc{FB-Johns55}: 
(a) Local NCPs for the seed nodes (out of the 50 nodes that we sampled) with the highest and lowest mean Spearman correlation over the sampled parameter 
values.  These NCPs highlight the difference in behavior for the two methods for 
large communities.
(b) Kamada-Kawai-like spring-embedding visualization \cite{viscoms} of the 2-neighborhoods 
of both seed nodes.  The one with the smallest difference in the bottom right and the one with the largest difference in the top left.  In these two visualizations, the node size decreases as a function of geodesic distance from the seed node. 
The smallest nodes are more than 2 steps away from the seed node, but 
they appear in at least one of the local communities. 
We color the nodes according to whether they belong to the local 
community that we obtained using the \textsc{AclCut} method, the one we obtained using the \textsc{MovCut} method, or both methods. 
\label{fig:local_Johns55}  
}
\end{figure*}

The larger deviations between the two methods occur for smaller values of 
$\epsilon$ in \textsc{CA-GrQc} and \textsc{FB-Johns55}; for these, the 
truncation effect is small, suggesting that the different way of 
representing a seed node is partially responsible for the difference between 
the results of the two methods for these networks. 
For larger values of $\epsilon$ (in particular, $\epsilon\geq 10^{-4}$), 
where the support of the approximate PPR vector from the \textsc{AclCut} 
method is small, the behavior of the two methods is very similar. 
Consequently, the differences in the choice of seed vector become more 
important for nodes that are ``far away'' from the seed node, in the sense 
that they are rarely visited by the personalized PageRank dynamics that 
underlie these methods. 
As a result, the ``local NCPs'' for the two methods in 
Figs.\nobreakspace \ref {fig:local_ncp_GR} and\nobreakspace  \ref {fig:local_ncp_johns} are largely identical for small 
community sizes but diverge for large community sizes.  
(We use the term \emph{local NCP} to refer to an NCP that we computed using 
only a single seed node without optimizing over the results from multiple 
seed choices;  see Ref.~\cite{Mahoney:2012wl} for details on the 
construction of local NCPs.)

For \textsc{US-Senate}, the two methods behave almost identically for small $\epsilon$ (see Table\nobreakspace \ref {spearman_comp_Senate}), so we conclude that the different ways of representing a seed node have only a small effect on this network. However, the truncation effect is more pronounced in this network compared with \textsc{CA-GrQc} or \textsc{FB-Johns55}. This feature manifests as larger deviations between \textsc{AclCut} and \textsc{MovCut} in Table\nobreakspace \ref {spearman_comp_Senate} for large $\epsilon$ and small $\alpha$ (i.e., where the truncation has the strongest impact). The discrepancy occurs because the \textsc{AclCut} method initially pushes
a large amount of probability to the interlayer neighbors of the seed node 
(i.e., to the same Senator in different Congresses).  This probability does not diffuse to other nodes for sufficiently large values of $\epsilon$.

\begin{figure*}[tb!]
\includegraphics[width=\linewidth]{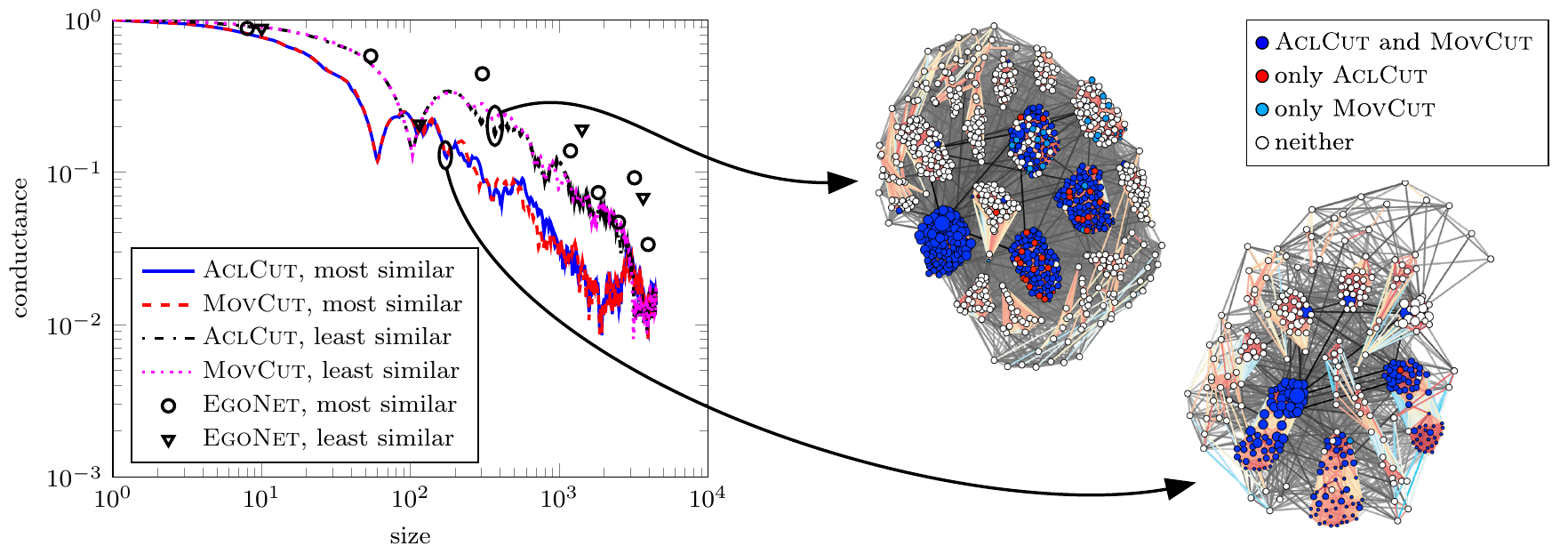}
\subfloat[Local NCP\label{fig:local_ncp_Senate}]{\hspace{0.5\linewidth}}
\subfloat[Spring-embedding visualizations\label{fig:spring_local_Senate}]{\hspace{0.5\linewidth}}
\caption{\textsc{US-Senate}: 
(a) Local NCP for the seed nodes out of the 50 nodes sampled with the 
highest and lowest average Spearman correlation over sampled parameter 
values.
These NCPs highlight the difference in behavior for the two methods for 
large communities.
(b) Kamada-Kawai-like spring-embedding visualization \cite{viscoms} of the 3-neighborhoods of both seed nodes. The one with the smallest difference in the bottom right and the one with the largest difference in the top left.  In these two visualizations, the node size decreases as a function of geodesic distance from the seed node.
We color the nodes according to whether they belong to the local 
community that we obtained using the \textsc{AclCut} method, the one we obtained using the \textsc{MovCut} method, or both methods. 
\label{fig:local_Senate} 
}
\end{figure*}

In Figs.\nobreakspace  \ref {fig:local_GR}--\ref {fig:local_Senate}, we illustrate the results from Tables\nobreakspace  \ref {spearman_comp_GR}--\ref {spearman_comp_Senate}. In these figures, we plot the local 
NCPs 
for  \textsc{CA-GrQc}, \textsc{FB-Johns55}, and \textsc{US-Senate} for the seed nodes (from the sample of 50) that yield the highest and lowest mean Spearman rank correlation between the 
\textsc{AclCut} and \textsc{MovCut} methods.  In these figures, we also include visualizations of example communities that we obtained from the \textsc{AclCut} and \textsc{MovCut} methods using
a Kamada-Kawai-like spring-embedding visualization~\cite{viscoms} of the 
$k$-ego-nets of these seed nodes. 

From the visualizations of the local communities, it seems for \textsc{CA-GrQc} (see Fig.\nobreakspace \ref {fig:local_GR}) and \textsc{FB-Johns55} (see Fig.\nobreakspace \ref {fig:local_Johns55}) that nodes included in local communities obtained from \textsc{AclCut} tend to be closer in geodesic distance than those obtained from \textsc{MovCut} to the seed node.  (To see this, observe that red nodes tend to be larger than light blue nodes in the visualization of the $k$-neighborhoods.)
If this observation holds more generally and is not just an artifact of the particular communities that we show in Figs.\nobreakspace \ref {fig:local_GR} and\nobreakspace  \ref {fig:local_Johns55}, then we should obtain higher Spearman rank correlations between \textsc{AclCut} and \textsc{EgoNet} than between \textsc{MovCut} and \textsc{EgoNet}.  Indeed, Tables\nobreakspace  \ref {spearman_comp_GR}--\ref {spearman_comp_Senate}  consistently show this effect for all choices of $\epsilon$ and $\alpha$ and for all three networks.  Note that this effect is also present in \textsc{US-Senate}, though it is less prominent in its $k$-neighborhood visualization than is the case for the other two networks.

Figures\nobreakspace  \ref {fig:local_GR}--\ref {fig:local_Senate}  also reveal that the 
three networks look very different from a local perspective. 
For \textsc{FB-Johns55} (see Fig.\nobreakspace \ref {fig:local_Johns55}), both seed nodes that 
we considered result in reaching a large fraction of all nodes after just 2 
steps. 
 This is consistent with known properties of the full Facebook graph 
(circa 2012) of individuals connected by reciprocal ``friendships.'' 
For example, the mean geodesic distance between pairs of nodes of the 
Facebook graph is very small: it was recently estimated by Facebook's Data 
Team and their collaborators to be about 4.74 \cite{ugander-fourdegrees}.  
Additionally, as reported by Facebook's Data Team, one can view Facebook as a collection of ego networks that have been patched together into a 
network whose global structure is sparse \cite{facebook-egonet} (and such 
structure is an important motivation for the locally-biased notion of 
community structure that we advocate in this paper).  

For \textsc{CA-GrQc}, we obtain very different neighborhoods starting from 
our two different seed nodes. 
The node that exhibits the largest difference in behavior for both the 
\textsc{AclCut} and \textsc{MovCut} methods appears to be better connected 
in the network in the sense that the $k$-neighborhood (for any $k$ until 
saturation occurs) is much larger than that of the node that showed the 
smallest difference. 
(That is, it is more in the ``core'' than in the ``periphery'' of the 
nested core-periphery structure of 
Refs.~\cite{Leskovec:2008vo,Leskovec:2009fy}.)
We observe a similar phenomenon for \textsc{FB-Johns55} and \textsc{US-Senate}. 
Furthermore, its 1-ego-net and 2-ego-net are highly clustered, in the sense 
that they contain many closed triangles. 
For the seed node that showed the smallest difference between the 
\textsc{AclCut} and \textsc{MovCut} methods, we need to consider the 
6-ego-net (which has 20 nodes) to obtain a network of similar size to the 
2-ego-net for the seed node with the largest difference (which has 15 nodes). 
In the case of the seed node in our sample that showed the least difference 
between the two methods, even the 6-ego-net appears rather tree-like; it 
contains few closed triangles and no larger cliques. 

For \textsc{US-Senate}, the 1-neighborhood of any seed node contains only the node itself and those corresponding to the same Senator in different Congresses~\footnote{This holds for all seed nodes because the maximum voting similarity between a pair of Senators is 1 (i.e., they voted the same way on every bill) and all interlayer edges (i.e., edges between the same Senator in different Congresses) have a weight of 1.  As one reduces the weight of interlayer edges, one needs to chose an increasingly large value of $k$ for nodes from interlayer edges to appear in the $k$-neighborhood. Hence, an increasing number of Senators from the same Congress can appear in a $k$-neighborhood that does not contain any nodes from interlayer edges.}.
As one begins to consider nodes that are further away, one first reaches corresponding Senators in other Congresses before reaching other Senators with similar voting patterns from the same Congress.  This behavior of the \textsc{EgoNet} method contrasts with the
(PageRank-based) \textsc{AclCut} and \textsc{MovCut} methods, which
tend to initially select all Senators from one Congress before reaching 
Senators from other Congresses.


\subsection{Meso-Scale Structure}
\label{sxn:mesoscale}

\begin{figure*}[tb]
\centering
\includegraphics[scale=0.5]{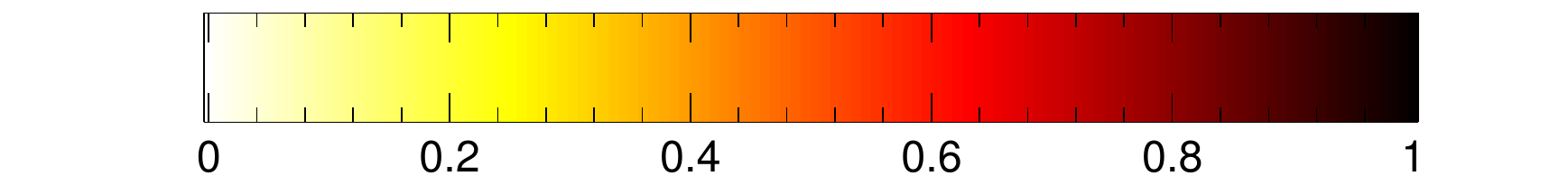}\\
\subfloat[\textsc{CA-GrQc} (\textsc{AclCut})\label{fig:GR_assoc_ACL}]{\includegraphics[height=5cm]{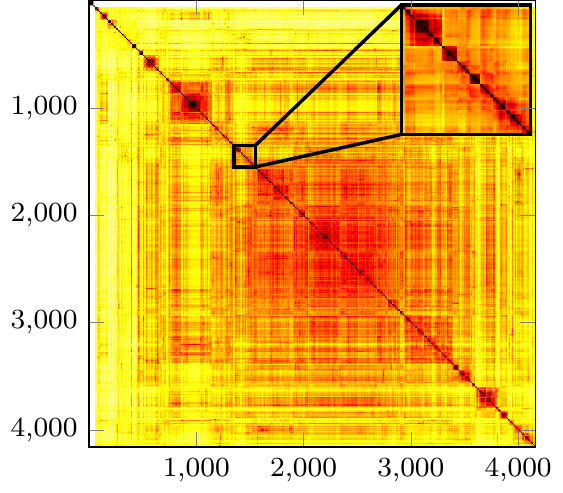}}\hfill
\subfloat[\textsc{FB-Johns55} (\textsc{AclCut}) \label{fig:Johns55_assoc_ACL}]{\includegraphics[height=5cm]{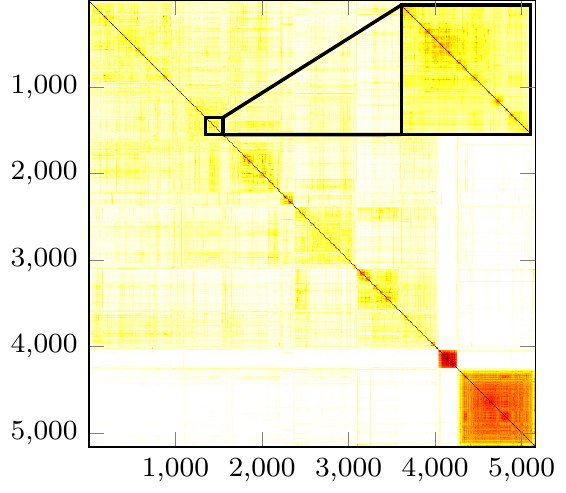}}\hfill
\subfloat[\textsc{US-Senate} (\textsc{AclCut}) \label{fig:Senate_assoc_ACL}]{\includegraphics[height=5cm]{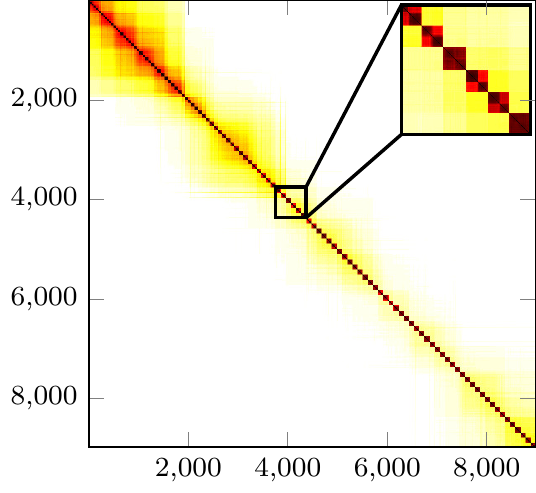}}
\\
\subfloat[\textsc{CA-GrQc} (\textsc{MovCut}) \label{fig:GR_assoc_MOV}]{\includegraphics[height=5cm]{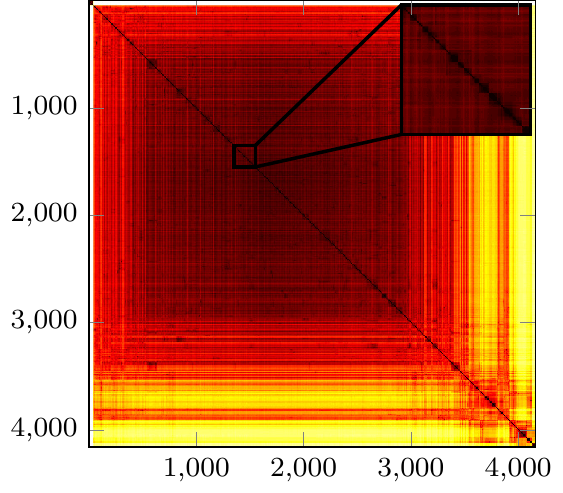}}
\hfill
\subfloat[\textsc{FB-Johns55} (\textsc{MovCut}) \label{fig:Johns55_assoc_MOV}]{\includegraphics[height=5cm]{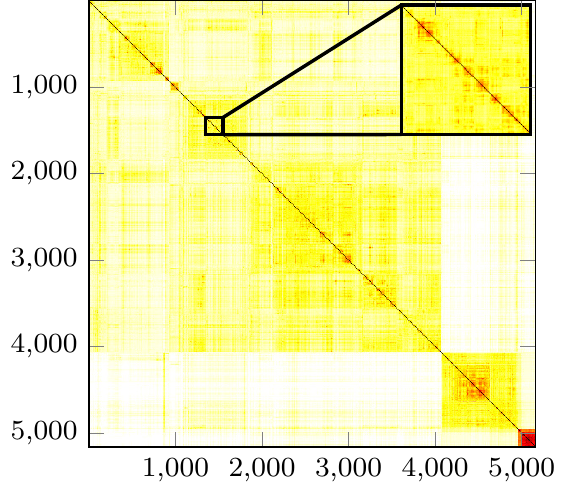}}
\hfill
\subfloat[\textsc{US-Senate} (\textsc{MovCut}) \label{fig:Senate_assoc_MOV}]{\includegraphics[height=5cm]{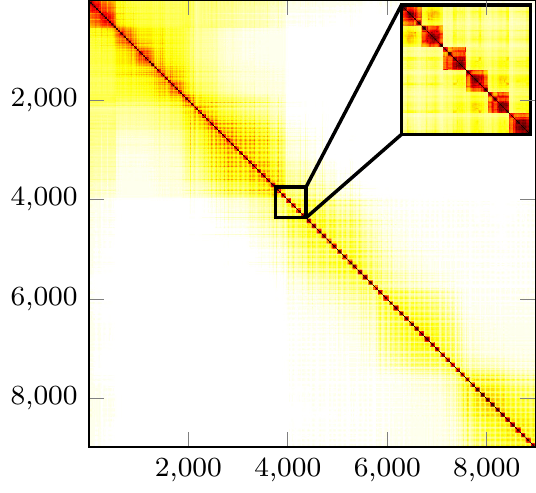}}
\\
\subfloat[\textsc{CA-GrQc} (\textsc{EgoNet})\label{fig:GR_assoc_EGO}]{\includegraphics[height=5cm]{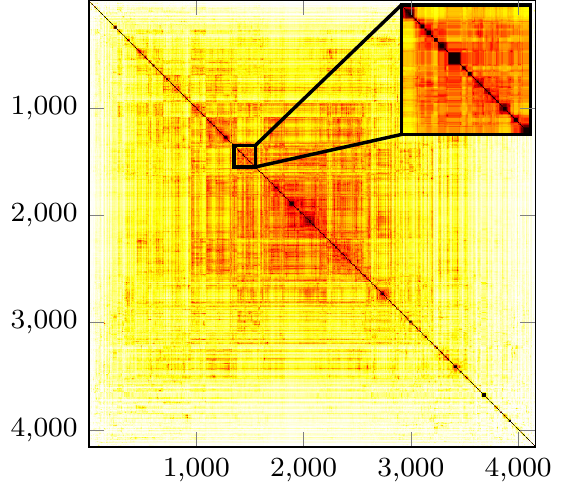}}\hfill
\subfloat[\textsc{FB-Johns55} (\textsc{EgoNet}) \label{fig:Johns55_assoc_EGO}]{\includegraphics[height=5cm]{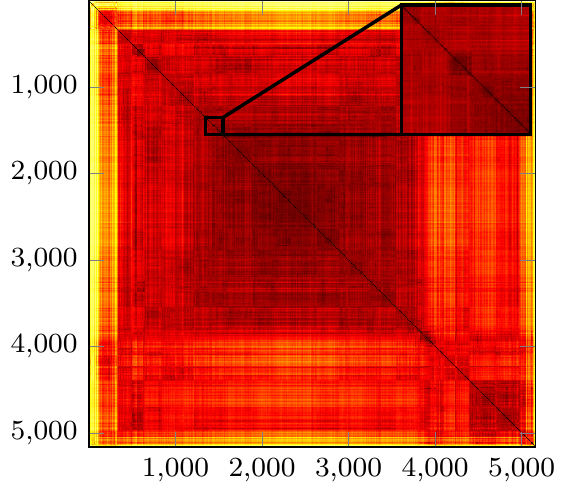}}\hfill
\subfloat[\textsc{US-Senate} (\textsc{EgoNet}) \label{fig:Senate_assoc_EGO}]{\includegraphics[height=5cm]{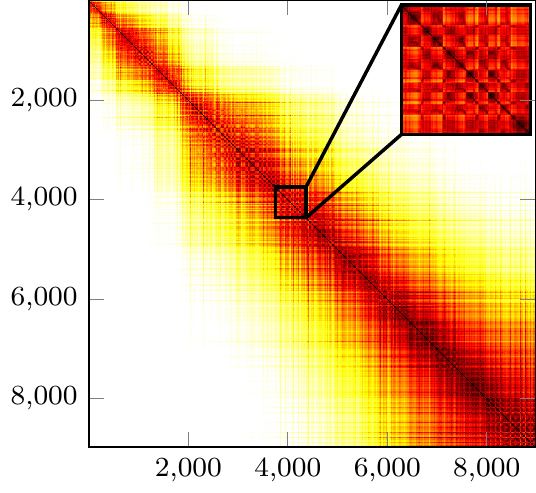}}

\caption{Visualizations of association matrices for \textsc{CA-GrQc}, 
\textsc{FB-Johns55}, and \textsc{US-Senate} illustrate how meso-scale and 
global structures emerge from the superposition and overlap of many local 
communities. See the main text for a description of how we construct the association 
matrices. For each of the three networks, we generate the subfigures using the same 
three sampling procedures that we use to generate the NCPs: 
we use \textsc{AclCut} for panels \sref{fig:GR_assoc_ACL}--\sref{fig:Senate_assoc_ACL},
we use \textsc{MovCut} for panels \sref{fig:GR_assoc_MOV}--\sref{fig:Senate_assoc_MOV},
and we use \textsc{EgoNet} for panels \sref{fig:GR_assoc_EGO}--\sref{fig:Senate_assoc_EGO}.
\label{fig:assoc}
}
\end{figure*}

\begin{figure}[tb]
\subfloat[$\epsilon=10^{-2}$]{\includegraphics[width=0.5\linewidth]{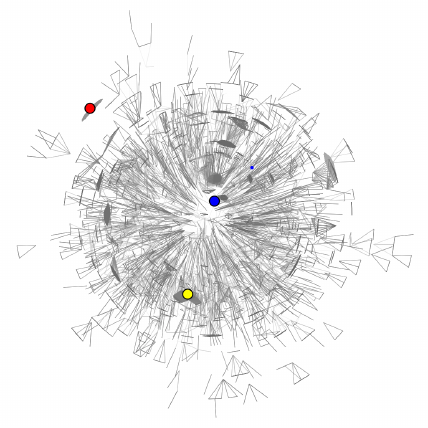}}\hfill
\subfloat[$\epsilon=10^{-3}$]{\includegraphics[width=0.5\linewidth]{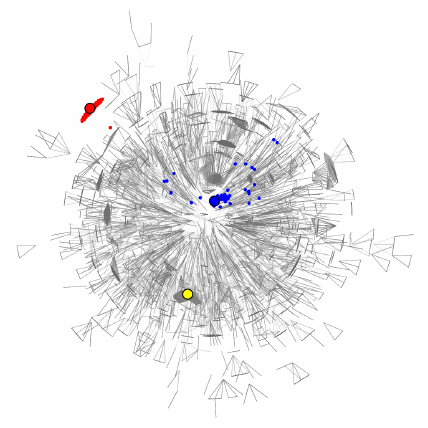}}\\
\subfloat[$\epsilon=10^{-4}$]{\includegraphics[width=0.5\linewidth]{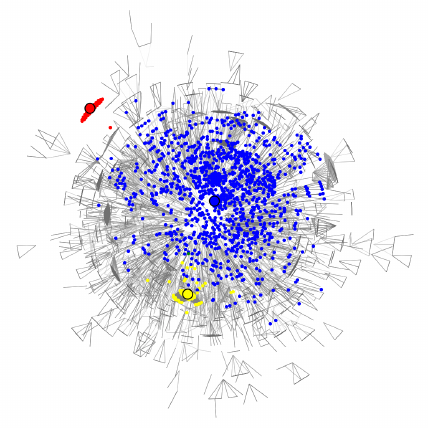}}\hfill
\subfloat[$\epsilon=10^{-5}$]{\includegraphics[width=0.5\linewidth]{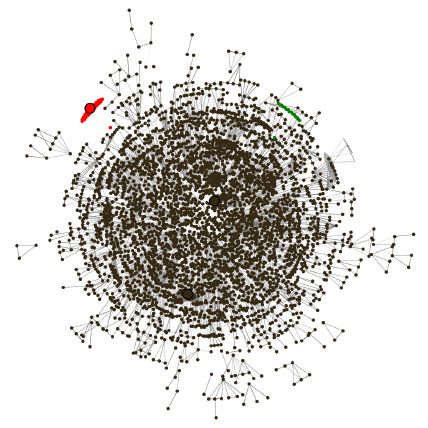}}
\caption{Visualization of global structure in \textsc{CA-GrQc}. 
We constructed the network layout by weighting each edge using the corresponding entry of the association matrix for the \textsc{AclCut} method (Fig.\nobreakspace \ref {fig:GR_assoc_ACL}). We then applied the spring embedding visualization algorithm~\cite{viscoms} to the resulting weighted network. For ease of visualization, we only plot edges with weight larger than the mean edge weight. Colored nodes correspond to local communities for three different seed nodes. Nodes that are a member of more than one community are drawn in a mixed color (e.g., blue and yellow become green; and blue, yellow, and red become blackish, in panel (d)). As we decrease the resolution parameter $\epsilon$, the different communities first explore local structure before merging and each covering most of the network, in panel (d).
\label{fig:GR_global_vis}}
\end{figure}

\begin{figure}[tb]
\subfloat[$\epsilon=10^{-3}$]{\includegraphics[width=0.5\linewidth]{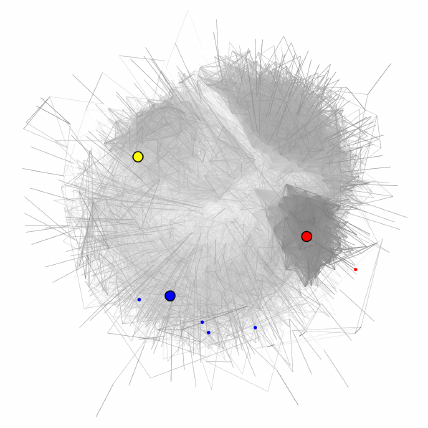}}\hfill
\subfloat[$\epsilon=10^{-4}$]{\includegraphics[width=0.5\linewidth]{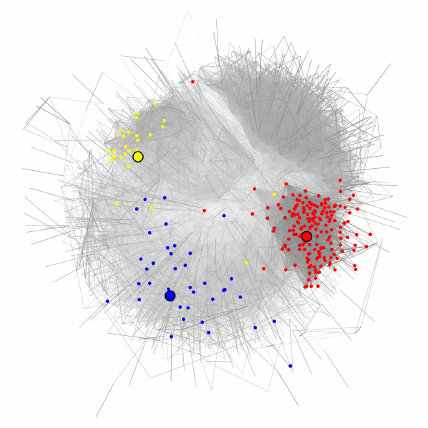}}\\
\subfloat[$\epsilon=10^{-5}$]{\includegraphics[width=0.5\linewidth]{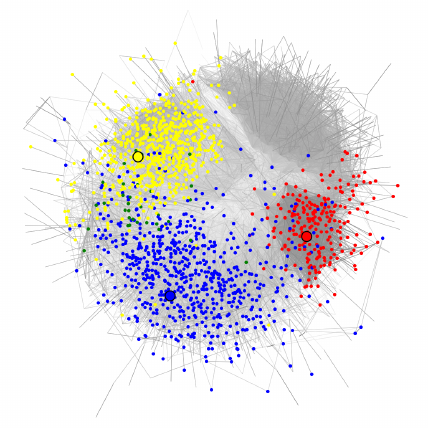}}\hfill
\subfloat[$\epsilon=10^{-6}$]{\includegraphics[width=0.5\linewidth]{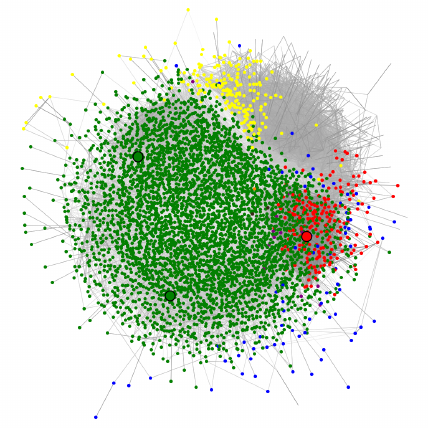}}
\caption{Visualization of global structure in \textsc{FB-Johns55}. We constructed the network layout by weighting each edge using the corresponding entry of the association matrix for the \textsc{AclCut} method (Fig.\nobreakspace \ref {fig:Johns55_assoc_ACL}) and then applying the same procedure as in Fig.\nobreakspace \ref {fig:GR_global_vis}. Colored nodes correspond to local communities for three different seed nodes. Note the difference in behavior for the red community versus the blue and yellow communities. The blue and yellow communities gradually spread as we decrease $\epsilon$, and they eventually merge to cover a large part of the network. However, the red community quickly spreads initially as we decrease $\epsilon$ but then remains localized as we decrease $\epsilon$ further. \label{fig:Johns55_global_vis}
}

\end{figure}

\begin{figure}[tb]
\subfloat[$\epsilon=10^{-3}$]{\includegraphics[width=0.5\linewidth]{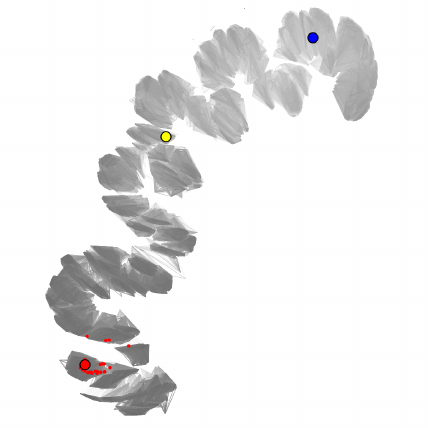}}\hfill
\subfloat[$\epsilon=10^{-4}$]{\includegraphics[width=0.5\linewidth]{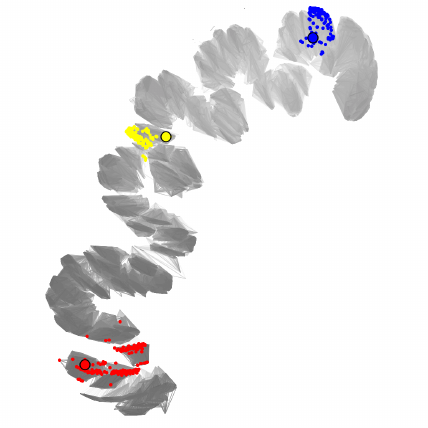}}\\
\subfloat[$\epsilon=10^{-5}$]{\includegraphics[width=0.5\linewidth]{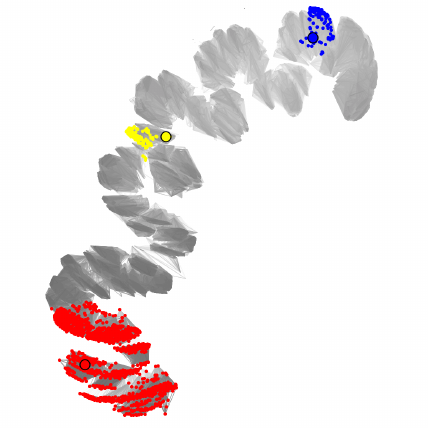}}\hfill
\subfloat[$\epsilon=10^{-6}$]{\includegraphics[width=0.5\linewidth]{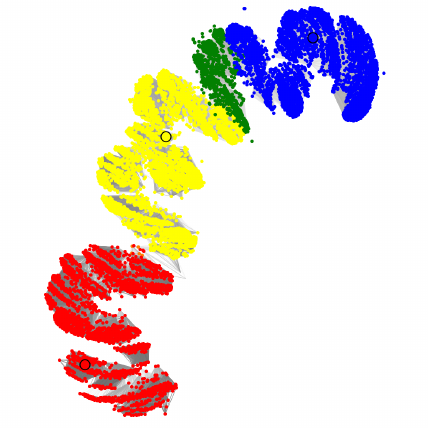}}
\caption{Visualization of global structure in \textsc{US-Senate}. We constructed the network layout by reweighting each edge using the corresponding entry of the association matrix for the \textsc{AclCut} method (Fig.\nobreakspace \ref {fig:Senate_assoc_ACL}) and then applying the same procedure as in Figs.\nobreakspace \ref {fig:GR_global_vis} and\nobreakspace  \ref {fig:Johns55_global_vis}. Colored nodes correspond to local communities for three different seed nodes. The spreading behavior of the different local communities largely follows the temporal structure of the network.  \label{fig:Senate_global_vis}
}
\end{figure}

From the perspective of the locally-biased community-detection methods that 
we use in this paper, one can view intermediate-sized (i.e., meso-scale) 
structures in networks as arising from collections of local features---e.g., 
via overlaps of local communities that one obtains algorithmically using 
locally-biased dynamics such as those that we consider.  
Such local features depend not only on the network adjacency matrix but also 
on the dynamical process under study, the initial seed(s) from which one is 
viewing a network, and the locality parameters of the method (which 
corresponds to the dynamical process) that determine how locally one is 
viewing the network.  
Although a full discussion of the relationship between local structure and 
meso-scale structure and global structures is beyond the scope of this 
paper, here we provide an initial example of such results.

To try to visualize meso-scale and global network structures that we obtain 
from the local communities that we identify, we define an $n \times n$ 
association matrix $\widetilde{A}$ (where $n$ is again the number of nodes 
in the network), which encodes pairwise relations between nodes based on a 
sample of local communities. 
For a given sample $\mathcal{S}$ of local communities (obtained, e.g., by 
running a given method with many seed nodes and values of a locality 
parameter), the entries of the association matrix are given by the number of 
times that a pair of nodes appear together in a local community, normalized 
by the number of times either of them appeared.  
That is, the elements of the association matrix are
\begin{equation}
	\widetilde{A}_{ij}=\frac{\left| \{S\in\mathcal{S} : i \in S \text{ and } j \in S\} \right|}{\left|\{S \in \mathcal{S} : i \in S \text{ or } j \in S\}\right|} \;.
\end{equation}

Our procedure for extracting global network structure from a sampled set of communities is 
similar in spirit to computing association (or ``co-classification'') matrices that have been constructed from sampling a landscape of the modularity objective function~\cite{SalesPardo:2007fn}, and one can in principle analyze these matrices further using the same methods. The additional normalization in our definition of association matrices is necessary to correct for the oversampling of large communities relative to small communities (which results from sampling nodes uniformly at random). At first 
glance, association matrices computed by sampling a modularity landscape appear to reveal much clearer 
community structure in these networks than what we obtain by sampling local communities. However, this is 
largely an artifact of the well-known resolution limit of modularity optimization~\cite{Fortunato:2007js}. One can
mitigate this effect by using one of the multi-resolution generalizations of modularity
~\cite{Reichardt:2006eh,Arenas:2008hq} to sample the modularity landscape across different values of the 
resolution parameter. This yields association matrices that are similar in appearance to the ones that we obtain 
by sampling local communities.

To visualize the association matrices in a way that reveals global network structure, it is important to find a 
good node order. We found the sorting method suggested in Ref.~\cite{SalesPardo:2007fn} to be impractically slow for the networks that we study. Instead, we sort the nodes based on the optimal
leaf ordering~\cite{BarJoseph:2001be} for the average-linkage hierarchical clustering tree 
of the association matrix. (For \textsc{US-Senate}, we do this procedure within a given Congress, and 
we then use the natural temporal ordering to define the inter-Congressional 
ordering.)

In addition, to see small-scale structure using samples $\mathcal{S}$ 
obtained from \textsc{MovCut}, we use a community-size parameter $c$ that 
limits the volume of the resulting community based on the desired 
correlation with the seed vector. 
In this paper, we use $c\in\{10^i : i=1,\ldots,5\}$. 
See Ref.~\cite{Mahoney:2012wl} for details.
We summarize our results in Figs.\nobreakspace  \ref {fig:assoc}--\ref {fig:Senate_global_vis}.

In Fig.\nobreakspace \ref {fig:assoc}, we show the result of applying this procedure with
communities that we sampled using the \textsc{AclCut}, \textsc{MovCut}, and 
\textsc{EgoNet} methods. 
In each case, we keep only the best conductance community for each sampled 
ranking vector.  
The most obvious feature of the visualizations in Fig.\nobreakspace \ref {fig:assoc} is 
that---except for \textsc{US-Senate}, for which there is a natural 
large-scale global structure defined by the one-dimensional temporal 
ordering---the visualizations are much more complicated than any of the 
idealized structures in Fig.\nobreakspace \ref {fig:stylized} (which suggests that the 
visualizations might be revealing at least as much about the inner workings 
of the visualization algorithm as about the networks being visualized).  
The structures in Fig.\nobreakspace \ref {fig:stylized} are trivially interpretable, whereas 
those in real networks (e.g., as illustrated in Fig.\nobreakspace \ref {fig:assoc}) are 
extremely messy and very difficult to interpret.  
In the paragraphs below, we will discuss the structural features in 
Fig.\nobreakspace \ref {fig:assoc} in more detail.

For \textsc{CA-GrQc} (see Fig.\nobreakspace \ref {fig:GR_global_vis} as well as 
Fig.\nobreakspace \ref {fig:assoc}), we observe many small communities that are composed of 
about 10--100 nodes. 
These communities, which correspond to the dark red blocks along the 
diagonal (see the inset in Fig.\nobreakspace \ref {fig:GR_assoc_ACL}), are responsible for the 
dips in the NCPs (see Figs.~\ref{NCP_ACL_small}, \ref{NCP_MOV_small}, and \ref{fig:NCP_EGO_small}) for this network. 
However, these small communities do not combine to form large communities, 
which would result in large diagonal blocks in the association matrices. 
Instead, the small communities appear to amalgamate into a single large 
block (or ``core''). 
In Fig.\nobreakspace \ref {fig:GR_global_vis}, we aim to make this observation more intuitive 
by showing how the local communities for three different seed nodes spread 
through the network as we change the resolution, i.e., the locality bias 
parameter. 
We construct the weighted network $\widetilde G=(V,E,\widetilde w)$ shown in 
Fig.\nobreakspace \ref {fig:GR_global_vis} from the unweighted \textsc{CA-GrQc} network 
$G=(V,E)$ using the association matrix for the \textsc{AclCut} method 
(Fig.\nobreakspace \ref {fig:GR_assoc_ACL}). 
We assign each edge $(i,j) \in E$ a weight based on the corresponding entry 
of the association matrix, i.e., $\widetilde w_{ij} = \widetilde A_{ij}$ if 
$(i,j)\in E$ and $\widetilde w_{ij}=0$ otherwise.
Based on our earlier results with the slowly-increasing NCP, as well as 
previous results in 
Refs.~\cite{Leskovec:2008vo,Leskovec:2009fy,Leskovec:2010uj}, we interpret
these features shown in Fig.\nobreakspace \ref {fig:GR_global_vis} in terms of a nested 
core-periphery structure, in which the network periphery consists of 
relatively good communities and the core consists of relatively densely 
connected nodes. 

For \textsc{FB-Johns55} (see Fig.\nobreakspace \ref {fig:Johns55_global_vis} as well as 
Fig.\nobreakspace \ref {fig:assoc}), we observe two relatively large communities, which 
correspond to the two large diagonal blocks in 
Figs.\nobreakspace \ref {fig:Johns55_assoc_ACL} and\nobreakspace  \ref {fig:Johns55_assoc_MOV} and which underlie the dips 
in the NCPs in Figs.\nobreakspace \ref {NCP_ACL_small} and\nobreakspace  \ref {NCP_MOV_small}.
Note, however, from the scale of the vertical axis in 
Figs.\nobreakspace \ref {NCP_ACL_small} and\nobreakspace  \ref {NCP_MOV_small} that the community quality of these
communities is very low, so one should actually construe the visualization 
in Figs.\nobreakspace \ref {fig:Johns55_assoc_ACL} and\nobreakspace  \ref {fig:Johns55_assoc_MOV} as highlighting a 
low-quality community that is only marginally better than the other 
low-quality communities that are present in that network.
Based on this visualization as well as our earlier results, the remainder of 
\textsc{FB-Johns55} does not appear to have much community structure (at 
least based on using the conductance diagnostic to measure internal versus 
external connectivity).
However, there do appear to be some remnants of highly 
overlapping communities that one could potentially identify using other 
methods (e.g., the one in Ref.~\cite{jure2013}). 
The \textsc{EgoNet} method (see Fig.\nobreakspace \ref {fig:Johns55_assoc_EGO}) is unable to 
resolve not only these small communities but also the larger low-quality 
communities. 
Figure\nobreakspace \ref {fig:Johns55_global_vis} shows how the local communities for two seed 
nodes that do not belong to one of the two large communities slowly spread 
and eventually merge (blue and yellow nodes), whereas the red community 
(which corresponds to the smaller of the two communities) is quickly 
identified and remains separate from the other communities. 

For \textsc{US-Senate} (see Fig.\nobreakspace \ref {fig:Senate_global_vis} as well as 
Fig.\nobreakspace \ref {fig:assoc}), we clearly observe the signature of temporal-based 
community structure at a large size scale. 
See Figs.\nobreakspace \ref {fig:Senate_assoc_ACL},  \ref {fig:Senate_assoc_MOV}, and  \ref {fig:Senate_assoc_EGO}. 
Using \textsc{AclCut} and \textsc{MovCut}, we also obtain partitions at the 
scale of individual Congresses (see the insets in 
Figs.\nobreakspace \ref {fig:Senate_assoc_ACL} and\nobreakspace  \ref {fig:Senate_assoc_MOV}), which sometimes 
split into two or occasionally three individual communities.  
These latter partitions have been discussed previously in terms of 
polarization between parties~\cite{Waugh:2009vz,Mucha:2010p2164,CM11_TR}.
Because we fixed the temporal order of Congresses for \textsc{US-Senate} and 
only sort Senators within the same Congress, this visualization reveals 
communities within each Senate as well as more temporally-disparate 
communities. 
In particular, for the \textsc{EgoNet} method, this ordering introduces a 
checkerboard pattern that correspond to temporal communities that contain 
Senators from several Congresses.  
Figure\nobreakspace \ref {fig:Senate_global_vis} clearly shows that this temporal structure also 
dominates the behavior of local communities for individual seed nodes.

An important point from these visualizations is that, for both 
\textsc{CA-GrQc} and \textsc{FB-Johns55}, the meso-scale and large-scale 
structures that result from the superposition of local communities does 
\emph{not} correspond particularly well to intuitive good-conductance 
communities.
Relatedly, it also does \emph{not} correspond particularly well to an 
intuitive low-dimensional structure or a nearly decomposable 
block-diagonal matrix of community assignments (see our illustration in 
Fig.\nobreakspace \ref {fig:stylized-hotdog}), one or both of which are often assumed 
(typically implicitly) by many global methods for algorithmically detecting 
communities in 
networks~\cite{Porter:2009we,Fortunato:2010iw,Mahoney:2010vt,Mahoney:2012wl,GM14_Subm}. 
Of the networks that we investigate, only the temporal structure in 
\textsc{US-Senate} (as well as in \textsc{US-House}, which is a related 
temporally-dominant network) closely resembles such an idealization. 
This is reflected clearly in its downward-sloping NCP (see 
Figs.\nobreakspace \ref {NCP_ACL_small},  \ref {NCP_MOV_small}, and  \ref {fig:NCP_EGO_small}) and in the visualizations in Fig.\nobreakspace \ref {fig:assoc}. 

Instead, in the other (e.g., collaboration, Facebook, and many many 
other realistic~\cite{Leskovec:2008vo,Leskovec:2009fy}) networks, community 
structure as a function of size is much more subtle and complicated.
Fortunately, our locally-biased perspective provides one means to try to 
resolve such intricacy.  
By averaging over results from different seed nodes, a local approach like 
ours leads naturally to the presence of strongly overlapping communities.
Overlapping community structure has now been studied for several years 
\cite{yy2010,yannis2011,bkn2011}, and recent observations continue to shed 
new light on the ubiquity of community overlap \cite{jure2013}.  
Overlap of communities in networks is a pervasive phenomenon 
\cite{sune-blog,jure2013}; and our expectation is that most large realistic 
networks have communities with significant overlap, rather than merely a 
small amount of overlap that would amount to a small perturbation of the 
idealized, nearly decomposable communities in Fig.\nobreakspace \ref {fig:stylized-hotdog}. 
Additionally, such overlaps imply that larger communities tend to have lower 
quality in terms of their internal versus external connectivity (i.e., in 
terms of how much they resemble the intuitive communities that many 
researchers know and love) than smaller communities---in agreement with our 
empirical results on both the collaboration networks and Facebook networks, 
but in strong disagreement with popular intuition.
In these latter cases, recent work that fits related networks with 
upward-sloping NCPs to hierarchical Kronecker graphs resulted in parameters 
that are consistent with the core-periphery structure that we illustrated 
in Fig.\nobreakspace \ref {fig:stylized-coreper}~\cite{Leskovec:2010wb}.


\begin{figure*}[tb]
\subfloat[\textsc{AclCut}, $\langle k \rangle = 10$, $k_{\max} = 100$, $\tau_1=-2$, $\tau_2=-3$, $c_{\min}=10$, $c_{\max}=50$]{\includegraphics[width=0.32\linewidth]{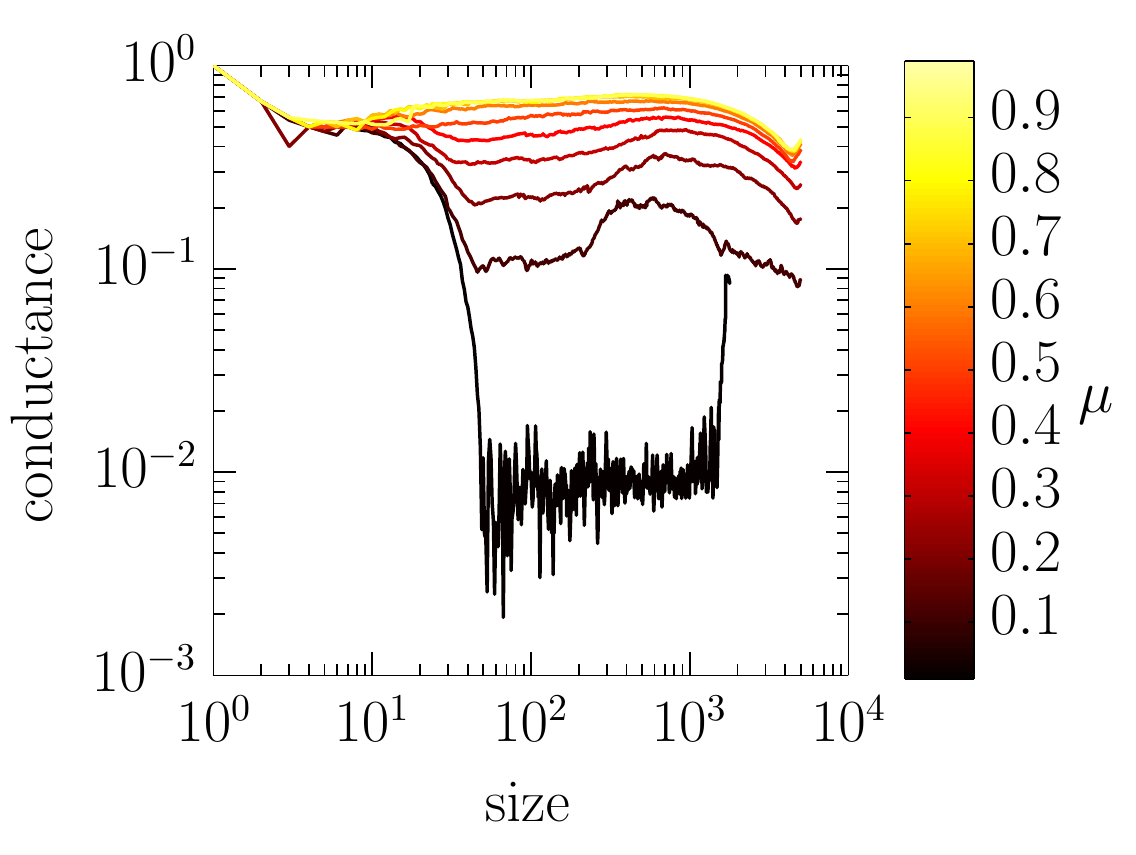}}\hfill
\subfloat[\textsc{AclCut}, $\langle k \rangle = 20$, $k_{\max} = 50$, $\tau_1=-2$, $\tau_2=-1$, $c_{\min}=10$, $c_{\max}=50$]{\includegraphics[width=0.32\linewidth]{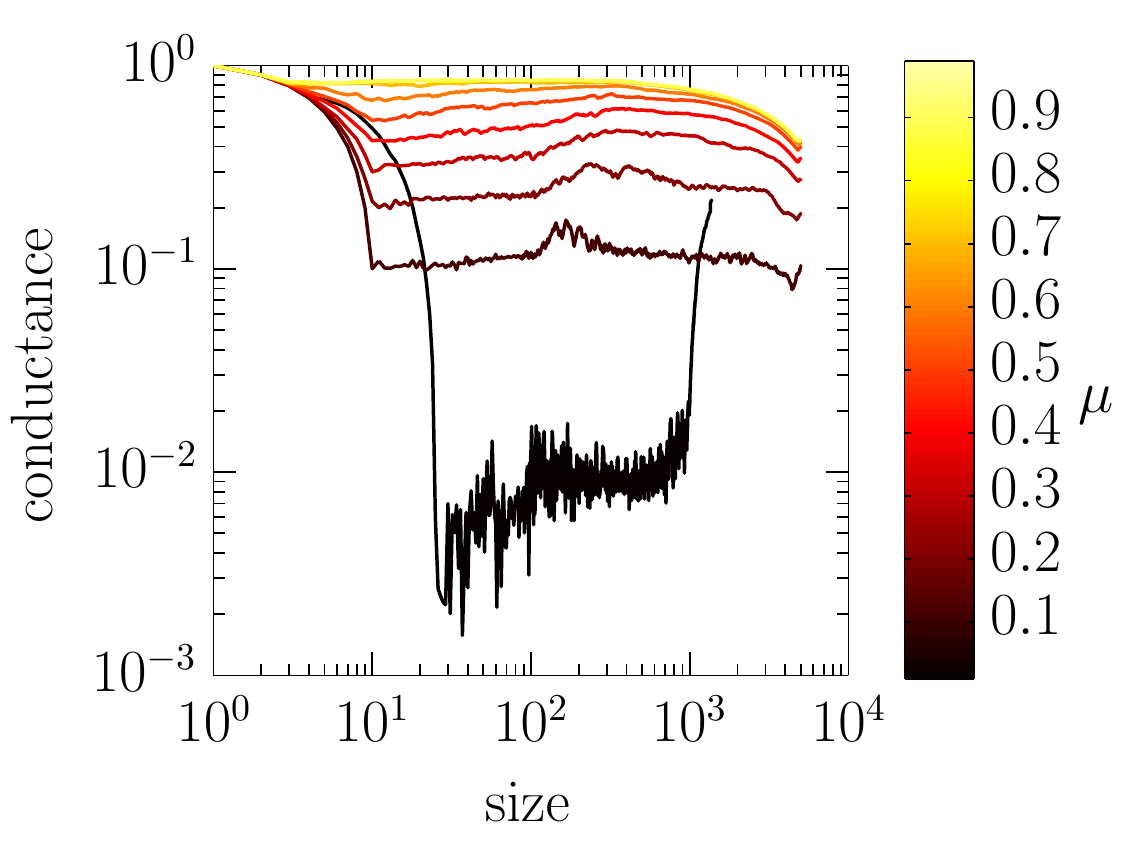}}\hfill
\subfloat[\textsc{AclCut}, $\langle k \rangle = 20$, $k_{\max} = 50$, $\tau_1=-2$, $\tau_2=-1$, $c_{\min}=20$, $c_{\max}=100$]{\includegraphics[width=0.32\linewidth]{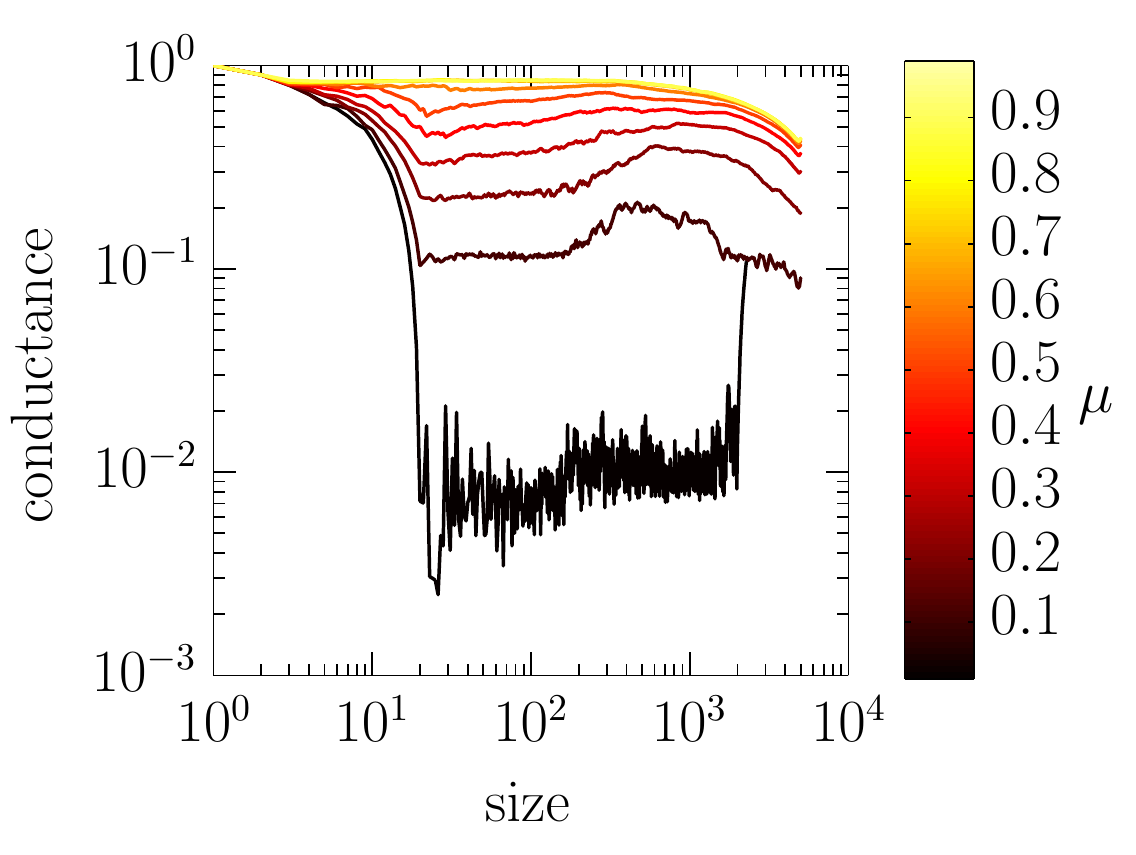}} 
\caption{NCPs of LFR synthetic benchmark networks \cite{Lancichinetti:2008ge} with $n=10000$ nodes. Colors correspond to different values of the mixing parameter $\mu$.
Our choices for the mean degree $\langle k \rangle$, maximum degree 
$k_{\max}$, exponent of the degree distribution $\tau_1$, exponent of the 
community size distribution $\tau_2$, minimum community size $c_{\min}$, 
and maximum community size $c_{\max}$ correspond to the ones used in Refs.~\cite{Lancichinetti:2012kx,Lancichinetti:2009bb} to benchmark community-detection algorithms. 
\label{fig:benchmarks}
}
\end{figure*}

\section{Empirical Results on Synthetic Benchmarks}
\label{sxn:benchmarks}

Synthetic benchmark networks with a known, planted community structure can 
be helpful for validating and gaining a better understanding of the behavior 
of community-detection algorithms. 
For such an approach to be optimally useful, it is desirable for the 
synthetic benchmarks to reproduce relevant features of real networks with 
community structure; and it is challenging to develop good benchmarks that 
reproduce community structure and other structural properties of 
medium-sized and larger realistic networks. 
An extremely popular---and in some ways useful---family of benchmark 
networks that aims to reproduce some features of real networks are the 
so-called \emph{LFR (Lancichinetti-Fortunato-Radicchi) networks} 
\cite{Lancichinetti:2008ge,Lancichinetti:2009gk}.
By design, LFR networks have power-law degree distributions as well as 
power-law community-size distributions, they are unweighted, and they have 
non-overlapping planted communities.  
Motivated by our empirical results on networks constructed from real data, 
we also apply our methods to LFR networks to test the extent to which they 
are able to reproduce the three classes of NCP behavior (upward-sloping, 
flat, and downward-sloping) that we have observed with real networks.

To parametrize the family of LFR networks, we specify its power-law degree distribution using its exponent $
\tau_1$, mean degree $\langle k \rangle$, and maximum degree $k_{\max}$. Similarly, we specify its power-law 
community size distribution using its exponent $\tau_2$, minimum community size $c_{\min}$, and maximum 
community size $c_{\max}$, with the additional constraint that the sum of community sizes should equal the size of 
the network $n$. Furthermore, we specify the strength of community memberships using a mixing parameter 
$\mu$, where each node shares a fraction $1-\mu$ of its edges with nodes in its own community. A simple 
calculation shows that this definition of the mixing parameter implies that each community in the planted partition 
has conductance $\mu$ (up to rounding effects). 

To construct a network with these parameters, we sample $n$ degrees from the degree distribution and sample community sizes from the community size distribution. We then assign nodes to communities uniformly at random, with the constraint that a node cannot be assigned to a community that is too small for the node to have the correct mixing-parameter value. We then construct inter-community and intra-community edges separately by connecting the corresponding stubs (i.e., ends of edges) uniformly at random. We use the implementation by Lancichinetti~\cite{LFR_code} to generate LFR networks.

In Fig.\nobreakspace \ref {fig:benchmarks}, we show representative NCPs for LFR networks for 
three choices of parameters for the degree distribution and community-size 
distribution that have been used previously to benchmark community-detection 
algorithms~\cite{Lancichinetti:2012kx,Lancichinetti:2009bb,Lancichinetti:2008ge}.
(We generated the results presented in Fig.\nobreakspace \ref {fig:benchmarks} using the 
\textsc{AclCut} method, but we obtain nearly identical NCPs using the 
\textsc{MovCut} method.) 
The three subfigures demonstrate that all three parameter choices yield 
networks with similar NCPs. 
In particular, we observe that---above a certain critical size---the best
communities have comparable quality, as a function of increasing size.
Depending on the particular parameter values, this can be of similar quality 
to or somewhat better than that which would be obtained by, e.g., a vanilla 
(not extremely sparse) ER random graph, across all larger size scales. 
That is, above the critical size, the NCP is approximately flat.
Increasing the topological mixing parameter $\mu$ in the LFR network 
generative mechanism at first shifts the entire NCP upwards because the 
number of inter-community edges increases. For $\mu \approx 1$, it levels 
off to the characteristic flat shape for an NCP of a network generated from 
the configuration model of random~graphs.

Importantly, the behavior for the LFR benchmark networks from 
Ref.~\cite{Lancichinetti:2008ge} that we illustrate in Fig.\nobreakspace \ref {fig:benchmarks} 
does \emph{not} resemble the NCPs for any of the real-world networks in 
either the present paper or in Ref.~\cite{Leskovec:2008vo,Leskovec:2009fy}.
In addition, we have been unable to find parameter values for which the 
qualitative properties of realistic NCPs---in particular, a relatively 
gradually upward-sloping NCP---are reproduced, which suggests that the 
community structure generated by the LFR benchmarks is \emph{not} realistic 
in terms of its size-resolved properties.

To verify that this behavior is not an artifact of the particular choices of 
parameters shown in Fig.\nobreakspace \ref {fig:benchmarks}, we sampled sets of 
parameters uniformly at random with $n\in\{1000,10\,000,50\,000\}$, 
$\tau_1,\tau_2 \in \{-1,-2,\ldots,-5\}$, 
$\langle k\rangle \in \{10,11,\ldots, 100\}$, 
$k_{\max}\in \{\langle k\rangle,\langle k \rangle +1, \ldots, 250\}$, $c_{\min} \in \{10,11,\ldots, 250\}$, and $c_{\max} \in \{\max(c_{\min},k_{\max}),\ldots,250\}$. 
The aggregate trends of the NCPs for the LFR benchmark networks with the 
different parameters we sample are similar to and consistent with the 
results shown in Fig.\nobreakspace \ref {fig:benchmarks}. 
Hence, although the LFR benchmark networks are useful as 
tests for community-detection techniques, our calculations suggest that they 
are unable to reproduce a fundamental feature of many real networks with 
respect to variation in community quality (and, in particular, worsening 
community quality) as a function of increasing community~size. 

Based on our empirical observations, our locally-biased perspective on 
community detection suggests a natural approach to determine whether 
synthetic benchmarks possess small-scale, medium-scale, and large-scale 
community structure that resembles that of large realistic networks: 
namely, a family of synthetic benchmark networks ought to include parameter 
values that generate networks with (robust) upward-sloping, flat, and 
downward-sloping NCPs (as observed in Figs.\nobreakspace \ref {fig:ncp-possiblencps} and\nobreakspace  \ref {NCP_ACL_small}).


\section{Conclusions and Discussion}
\label{sxn:conc}

In this paper, we have conducted a thorough investigation of community 
quality as a function of community size in a suite of realistic networks, 
and we have reached several conclusions with important implications for the 
investigation of realistic medium-sized to large-scale networks. 
Our results build on previous work on using network community profiles 
(NCPs) to study large-scale 
networks~\cite{Leskovec:2008vo,Leskovec:2009fy,Leskovec:2010uj}.  
In this paper, we have employed a wider class of community-identification 
procedures, and we have discovered a wider class of community-like 
behaviors (as a function of community size) in realistic networks than what 
had been reported previously in the literature~\footnote{To give one example, the coauthorship networks that we examined possess a familiar upward-sloping NCP and a nested core-periphery structure with small, well-separated communities in the periphery. However, the small and large communities that we observed in the Facebook networks do not lead to bottlenecks for diffusion processes.  This novel NCP behavior in the Facebook graphs is consistent both with the larger edge density of the Facebook networks relative to the SNAP graphs~\cite{Leskovec:2008vo,Leskovec:2009fy,Leskovec:2010uj} as well as with previous work on simple dynamical processes on these Facebook networks~\cite{Melnik:2011fn}.}. 
In addition, using NCPs, we have discovered that the popular LFR synthetic 
benchmark networks, which are often used to validate community-detection 
algorithms---and which are the most realistic synthetic benchmark networks that have been produced to test methods for community detection \cite{lfr-real}---exhibit behavior that is markedly different from many realistic 
networks.  
Our result thus underscores the importance of developing realistic 
benchmark graphs whose NCPs are qualitatively similar to those of real 
networks.  
Taken together, our empirical results yield a much better understanding of 
realistic community structure in large realistic networks than was previously available, and they provide 
promising directions for future work. 
More generally, because our approach for comparing community structures in 
networks (using NCPs and conductance ratio profiles) is very general---e.g., 
one can follow an analogous procedure with other community-quality 
diagnostics, other procedures for community generation, etc.---our 
locally-biased and size-resolved methodology is an effective way to 
investigate size-resolved meso-scale network structures much more generally.

The main conclusion of our work is that community structure in real networks is 
much more intricate than what is suggested by the block-diagonal assumption that is (either implicitly 
or explicitly) made by most community-detection methods (including ones that allow overlapping communities \cite{yy2010}) and when using the synthetic benchmark networks that have been developed to test those methods. Community structure interplays with other meso-scale features, such as core-periphery structure \cite{Rombach:2012uy,jure2013,cp-review}, and investigating only community structure without consideration of other structures can lead to misleading results. A local perspective on community detection, like the one that we have advocated in the present paper, allows pervasive community overlap in a natural way---which is an important feature to capture when considering real social networks. Additionally, the large-scale consensus community structure that we obtain subsequently by ``pasting together'' local communities is not constrained to resemble a global block-diagonal structure. This is a key consideration in the study of meso-scale structures in real networks.

Although most algorithmic methods for community detection take a different approach from ours, the observation that network community structure depends not 
only on the network structure per se but also on the dynamical processes 
that take place on a network and the initial conditions (i.e., seed node or 
nodes) for those processes, is rather traditional in many ways.
Recall, for example, Granovetter's observation that a node with many weak 
ties is ideally suited to initialize a successful social contagion 
process~\cite{Granovetter:1973wj}. 
Our perspective also meshes better than global ones with real-life 
experience in our own networks.  
Both of these observations underscore our point that whether particular 
network structures form bottlenecks for a dynamical process depends not only 
on the process itself but also on the initial conditions of that~process.  

More generally, one might hope that our size-resolved and locally-biased 
perspective on community detection can be used to help develop new 
diagnostics that complement widely-used and intuitive concepts such as 
closeness centrality, betweenness centrality, and the many other existing 
global notions.
These will be of particular interest for investigating large networks---or 
even modestly-sized networks such as those that we have considered---where 
traditional algorithmic and visualization methods have serious difficulties.  
Because the study of meso-scale structure in networks is important for 
understanding how local and small-scale properties of a network interact 
with global or large-scale properties, we expect that taking a 
locally-biased perspective on community detection and related problems will yield interesting and novel insights on these and related 
questions.

\appendix

\section{Expander Graphs}
\label{sxn:expanders}

In this section, we provide a brief introduction to the concept of an 
\emph{expander graph} (or, more simply, an \emph{expander}) \cite{expander-whatis}.
Essentially, expanders are graphs that are very well-connected and thus do 
not have any good 
communities (when measured with respect to 
diagnostics such as conductance).
Because our empirical results indicate that many large social and information 
networks are expanders---at least when viewed at large size-scales---it is 
useful to review basic properties about expander graphs.
Although most of the technical aspects of expander graphs are beyond the 
scope of this paper, Ref.~\cite{Hoory:2006kj} provides an excellent overview 
of this topic.

Let $G=(V,E)$ be a graph, which we assume for simplicity is undirected and 
unweighted. 
For the moment, we assume that all nodes have the same degree $d$ (i.e., $G$ 
is $d$-regular).
For $S_1,S_2 \subset V$, the set of edges from $S_1$ to $S_2$ is then
\begin{equation}
	E(S_1,S_2) = \{ (u,v) : u \in S_1, v \in S_2, (u,v) \in E \}  \,.
\end{equation}
In this case, the number $|S|$ of nodes in $S$ is a natural measure of the 
size of $S$.  
Additionally, the quantity $|E(S,\wbar S)|$, which indicates the number of 
edges that cross between $S$ and $\wbar S$, is a natural measure of the size 
of the boundary between $S$ and $\wbar S$.

We also define the \emph{edge expansion of a set of nodes $S \subset V$} as
\begin{equation}
	h(S) = \frac{|E(S,\wbar S)|}{|S|}  \,,
\label{eqn:expansion-set}
\end{equation}
in which case the \emph{edge expansion of a graph $G$} is the minimum edge 
expansion of any subset (of size no greater than $n/2$) of nodes:
\begin{equation}
	h(G)=\min_{S\subset V : |S|\le\frac{n}{2}} h(S)\,.
\label{eqn:expansion-graph}
\end{equation}
A sequence of $d$-regular graphs $\{G_i \}_{i \in \mathbb{N}}$ is a 
\emph{family of expander graphs} if there exists an $\epsilon>0$ such that 
$h(G_i) \ge \epsilon$ for all $i\in\mathbb{N}$.
Informally, a given graph $G$ is an expander if its edge expansion is large.

As reviewed in Ref.~\cite{Hoory:2006kj}, one can view expanders from 
several complementary viewpoints.
From a combinatorial perspective, expanders are graphs that are highly 
connected in the sense that one has to sever many edges to disconnect a 
large part of an expander graph. 
From a geometric perspective, this disconnection difficulty implies that 
every set of nodes has a relatively very large boundary. 
From a probabilistic perspective, expanders are graphs for which the natural 
random-walk process converges to its limiting distribution as rapidly as 
possible. 
Finally, from an algebraic perspective, expanders are graphs in which the 
first nontrivial eigenvalue of the Laplacian operator is bounded away from $0$.
(Because we are talking here about $d$-regular graphs, note that this 
statement holds for both the combinatorial Laplacian and the normalized Laplacian.)
In addition, constant-degree (i.e., $d$-regular, for some fixed value of $d$)
expanders are the metric spaces that (in a very precise and strong 
sense~\cite{Hoory:2006kj}) embed least well in low-dimensional spaces (such 
as those discussed informally in Section\nobreakspace \ref {sxn:prelim-looklike}).
All of these interpretations imply that smaller values of expansion 
correspond more closely to the intuitive notion of better communities 
(whereas larger values of expansion correspond, by definition, to better 
expanders.) 

Note the similarities between Eq.\nobreakspace \textup {(\ref {eqn:expansion-set})} and 
Eq.\nobreakspace \textup {(\ref {eqn:expansion-graph})}, which define expansion, with 
Eq.\nobreakspace \textup {(\ref {eqn:conductance-set})} and Eq.\nobreakspace \textup {(\ref {eqn:conductance-graph})}, which define 
conductance. 
These equations make it clear that the difference between expansion and 
conductance simply amounts to a different notion of the size (or volume) of 
sets of nodes and the size of the boundary (or surface area) between a set 
of nodes and its complement. 
This difference is inconsequential for $d$-regular graphs. 
However, because of the deep connections between expansion and 
rapidly-mixing random walks, the latter notion (i.e., conductance) is much 
more natural for graphs with substantial degree heterogeneity.
The interpretation of failing to embed well in low-dimensional spaces (like 
lines or planes) is not as extremal in the case of conductance and 
degree-heterogeneous graphs as it is in the case of expansion and 
degree-homogeneous graphs; but the interpretations of being well-connected, 
failing to provide bottlenecks to random walks, etc. all hold for 
conductance and degree-heterogeneous graphs such as those that we consider 
in the main text of the present paper.  
Accordingly, it is insightful to interpret our empirical results on 
small-scale versus large-scale structures in networks should be in light of 
known facts about expanders and expander-like graphs.

\section{Community Quality, Dynamics on Graphs, and Bottlenecks to Dynamics}
\label{sxn:measures}

In this section, we describe in more detail how we algorithmically identify 
possible communities in graphs.
Because we are interested in local properties and how they relate to 
meso-scale and global properties, we take an operational approach and 
view communities as the output of various dynamical processes (e.g., 
diffusions or geodesic hops), and we discuss the relationship between the 
output of those procedures to well-defined optimization problems.
The idea of using dynamics on a network has been exploited successfully by 
many methods for finding ``traditional'' communities (of densely connected 
nodes)~\cite{Arenas:2006ww,Lambiotte:2010vb,Lambiotte:2008p2891,Delvenne:2010vx,Mucha:2010p2164,Rosvall:2008fi,Estrada:2011kd} 
as well as for finding sets of nodes that are related to each other in other 
ways~\cite{Arenas:2006ww,Lambiotte:2011gk,Lerman:2012cq,Anderson:2012kx,mariano2013}.  

In this paper, we build on the idea that random walks and related 
diffusion-based dynamics, as well as other types of local dynamics (e.g., 
ones, like geodesic hops, that depend on ideas based on egocentric 
networks), should get ``trapped'' in good communities. 
In particular, we consider the following three dynamical methods for 
community identification.
\subsection{Dynamics Type 1: Local Diffusions (the ``{\sc AclCut}'' method).}
In this procedure, we consider a random walk that starts at a given seed 
node $s$ and runs for some small number of steps. 
We take advantage of the idea that if a random walk starts inside a good 
community and takes only a small number of steps, then it should become 
trapped inside that community. 
To do this, we use the locally-biased personalized PageRank (PPR) procedure 
of Refs.~\cite{Andersen:2006we,Andersen:2006gp}. 
Recall that a PPR vector is implicitly defined as the solution of the equation
\begin{equation}
\label{eqn:PPR}
	\pr(\alpha,\vec s)=\alpha D^{-1} A \pr(\alpha,\vec s) + (1-\alpha) \vec s\,, 
\end{equation}
where $1-\alpha$ is a ``teleportation'' probability and $\vec s$ is a seed 
vector. 
From the perspective of random walks, evolution occurs either by the walker 
moving to a neighbor of the current node or by the walker ``teleporting'' to 
a random node (e.g., determined uniformly at random as in the usual PageRank 
procedure, or to a random node that is biased towards $\vec s$ in the PPR 
procedure). 
In general, teleportation results in a bias to the random walk, which one 
usually tries to minimize when detecting communities. 
(See Ref.~\cite{Lambiotte:2011fe} for clever ways to choose $\vec s$ with 
this goal in mind.)  

The algorithm of Refs.~\cite{Andersen:2006we,Andersen:2006gp} deliberately 
exploits the bias from teleportation to achieve localized results. 
It computes an approximation to the solution of Eq.\nobreakspace \textup {(\ref {eqn:PPR})} (i.e., it 
computes an \emph{approximate PPR vector}) by strategically ``pushing'' mass 
between the iteratively-updated approximate solution vector and a residual 
vector in such a way that most of the nodes in the original network are 
\emph{not} reached. 
Consequently, this algorithm is typically \emph{much} faster for 
moderately-large to very large graphs than is the na\"{i}ve algorithm to 
compute a solution to Eq.\nobreakspace \textup {(\ref {eqn:PPR})}. 
The algorithm is parametrized in terms of a ``truncation'' parameter 
$\epsilon$ where 
larger values of $\epsilon$ correspond to more locally-biased solutions.
We refer to this procedure as the \textsc{AclCut} method.

\subsection{Dynamics Type 2: Local Spectral Partitioning (the ``{\sc MovCut}'' method).}
In this procedure, we formalize the idea of a locally-biased version of the 
leading nontrivial eigenvector of the normalized Laplacian $\mathcal{L}$ 
that can be used in a locally-biased version of traditional spectral graph 
partitioning.

Following Ref.~\cite{Mahoney:2012wl}, consider the following optimization 
problem:
\begin{equation}
\begin{aligned}
  & \underset{\vec x}{\text{minimize}}
  & & \vec{x}^T\mathcal{L}\vec{x} \\ 
  & \text{subject to}
  & & \vec{x}^T\vec{x} = 1\,, \\
  & & & \vec{x}^T D^{1/2} \vec{1} = 0  \\
  & & & (\vec{x}^T D^{1/2} \vec{s})^2 \geq \kappa \, ,
\end{aligned}
\label{eqn:mov-vp}
\end{equation}
where $\kappa$ is a locality parameter and $\vec{s}$ is a vector, which 
satisfies the constraints $\vec{s}^T D \vec{s}=1$ and 
$\vec{s}^T D \vec{1} =0$, and which represents a seed set of nodes. 
That is, in the norm defined by the diagonal $D$ matrix, the seed vector 
$\vec{s}$ is unit length and is exactly orthogonal to the all-ones vector.
This \emph{locally-biased} version of standard spectral graph partitioning 
(which becomes the usual global spectral-partitioning problem if the 
locality constraint $(\vec{x}^T D^{1/2} \vec{s})^2 \geq \kappa$ is removed) was introduced 
in~\cite{Mahoney:2012wl}, where it was shown that the solution vector 
$\vec{x^*}$ inherits many of the nice properties of the solution to the usual 
global spectral-partitioning problem. 
The solution $\vec{x}^*$ is of the form
\begin{equation}
	\vec{x}^{*} = c \left( L_G - \gamma D_G \right)^{+} D_G \vec{s} \,,
\label{eqn:ppr-soln-form}
\end{equation}
where the parameter $\gamma \in (-\infty,\lambda_2(G))$ is related to the 
teleportation parameter $\alpha$ via the relation 
$\gamma=\frac{\alpha-1}{\alpha}$ (see~\cite{Mahoney:2012wl}) and 
$c \in [0,\infty]$ is a normalization constant. 

As one can see from Eq.\nobreakspace \textup {(\ref {eqn:ppr-soln-form})}, the solution $\vec{x}^{*}$ of 
Eq.\nobreakspace \textup {(\ref {eqn:mov-vp})} is an \emph{exact PPR vector} with personalized 
teleportation vector $\vec{s}$. 
Consequently, it can be computed as the solution to a system of linear 
equations. 
In addition, if one performs a sweep cut (see the discussion below) of this 
solution vector to obtain a locally-biased network partition, then one 
obtains Cheeger-like guarantees on approximation quality for the associated 
network community. 
Moreover, if the seed vector $\vec{s}$ corresponds to the indicator vector 
of a single node $i$, then this is a relaxation of the following 
\emph{locally-biased graph partitioning problem}: 
given as input a graph $G=(V,E,w)$, an input node $u$, and a positive integer
$k$; find a set of nodes $S \subseteq V$ that is the best conductance set of 
nodes of volume no greater than $k$ that contains the input 
node~$i$~\cite{Mahoney:2012wl}.
We refer to this procedure (with a seed vector corresponding to a single 
seed node) as the \textsc{MovCut} method.

\subsection{Dynamics Type 3: Local Geodesic Spreading (the ``{\sc EgoNet}'' method).}
In this procedure, we perform a geodesic-based (i.e., ego-network-based) 
dynamics that is analogous to the local random walks that we described 
above. 
This method is similar to the technique for finding local communities that 
was introduced in Ref.~\cite{Bagrow:2005et} and that was generalized to 
weighted networks in Ref.~\cite{Porter:2007gd}. 
Starting with a seed node $s$ and a distance parameter $k$, this method 
considers all nodes $j$ whose geodesic distance from $s$ is at most $k$ 
away---i.e., all nodes $j$ such that $\Delta_{sj} \le k$---to form a local 
community.
In the unweighted case, the \emph{egocentric network} (i.e., 
\emph{ego network} or \emph{ego-net} \cite{Wasserman:1994wu}) for a seed 
node (the ego) is the subgraph induced by the seed node's 
$1$-neighborhood---i.e., the network that consists of all nodes that are in 
the $1$-neighborhood (including the seed node) and all edges between these 
nodes that are present in the original network. 
(The traditional definition of an ego-net excludes the seed node and its 
edges, but we specifically include them.) 
We use the term \emph{$k$-ego-net} for the subgraph that is induced by the 
$k$-neighborhood of a seed node. 
Consequently, the local communities that we obtain using this method are 
simply the $k$-ego-nets of the seed node. 
For consistency with the other two methods, it is useful to think of this 
method as inducing a ranking of the nodes:
\begin{equation}
   \text{EgoRank}_i(s)=\frac{1}{1+\Delta_{is}}   \,,
   \label{eq:egorank}
\end{equation}
where $i$ is some node in a network. 
Given the ranking interpretation in Eq.\nobreakspace \textup {(\ref {eq:egorank})}, we recover local 
geodesic-based communities from the EgoRank vector by using the sweep cut 
procedure that we describe below. 
The underlying dynamics for this method is analogous to the extreme case of 
a susceptible-infected (SI) spreading 
process~\cite{gleeson2013PRX,mikko-slow}, in which an infected node infects 
all its neighbors with probability 1 at the time step following the one in 
which it is infected.
One can then interpret the EgoRank of node $i$ for a seed node $s$ as the 
inverse of the time that it takes for node $i$ to first become infected when 
only the seed node $s$ is infected initially. 
We refer to this procedure as the \textsc{EgoNet} method. 

\subsection{Sampling Procedures and Parameter Choices}
\label{sxn:procedures}

To obtain an accurate picture of local community structure at different 
size scales throughout a network, we run each of the above 
community-identification procedures many times, starting at different seed 
nodes and running for different numbers of steps, and we then examine which 
nodes get visited as the dynamical processes unfold.  
For each seed node and value of the parameters, each of the \textsc{AclCut}, 
\textsc{MovCut}, and \textsc{EgoNet} methods returns a vector that can be 
used to ``rank'' the nodes of a network (in a locally-biased and 
size-resolved manner): \textsc{AclCut} and \textsc{MovCut} return a variant 
of the PPR vector, and \textsc{EgoNet} returns the EgoRank vector in 
Eq.\nobreakspace \textup {(\ref {eq:egorank})}. 
Then, given a ranking vector $\vec{p}$, the so-called ``sweep sets'' are 
given by {$S_t=\{i\in V : \vec{p}_i\geq t\}$}; and thus there are at most 
$n+1$ distinct sweep sets (where we recall that $n$ is the number of nodes 
in the graph). 
A corresponding ``sweep cut'' is then the partition of the network obtained 
from a sweep set that has minimal conductance, over all $n+1$ possible sweep 
set partitions. 
By computing the conductance for each of the sweep sets, one obtains a 
locally-biased estimate for an NCP, centered around a seed node. 
One can then estimate a global NCP by taking the lower envelope over local 
NCPs for different seed nodes and parameter values. Our {\sc Matlab} code that implements these methods is available at \cite{LocalCode}.

Recall that \textsc{AclCut} has two parameters (the teleportation parameter $\alpha$ 
and the truncation parameter $\epsilon$), but that 
\textsc{MovCut} only has a single parameter (a teleportation parameter). 

For \textsc{AclCut}, theoretical results~\cite{Andersen:2006we} suggest that 
the method should find good communities of volume roughly $\epsilon^{-1}$, 
where we have ignored constants and logarithmic factors. 
Furthermore, for a seed node $i$ with strength $k_i$, \textsc{AclCut} 
returns empty communities for $\epsilon<k_i^{-1}$.  
This suggests that sampling using 
$\epsilon \in \left[k_\text{max}^{-1},\vol(G)^{-1}\right]$ gives good 
coverage of different size scales in practice. 
In this paper, we use 20 logarithmically-spaced points in 
$\left[k_\text{max}^{-1},\vol(G)^{-1}\right]$ (including the endpoints) to 
generate Figs.\nobreakspace \ref {ACL_small},  \ref {ACL_large}, and  \ref {fig:benchmarks}. 
In addition, we use $\tilde \alpha=0.001$, where $\tilde \alpha$ is the 
teleportation parameter of the ``lazy random walk'' defined 
in~\cite{Andersen:2006we}. 
The (conventional) teleportation parameter that we use satisfies 
$\alpha=1-\frac{2 \tilde \alpha}{1 + \tilde \alpha}$, so that 
$\alpha \approx 0.998$ in Eq.\nobreakspace \textup {(\ref {eqn:PPR})}. 
In our computations, we observed that increasing $\alpha$ leads to more 
accurate NCPs at the cost of longer computation times. 

For \textsc{MovCut}, we use 20 equally-spaced values of $\alpha$ in the 
interval $\left[0.7, (1-\lambda_2)^{-1}-10^{-10}\right]$ (including the 
endpoints), where $(1-\lambda_2)^{-1}$ is the theoretical maximum for 
$\alpha$ (see~\cite{Mahoney:2012wl}).

To sample seed nodes, we modified the strategy described in Ref.~\cite{Leskovec:2009fy} to be applicable to the \textsc{MovCut} method as well as the \textsc{AclCut} method. For each choice of parameter values, we sampled nodes uniformly at random without replacement and stopped the sampling process either when all nodes were sampled or when the sampled local communities sufficiently covered the entire network. To determine sufficient coverage, we tracked how many times each node was included in the best local community that we obtained from the sweep sets and stopped the procedure once each node was included at least 10 times. This procedure ensures that good communities are sampled consistently. 

\begin{figure*}[ptb!]

\subfloat[NCP \label{fig:NCP_EGO_small}]{\includegraphics[width=.47\linewidth]{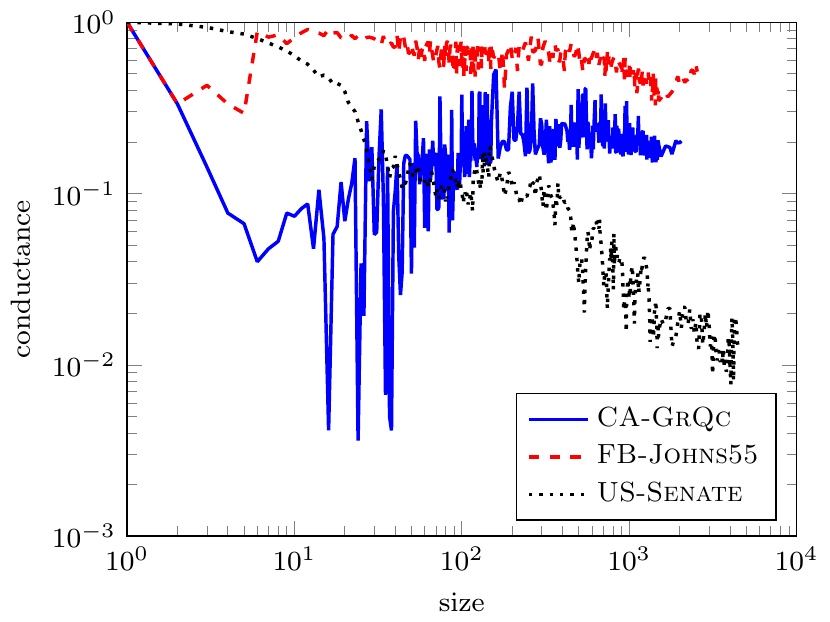}}
\hfill
\subfloat[CRP \label{fig:condratio_EGO_small}]{\includegraphics[width=.47\linewidth]{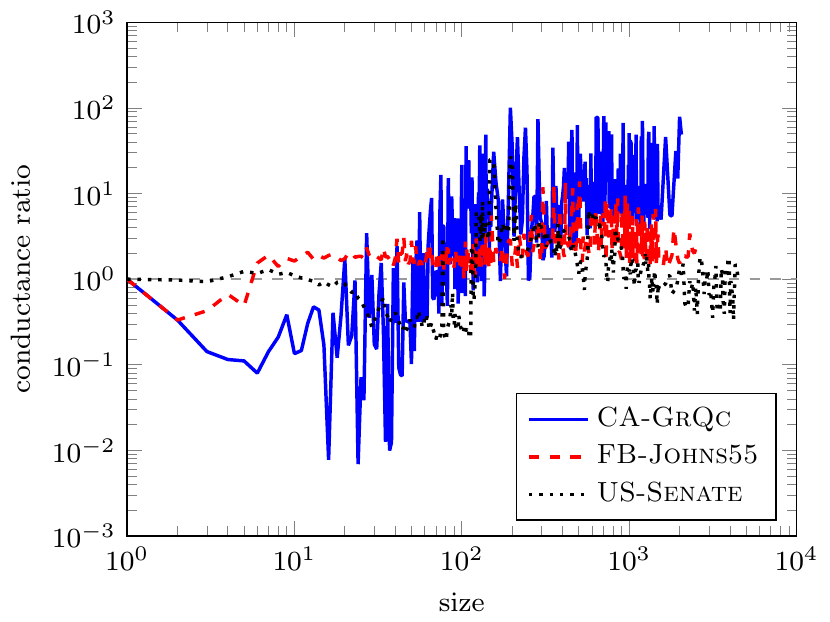}}
\caption{
NCP plots [in panel \sref{fig:NCP_EGO_small}] and CRP plots [in panel \sref{fig:condratio_EGO_small}] for \textsc{CA-GrQc}, \textsc{FB-Johns55}, and \textsc{US-Senate} (i.e., the 
smaller network in each of the three pairs of networks from 
Table\nobreakspace \ref {tab:data_summary}) using the \textsc{EgoNet} method. We find qualitatively similar behavior as with the other two methods, 
although the NCPs are shifted upwards and some of the large-scale structure is no longer present
(especially in the Facebook network). 
\label{fig:NCP-EGO-small}}
\end{figure*}

\begin{figure*}[ptb!]

\subfloat[NCP \label{fig:NCP_EGO_large}]{\includegraphics[width=.47\linewidth]{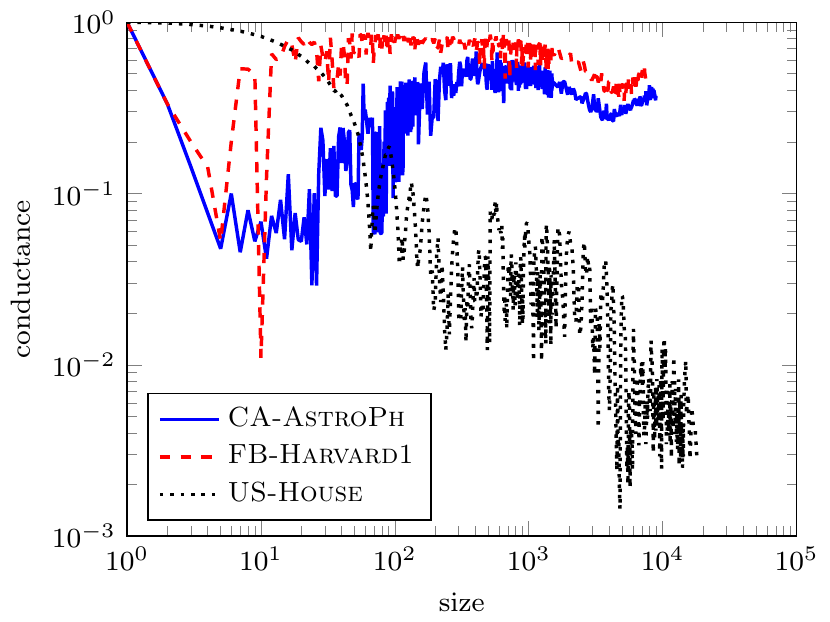}}
\hfill
\subfloat[CRP \label{fig:condratio_EGO_large}]{\includegraphics[width=.47\linewidth]{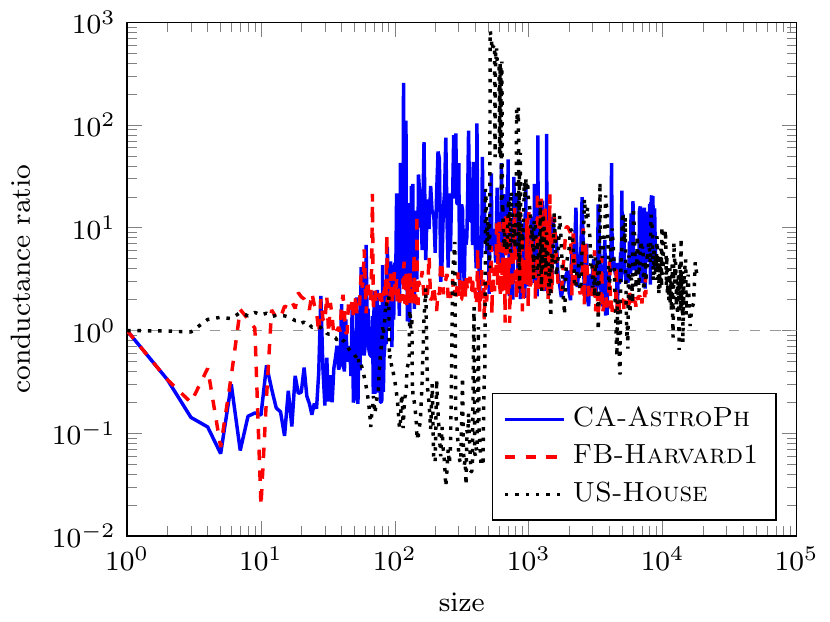}}

\caption{
NCP plots [in panel \sref{fig:NCP_EGO_large}] and CRP plots [in panel \sref{fig:condratio_EGO_large}] for \textsc{CA-AstroPh}, \textsc{FB-Harvard1}, and \textsc{US-House} (i.e., the larger networks in each of the three pairs of networks from Table\nobreakspace \ref {tab:data_summary}) using the \textsc{EgoNet} method. We find qualitatively similar behavior as with the other two methods, 
although the NCPs are shifted upwards and some of the large-scale structure is no longer present
(especially in the Facebook network). 
\label{fig:NCP-EGO-large}}
\end{figure*}

The \textsc{EgoNet} method does not have any size-scale parameters. 
For the network sizes that we consider, it is feasible to use all nodes 
rather than sampling them. 
We use this approach to generate Figs.\nobreakspace \ref {fig:NCP-EGO-small} and\nobreakspace  \ref {fig:NCP-EGO-large}. 

Finally, for readability, we only plotted the NCPs for communities that 
contain at most half of the nodes in a network. 
The symmetry in the definition of conductance 
(see Eq.\nobreakspace \textup {(\ref {eqn:conductance-set})}) implies that the complement of a good small 
community is necessarily a good large community and vice versa. 
Hence, a sampled NCP is roughly symmetric, though this is hard to see on a 
logarithmic scale, and an NCP without sampling is necessarily symmetric.


\begin{figure*}[ptb]
\subfloat[NCP\label{NCP_MOV_small}]{
\includegraphics[width=.47\linewidth]{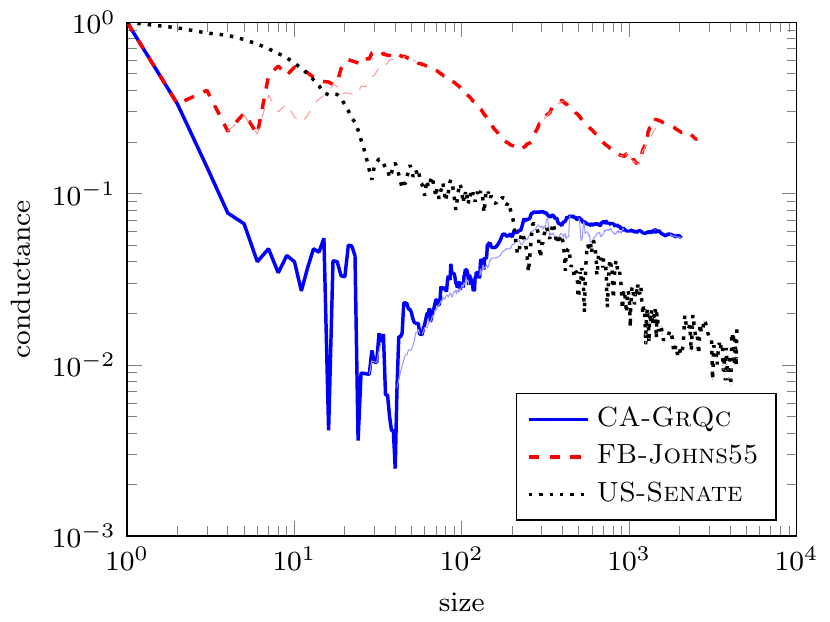}
}
\hfill
\subfloat[CRP\label{condratio_MOV_small}]{
\includegraphics[width=.47\linewidth]{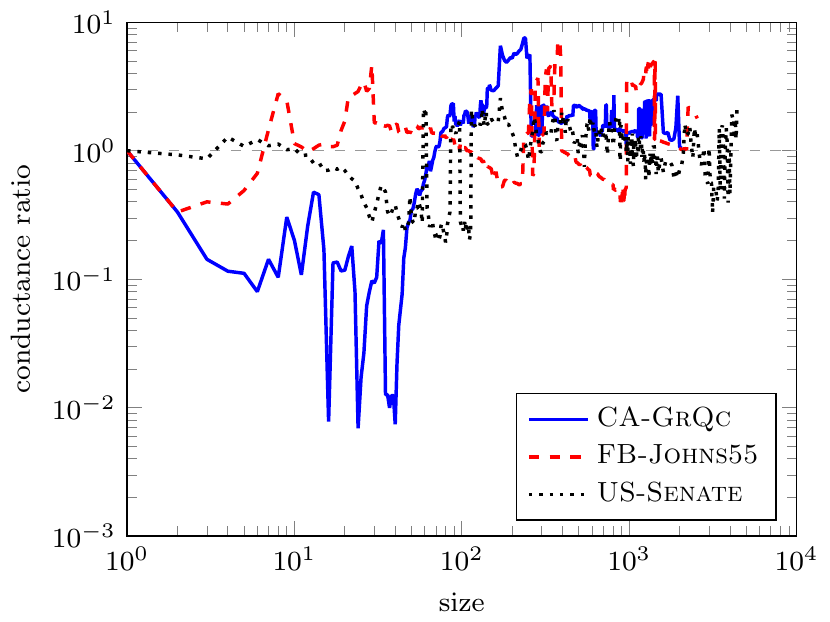}
}\\
\caption{
NCP plots [in panel \sref{NCP_MOV_small}] and CRP 
plots [in panel \sref{condratio_MOV_small}] for
\textsc{FB-Johns55}, \textsc{CA-GrQc}, and \textsc{US-Senate} (i.e., the 
smaller network in each of the three pairs of networks from 
Table\nobreakspace \ref {tab:data_summary}) generated using the \textsc{MovCut} method. The thin curves are the NCPs that we obtain when also consider disconnected sweep sets.
}
 \label{MOV_small}
\end{figure*}

\begin{figure*}[ptb]
\subfloat[NCP\label{NCP_MOV_large}]{
\includegraphics[width=.47\linewidth]{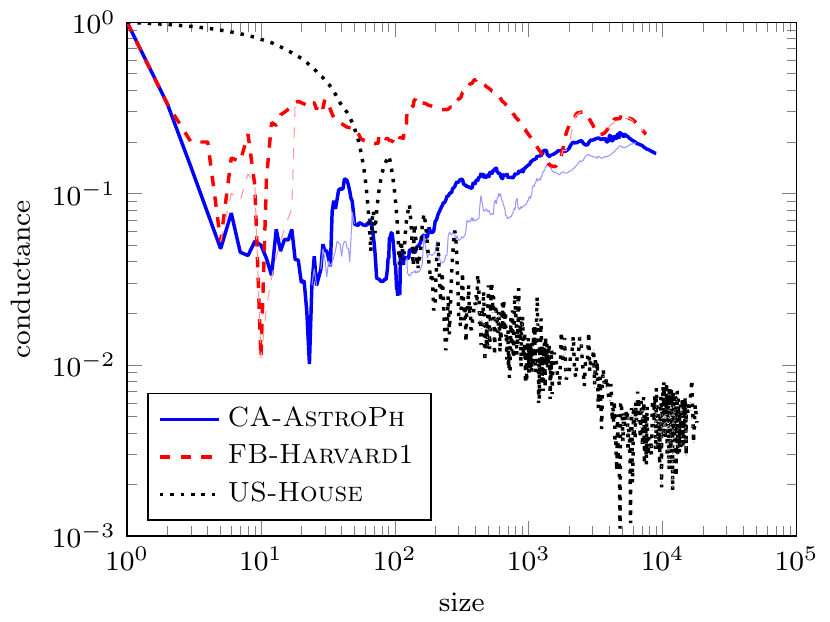}
}
\hfill
\subfloat[CRP\label{condratio_MOV_large}]{
\includegraphics[width=.47\linewidth]{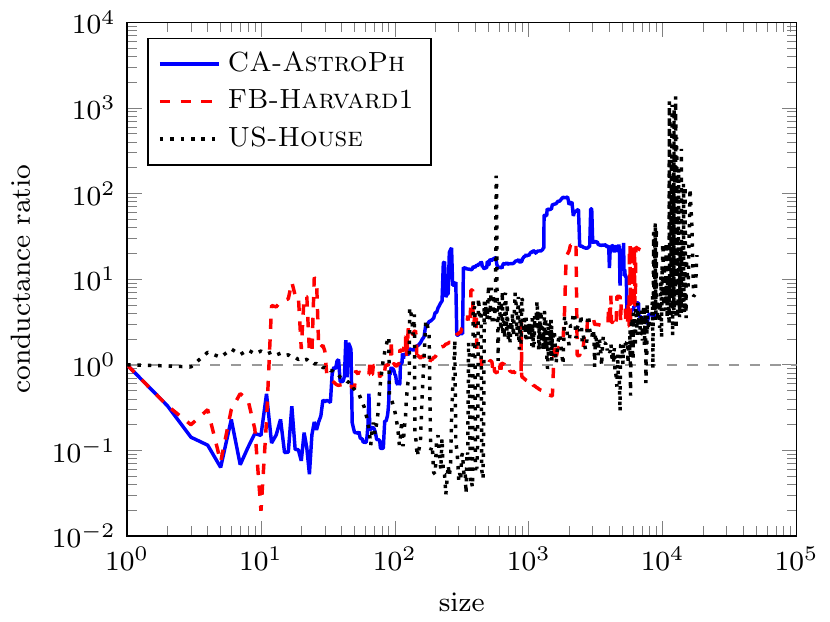}
}
\caption{
NCP plots [in panel \sref{NCP_MOV_large}] and CRP
plots [in panel \sref{condratio_MOV_large}] for
\textsc{CA-AstroPh}, \textsc{FB-Harvard1}, and \textsc{US-House} (i.e., the larger network in each of the three pairs of networks 
from Table\nobreakspace \ref {tab:data_summary}) generated using the \textsc{MovCut} method. The thin curves are the NCPs that we obtain when also consider disconnected sweep sets.
}
 \label{MOV_large}
\end{figure*}

\section{Detailed Results for the {\sc MovCut} Method}
\label{sxn:NCP-MOV}

The \textsc{MovCut} method provides an alternative way of sampling local 
community profiles to construct an NCP. 
Unlike \textsc{AclCut}, which uses \emph{only} local information to obtain 
good communities, \textsc{MovCut} also incorporates some global information 
about a network to construct local communities around a seed node. 
In particular, this implies that there can be sweep sets and thus 
communities that consist of disconnected components of a network.  
Such communities have infinitely large conductance ratios. 
We observe this phenomenon often for the coauthorship and Facebook networks, 
but it almost never occurs for the Congressional voting networks. 
Upon examination, these sweep sets consist of several small sets of 
peripheral nodes, each of which has moderate to very low conductance, but 
which are otherwise unrelated. 
Although one would not usually think of such a set of nodes as a single good 
community, optimization-based algorithms often clump several unrelated 
communities into a single community for networks with a global 
core-periphery structure.
For completeness and comparison, we include our results both when we keep 
the disconnected sweep sets and when we restrict our attention to connected 
communities. 
As we discuss below, the NCP does not change substantially, although there 
are some small differences.

The resulting NCPs for the \textsc{MovCut} method (see 
Figs.\nobreakspace \ref {NCP_MOV_small} and\nobreakspace  \ref {NCP_MOV_large}) are similar to those that we obtained 
for the \textsc{AclCut} method (see Figs.\nobreakspace \ref {NCP_ACL_small} and\nobreakspace  \ref {NCP_ACL_large}), 
although there are a few differences worth discussing.  
The CRP plots are also very similar (compare 
Figs.\nobreakspace \ref {condratio_MOV_small} and\nobreakspace  \ref {condratio_MOV_large} to 
Figs.\nobreakspace \ref {condratio_ACL_small} and\nobreakspace  \ref {condratio_ACL_large}).
For the coauthorship networks (\textsc{CA-GrQc} and \textsc{CA-AstroPh}), 
as well as \textsc{FB-Harvard1}, both \textsc{MovCut} and \textsc{AclCut} 
identify the same good small communities that are responsible for the spikes 
in the NCP plots. 
In addition, the communities that yield the dips in the NCPs for 
\textsc{FB-Johns55} near 220 and 1100 nodes, and for \textsc{FB-Harvard1} near 
1500 nodes, all share more than 98\% of their nodes. 
This indicates that both methods are able to find roughly the same 
community-like structures.  
However, the results from the \textsc{MovCut} NCP for \textsc{CA-GrQc} is 
higher and less choppy than the one that we computed using 
\textsc{AclCut}---because the truncation employed by \textsc{AclCut} 
performs a form of implicit sparsity-based regularization that is absent 
from \textsc{MovCut}.  
See Refs.~\cite{Mahoney:2010vt,Mah12,GM14_Subm} for a discussion and precise 
characterization of this regularization.
For the coauthorship and Facebook networks, we also note that there are 
regions of the computed NCPs, when using the \textsc{MovCut} method, in which 
one finds disconnected sweep sets (see the thin curves) with lower 
conductance than that for the best connected sets of the same size. 
At other sizes, we see some differences between the NCPs from 
\textsc{MovCut} and \textsc{AclCut}. 
This illustrates that the two methods can have somewhat different local 
behavior, although both methods produce similar insights regarding the 
large-scale structure in these networks. 
In Section\nobreakspace \ref {sxn:local-comp}, we discuss some of these differences 
between our results from the two methods in more detail.


\section{Detailed Results for the {\sc EgoNet} Method}
\label{sxn:NCP-EGO}

The \textsc{EgoNet} method was not originally developed to optimize 
conductance, although there is some recent evidence that $k$-neighborhoods 
can be good conductance communities \cite{Gleich:2011wv}. 
The assumption that underlies the \textsc{EgoNet} method is that nodes in the 
same community should be connected by short paths. 
However, unlike the spectral-based methods (\textsc{AclCut} and 
\textsc{MovCut}), the \textsc{EgoNet} method does not take into account the 
number of paths between nodes. 
In contrast to Ref.~\cite{Gleich:2011wv}, which considered only 
$1$-neighborhoods, here we also examine $k$-neighborhoods with $k>1$. 
We can then use this method to sample a complete NCP for a network.

Despite its simplicity, and in agreement with Ref.~\cite{Gleich:2011wv}, the 
\textsc{EgoNet} method produces NCP's that are qualitatively similar to 
those from both the \textsc{AclCut} and \textsc{MovCut} methods, for all of 
the networks that we considered; 
see Figs.\nobreakspace \ref {fig:NCP-EGO-small} and\nobreakspace  \ref {fig:NCP-EGO-large}. 
The NCPs  for the \textsc{EgoNet} method are shifted upwards compared to 
those for the \textsc{AclCut} and \textsc{MovCut} methods; and this is 
particularly noticeable at larger community size.  
This is unsurprising, because the latter two methods more aggressively 
optimize the conductance objective.
However, for all six of our networks, this method preserves an NCP's 
small-scale structure as well as the global tendency to be upward-sloping, 
flat, or downward-sloping.  
This provides further evidence that the qualitative features of an NCP 
provide a signature of community structure in a network and are not just an 
artifact of a particular way to sample communities. 
In Section\nobreakspace \ref {sxn:local-comp}, we give a more detailed comparison between 
the results of these methods.


\section*{Acknowledgements}

LGSJ acknowledges a CASE studentship award from the EPSRC (BK/10/039).  
MAP was supported by a research award (\#220020177) from the James S. 
McDonnell Foundation, the EPSRC (EP/J001759/1), and the FET-Proactive 
project PLEXMATH (FP7-ICT-2011-8; grant \#317614) funded by the European 
Commission; MAP also thanks SAMSI for supporting several visits and MWM 
for his hospitality during his sabbatical at Stanford. PJM was funded by the NSF (DMS-0645369) and by Award Number R21GM099493 from the National Institute Of General Medical Sciences. 
 MWM acknowledges funding from the Army Research Office and from the 
Defense Advanced Research Projects Agency.
The content is solely the responsibility of the authors and does not necessarily represent the official views of the funding agencies.
In addition, we thank Adam D'Angelo and Facebook for providing the 
Facebook data, Keith Poole for providing the Congressional voting 
data (which is available from Ref.~\cite{voteview}), and Jure 
Leskovec for making many large network data sets publicly available 
as part of SNAP~\cite{snap}. 


%

\end{document}